\documentclass[aps,prd,showpacs,superscriptaddress,amsmath,amssymb]{revtex4}
\usepackage{amsmath}
\usepackage{amssymb}
\usepackage{graphicx}
\usepackage[english]{babel}

\def\kt{{k_{_\perp}}}
\def\cO#1{{{\cal{O}}}\left(#1\right)}

\newcommand{\lqcd}{\Lambda_{_{_{\rm QCD}}}}
\newcommand{\dd}{{\rm d} }
\newcommand{\lmin}{\ell_{\rm min}}

\begin{document}
%%%%%%%%%%%%%%%%%%%%%%%%%%%%%%%%%%%%%%%%%%%%%%%%%%%%%%%%%%%%%%%%%%%%%%%%%%%%

\title{Next-to-MLLA corrections to single inclusive $\kt$-distributions
and 2-particle correlations in a jet}

\author{Redamy P\'erez-Ramos}
\affiliation{Max-Planck-Institut f\"ur Physik,
Werner-Heisenberg-Institut,
F\"ohringer Ring 6,
D-80805 M\"unchen (Germany)}
%\email[E-mail: ]{redamy@mppmu.mpg.de}
\affiliation{Present address: II.Institut f\"ur Theoretische Physik,
Universit\"at Hamburg,
Lupurer Chaussee 149,
D-22761 Hamburg (Germany)}
\email[E-mail: ]{redamy@mail.desy.de}
\author{Fran\c{c}ois Arleo}
\affiliation{LAPTH\footnote{Laboratoire de Physique Th\'eorique
d'Annecy-le-Vieux, UMR~5108}, 
Universit\'e de Savoie et CNRS,
9 chemin de Bellevue, BP110,
74941 Annecy-le-Vieux Cedex, France}
\email[E-mail: ]{arleo@lapp.in2p3.fr}
\author{Bruno Machet}
\affiliation{Laboratoire de Physique Th\'eorique et Hautes \'Energies
\footnote{LPTHE, 
UMR~7589 du CNRS associ\'ee \`a l'Universit\'e P. et M. Curie - Paris\,6},
UPMC Univ Paris 06, BP 126,
4 place Jussieu\\ F-75252 Paris Cedex 05 (France)}
\email[E-mail: ]{machet@lpthe.jussieu.fr}

\date{January 15th 2008; revised March 31st 2008}

\begin{abstract}
The hadronic $\kt$-spectrum inside a high energy
 jet is determined including corrections of relative magnitude
 $\cO{\sqrt{\alpha_s}}$ with respect to the Modified Leading Logarithmic
Approximation (MLLA),  in the limiting spectrum approximation
(assuming an infrared cut-off $Q_0 =\lqcd$) and beyond ($Q_0\ne\lqcd$).
The  results in the limiting spectrum approximation
are found to be, after normalization, in
impressive agreement with preliminary measurements by the CDF
collaboration, unlike what occurs at MLLA, pointing out small overall
non-perturbative contributions.
Within the same framework, 2-particle  correlations inside a jet
are also predicted at NMLLA and compared to previous MLLA calculations.
\end{abstract}


\maketitle

\setcounter{footnote}{0}
\renewcommand{\thefootnote}{\arabic{footnote}}

%%%%%%%%%%%%%%%%%%%%%%%%%%%%%%%%%%%%%%%%%%%%%%%%%%%%%%%%%%%%%%%%%%%%%%%%%%%%%
\section{Introduction}
\label{section:intro}
%%%%%%%%%%%%%%%%%%%%%%%%%%%%%%%%%%%%%%%%%%%%%%%%%%%%%%%%%%%%%%%%%%%%%%%%%%%%%

The production of jets --~a collimated bunch of hadrons~--
in $e^+e^-$, $e^-p$ and hadronic collisions is an ideal playground
to investigate the parton evolution process in perturbative QCD (pQCD).
One of the great successes of pQCD is the existence of the hump-backed
shape of inclusive spectra, predicted in~\cite{HBP} within the Modified
Leading Logarithmic Approximation (MLLA),
and later discovered experimentally~(for review, see e.g.~\cite{KhozeOchs}).
Refining the comparison of pQCD calculations
with jet data taken at LEP, Tevatron and LHC will ultimately allow for
a crucial test of the Local Parton Hadron Duality (LPHD)
hypothesis~\cite{LPHD} and for a better understanding of color
neutralization processes. 

Progress towards this goal has been achieved recently.
On the theory side, the inclusive $\kt$-distribution of particles
inside a jet has been computed at MLLA accuracy~\cite{PerezMachet},
as well as  correlations between two particles in a jet~\cite{RPR2}.
Analytic calculations have first been done in the limiting spectrum
approximation, {\it i.e.} assuming an infrared cutoff $Q_0$ equal to
 $\lqcd$ ($\lambda\equiv\ln Q_0/\lqcd= 0$).
Subsequently, analytic approximations for correlations
 were obtained beyond the limiting
spectrum using the steepest descent method~\cite{RPR3}.
Experimentally, the CDF collaboration at Tevatron reported on
$\kt$-distributions of unidentified hadrons in jets produced in
$p\bar{p}$ collisions at $\sqrt{s}=1.96$~TeV~\cite{CDF}.

MLLA corrections, of relative magnitude $\cO{\sqrt{\alpha_s}}$ with respect
to the leading double logarithmic approximation (DLA), were shown to be quite
substantial for single-inclusive distributions and 2-particle correlations
~\cite{PerezMachet,RPR2}. Therefore, it appears legitimate
to wonder whether corrections of order $\cO{\alpha_s}$, that is
next-to-next-to-leading or next-to-MLLA (NMLLA), are negligible or not.

The starting point of this analysis is the MLLA evolution equation for
the generating functional of QCD jets \cite{Basics}.
Together with the initial condition at threshold, it determines
jet properties at all energies. At high energies one can represent
the solution as an expansion in $\sqrt{\alpha_s}$. Then,
the leading (DLA) and next-to-leading (MLLA) approximations are complete.
The next terms (NMLLA) are not complete but they include an important
contribution which takes into account energy conservation and an
improved behavior near threshold.
An example of a solution for
the single inclusive spectrum from the MLLA equation is the so-called
``limiting spectrum'' (for a review, see \cite{Basics})
which represents a perturbative computation
of the spectrum at $\lambda=0$ with complete leading and
next-to-leading asymptotics. Some results for such NMLLA terms have been
studied previously for global observables and have been found to
better account for recoil effects. They were shown to
drastically affect multiplicities and particle correlations in jets:
this is in particular the case in  \cite{CuypersTesima}, which 
deals with multiplicity correlators of order 2, and in \cite{DokKNO},
where  multiplicity correlators involving a higher number of partons are
studied; in particular, the higher
this number, the larger turn out to be  NMLLA corrections.

The present study makes use of this evolution equation to estimate NMLLA
contributions to our differential observables.
It presents the complete calculations of the single
inclusive $k_\perp$-distribution leading to the main results
published in \cite{PRL}, and extends them to 2-particle correlations
inside a high energy jet.

The paper is organized as follows.
First, Section~\ref{section:sis} presents
a system of evolution equations including $\cO{\alpha_s}$ corrections,
which allows for the computation of the inclusive spectrum, $G$,
beyond MLLA accuracy.
Section \ref{section:kperp} is devoted to the NMLLA evaluation of
the color currents of quark and gluon jets and, from them, to the
inclusive $k_\perp$-distribution in the limiting spectrum approximation.
These predictions are also compared to preliminary measurements performed
recently by the CDF collaboration.
Going beyond the limiting spectrum is the subject of
Section~\ref{section:numer}, in which inclusive $\kt$-distributions
are computed at an arbitrary $\lambda$.
The 2-particle correlations including NMLLA corrections are determined
in Section \ref{section:TPC}.
Finally, the present approach and the results obtained in this paper are
discussed in detail and summarized in Section~\ref{sec:CONCL}.

%%%%%%%%%%%%%%%%%%%%%%%%%%%%%%%%%%%%%%%%%%%%%%%%%%%%%%%%%%%%%%%%%%%%%%%
\section{Evolution equations}
\label{section:sis}
%%%%%%%%%%%%%%%%%%%%%%%%%%%%%%%%%%%%%%%%%%%%%%%%%%%%%%%%%%%%%%%%%%%%%%%

\subsection{Logic and energy conservation}
\label{subsection:enercons}
%%%%%%%%%%%%%%%%%%%%%%%%%%%%%%%%%%%%%%%%%%

As a consequence of the probabilistic shower picture, 
the notion of {\em Generating Functional} (GF) was proved suitable
to understand and include higher order 
corrections to DLA asymptotics 
(see \cite{Basics} and references therein). 

The single inclusive spectrum and the $n$-particle momentum
correlations can be derived from the
MLLA Master Equation for the GF $Z=Z(u)$ \cite{Basics}  after successively  
differentiating with respect to a certain {\em probing} function $u=u(k)$;
$k$ denotes the quadri-momentum of one parton inside the shower
and the solution of the equations are written as a perturbative
expansion in $\alpha_s$.
At high energies this expansion can be resummed and the leading contribution
be represented as an exponential of the anomalous dimension 
$\gamma(\alpha_s)$; since further details to this logic
can be found in \cite{Basics,RPR2}, we only
give the symbolic structure of the equation for the GF 
and its solution as
\begin{equation}\label{eq:symbGF}
\frac{dZ}{dy}\simeq\gamma_0(y)Z\quad\Rightarrow\quad
Z\simeq\exp{\left\{\int^y\gamma(\alpha_s(y'))dy'\right\}}
\end{equation}  
where $\gamma(\alpha_s)$ can be written as an expansion in 
powers of $\sqrt\alpha_s$
\begin{equation}\label{eq:anomdim}
\gamma(\alpha_s)=\,\sqrt\alpha_s+\,\alpha_s+\,\alpha_s^{3/2}
+\,\alpha_s^2+\ldots
\end{equation}
The equation in (\ref{eq:symbGF}) applies to each vertex
of the cascade and its solution represents the fact that 
successive and independent partonic 
splittings inside the shower, such as the one displayed in Fig.~\ref{fig:spplit}, exponentiate with respect to the {\em evolution-time} parameter 
$dy=d\Theta/\Theta$; $\Theta\ll1$ is the angle
between outgoing couples of partons. The choice of $y$
follows from Angular Ordering (AO) in intrajet cascades;
it is indeed the suited variable for describing {\em time-like}
evolution in jets. Thus, Eq.~(\ref{eq:symbGF}) incorporates the Markov chains
of sequential angular ordered partonic decays which are singular in $\Theta$
and $\gamma(\alpha_s)$ determines the rate of inclusive quantities
growth with energy.

While DLA treats the emission of both particles as independent
by keeping track of the first term $\sim\sqrt\alpha_s$
in (\ref{eq:anomdim}) without constraint,
the exact solution of the MLLA evolution equation (partially) fulfills
the energy 
conservation in each individual splitting process ($z+(1-z)=1$) by
incorporating higher order ($\alpha_s^{n/2},n>1$) terms to the anomalous 
dimension. Symbolically; the first two analytical steps towards a better
account of these corrections in the MLLA, NMLLA evolution,
which we further discuss in \ref{subsection:logic}, 
can be represented in the form
$$
\Delta\gamma\simeq\int(\alpha_s+\alpha_s\ell^{-1}\ln z)dz\sim\alpha_s+
\alpha_s^{3/2},
$$
where $\ell=\ln(1/x)\sim\alpha_s^{-1/2}$ with $x\ll1$ (fraction of the
jet energy taken away by one hadron), $z\sim1$ for
hard partons splittings such as $g\to q\bar q$\ldots (this is
in fact the region where the two partons are strongly
correlated).

Energy conservation is particularly important for energetic particles
as the remaining phase space is then very limited.
On the other hand, a soft particle can be emitted with little impact on 
energy conservation. Some consequences of this behavior have also been
noted in \cite{KLO}:

\begin{itemize}
\item[$(i)$] the soft particles follow the features expected from DLA;
\item[$(ii)$] there is no energy dependence of the soft spectrum;
\item[$(iii)$] the ratio of soft particles $r=N_g/N_q$
in gluon and quark jets is consistent with the DLA prediction $N_c/C_F=9/4$
(see the measurement by DELPHI \cite{lepdelphi}).
\end{itemize}
 
This is quite different from the ratio of global 
multiplicities which acquires large corrections beyond DLA
(see, for example Fig.18 in the second reference given in \cite{Dremin2}). 
For this quantity the HERWIG parton shower model corresponding to 
MLLA and exact energy conservation (same Fig.~18) and the full
summation of the perturbative series of MLLA evolution equation
(see also \cite{LO})
come close to the data at $r = N_g/N_q \approx 1.5$ at LEP energies.
As an intermediate example, we can mention the successful description of the 
semi-soft particle $\ln(1/x)$ distribution (``hump-backed plateau'') where 
the first correction (MLLA), despite the large value of the expansion parameter
$\sqrt\alpha_s\approx0.35$,
already gives a good description of the data at the $Z^0$ peak 
($Q=91.2$ GeV) of the $e^{+}e^{-}$ annihilation into a
$q\bar q$ pair \cite{OPALZ0}.

\subsection{MLLA evolution}
\label{subsec:eveq}
%%%%%%%%%%%%%%%%%%%%%%%%%%%%

\begin{figure}[ht]
\begin{center}
  \includegraphics[height=4.2cm]{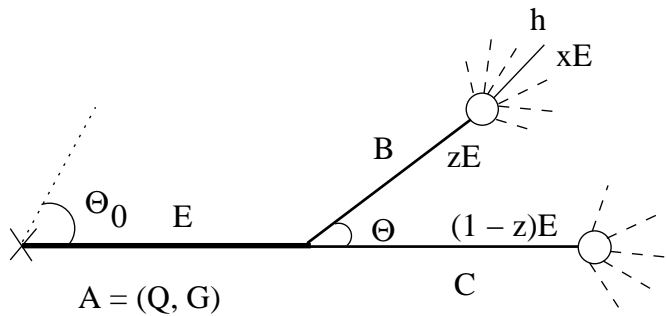}
\end{center}
\caption{Parton {\tt A} with energy $E$ splits into parton {\tt B}
(respectively, {\tt C}) with energy $zE$ (respectively, $(1-z)E$)
which fragments into a hadron $h$ with energy $xE$.}
\label{fig:spplit}
\end{figure}

We study the formation of hadrons inside a jet produced in
high-energy scattering processes, such as $e^+e^-$ annihilation
or $pp$ and $p\bar{p}$ collisions. A jet of total opening angle
$\Theta_0$ is initiated by a parton {\tt A} (either a quark, $Q$,
or a gluon, $G$) with energy $E$; {\tt A} then splits into partons {\tt B}
and {\tt C}, with energy fractions $z$ and $(1-z)$ respectively,
forming a relative angle $\Theta$ (see  Fig.~\ref{fig:spplit}).
At the end of the cascading process, the parton {\tt B} fragments
into a hadron $h$ with energy $x E$, with the fragmentation
function
\begin{equation}\label{eq:dfunction}
 B(z)=\frac{x}{z}D_{\tt B}^{h}\left(\frac{x}{z}, 
zE\Theta_0,Q_0\right),\qquad (B = Q, G)
\end{equation}
which describes the distribution of the hadron $h$ inside the sub-jet {\tt B}
with an energy-fraction $x/z$.
As a consequence of AO in parton cascades, the functions
$Q(z)$ and $G(z)$ satisfy the system of two-coupled integro-differential
 evolution equations \cite{RPR2}:
\begin{eqnarray}
 Q_{y}\equiv \frac{d Q}{dy} &=&\ \int_0^1 \dd{z}\>  \frac{\alpha_s}{\pi} \>\Phi_q^g(z)\>
\bigg[ \Big( Q(1-z)-Q\Big) +  G(z) \bigg],
\label{eq:qpr}\\
 G_{y}\equiv \frac{d G}
{dy} &=&\ \int_0^1 \dd{z}\> \frac{\alpha_s}{\pi} \>
\bigg[\Phi_g^g(z)(1-z) \Big( G(z) + G(1-z) - G\Big)\nonumber\\
&&\qquad\qquad\qquad\qquad\qquad +\ n_f\ \Phi_g^q(z)\ \Big(2 Q(z)- G\Big) \bigg],
\label{eq:gpr}
\end{eqnarray}
with $\alpha_s$, the running coupling constant of QCD, given by
\begin{equation}\label{eq:alphas}
\alpha_s\equiv\alpha_s(\ell,y) = \frac{2\pi}{4N_c\beta_0
(\ell+y+\lambda)},
\end{equation}
and where we define
\begin{equation}\label{eq:var}
\qquad \ell = \ln\left(1/x\right),\quad
 y = \ln\frac\kt{Q_0}= \ln\frac{xE\Theta_0}{Q_0}, \quad
 \lambda=\ln\frac{Q_0}{\lqcd},
\end{equation}
following the notations of Ref.~\cite{Basics}; the MLLA equations
above follow from the GF logic commented in the introductory
paragraph. The scale $Q_0$ appearing in (\ref{eq:var}) is the collinear cut-off
parameter, $\lqcd$ is the non-perturbative scale of QCD which we set
to 250 MeV in this work~\footnote{NMLLA 
corrections arising from varying $\lqcd$ are studied
in detail in subsection \ref{subsec:uncertain}.}
, and
\begin{equation}
\beta_0=\frac1{4N_c}\left(\frac{11}3N_c-\frac43T_R\right)
\label{eq:beta0}
\end{equation}
is the first term in the perturbative expansion of the $\beta$-function
($N_c$ is the number of colors, $T_R=n_f/2$ where $n_f=3$ is the number of
light quark flavors).
We only consider in this work the 1-loop expression for the running
coupling constant, assuming that the role of the conservation of energy is
much more important than the effects of 2-loop corrections to $\alpha_s$,
as seen for instance in the case of multiplicity
distributions~\cite{Dremin2};
we shall discuss this further
in Section~\ref{se:discussion}. The coupling constant $\alpha_s$ is
also linked to the DLA anomalous dimension $\gamma_0$ of twist-2
operators by
\begin{equation}\label{eq:anodim}
\gamma_0^2(\ell,y)\ =\ 2N_c\ \frac{\alpha_s(\ell,y)}{\pi}\ =\
\frac1{\beta_0(Y_{\Theta}+\lambda)},\qquad
Y_{\Theta}\ =\ \ell+y\ =\ \ln\frac{E\Theta}{Q_0}.
\end{equation}
In Eqs.~(\ref{eq:qpr}) and (\ref{eq:gpr}), $\Phi_{\tt A}^{\tt B}(z)$
represent the one-loop DGLAP splitting functions \cite{Basics} and we note:
\begin{equation*}\label{eq:GQ1}
Q\equiv
Q(1) = xD_{q}^{h}(x,E\Theta_0,Q_0),\quad G\equiv  G(1) = xD_{g}^{h}(x,E\Theta_0,Q_0).
\end{equation*} 
In the small $x\ll z$ limit which we consider here, the fragmentation
functions behave as
\begin{equation}\label{eq:DBh}
 B(z)\ \stackrel{x\ll z}{\approx}\
\rho_{\tt B}^{h} \left(\ln\frac{z}{x}, \ln\frac{zE\Theta_0}{Q_0}\right)\
=\ \rho_{\tt B}^{h}\left(\ln z + \ell, y\right),
\end{equation}
where $\rho_{\tt B}^{h}$ is a slowly
varying function of the two logarithmic variables $\ln (z/x)$ and
$y$~\cite{HBP} that describes the hump-backed plateau.

\subsection{Taylor expansion}
\label{subsection:logic}
%%%%%%%%%%%%%%%%%%%%%%%%%%%%%%

The resummation scheme at MLLA is discussed in \cite{RPR2},
in which $G(z)$ and $G(1-z)$ were replaced by $G(1)$
in the non-singular part of the integrands in  Eqs.~(\ref{eq:qpr})
and (\ref{eq:gpr}).
In the present work, we calculate next-to-MLLA (NMLLA) corrections
from the Taylor expansion of $\rho_{\tt B}^{h}$ in the variables
$\ln z$ and $\ln(1-z)$ in the domain:
\begin{equation*}\label{eq:hardregion}
z \sim 1-z \sim 1,\qquad x \ll 1 \Rightarrow \ell \gg |\ln z| \sim |\ln (1-z)|
\end{equation*}
corresponding to hard parton splittings.  To first order,
\begin{eqnarray}
\label{eq:logic}
\rho(\ln z)&=& \rho(\ln z =0)
+ \frac{\partial \rho(\ln z)}{\partial \ln z}\Big |_{\ln z=0} \ln z + \cO{\ln^2z},\\
\rho(\ln (1-z))&=& \rho\big(\ln (1-z) =0\big)
+ \frac{\partial \rho(\ln (1-z))}{\partial \ln (1-z)}\Big |_{\ln (1-z)=0}
\ln (1-z) + \cO{\ln^2(1-z)},\nonumber\\
&&
\end{eqnarray}
or, equivalently, for the function $B(z)$:
\begin{eqnarray}\label{eq:logic1}
B(z) &\stackrel {|\ln z| \ll \ell}{\approx}&
B(1) + B_\ell(1) \ln z + \cO{\ln^2z},\\
B(1-z) &\stackrel {|\ln (1-z)| \ll \ell}{\approx}&
B(1) + B_\ell(1) \ln(1-z) + \cO{\ln^2(1-z)}.
\end{eqnarray}
The derivative with respect to $\ln z$ or $\ln(1-z)$ has been replaced by
the one with respect to $\ell$  because of (\ref{eq:DBh})
 and the property that, at low $x$, $B$ is a function of $(\ln z + \ell)$ or
$(\ln (1-z) + \ell)$.
Since $\ell=\cO{1/\sqrt{\alpha_s}}$ (see \cite{Basics})
the above expansion can be written symbolically    
$$
B\left(z\right)\sim B\left(1-z\right)\simeq
c_1+c_2(\sqrt{\alpha_s})+{\cal O}(\alpha_s), \qquad c_1,c_2 ={\cal O}(1).
$$
The terms proportional to $B_\ell$  thus provide NMLLA corrections to the solutions
of the MLLA evolution equations  (\ref{eq:qpr}) and (\ref{eq:gpr}).

\subsection{Evolution equations including NMLLA corrections}
%%%%%%%%%%%%%%%%%%%%%%%%%%%%%%%%%%%%%%%%%%%%%%%%%%%%%%%%%%%%%%%%%%%%

\subsubsection{Quark jet}
%%%%%%%%%%%%%%%%%%%%%%%%%

In order to determine NMLLA corrections to the evolution equation
(\ref{eq:qpr}), the 1-loop splitting functions (see \cite{Basics})
are written
$$
\Phi_q^g(z)=C_F\left(\frac2z+\phi_q^g(z)\right), \qquad
(1-z)\Phi_g^g(z)= 2N_c\left(\frac1z+\phi_g^g(z)\right),
$$
where $\phi_q^g(z)=(z-2)$ and $\phi_g^g(z)=(z-1)\left(2-z(1-z)\right)$
are regular functions of $z$.
The term proportional to $G(z)$ in the integrand of (\ref{eq:qpr}) becomes
\begin{equation}\label{eq:nmlla1}
\int_0^1 \dd{z}\>  \frac{\alpha_s}{\pi} \>\Phi_q^g(z)\;G(z)=2C_F\int_0^1
\frac{\dd{z}}z\>\frac{\alpha_s}{\pi}\>G(z)
+ C_F\int_0^1 \dd{z}\>\frac{\alpha_s}{\pi}\ \phi_q^g(z)\ G(z),
\end{equation}
the second part of which is expanded according to (\ref{eq:logic1}).
Replacing  $\alpha_s/\pi=\gamma_0^2/2N_c$ (see \ref{eq:anodim}), one gets
\begin{eqnarray}
\int_0^1 \dd{z}\frac{\alpha_s}{\pi}\Phi_q^g(z)\ G(z)&\approx&
\frac{C_F}{N_c}\left[\left(\int_0^1 \frac{\dd{z}}z\gamma_0^2 G(z)\right)
-\frac{3}{4}\gamma_0^2\ G + \frac78 \gamma_0^2\ G_\ell+\ldots\right]\!,
\label{eq:nmlla2}
\end{eqnarray}
where $G_\ell \equiv G_\ell(1)$ and $Q_\ell \equiv
Q_\ell(1)$. The first integral in the r.h.s of (\ref{eq:nmlla2})
provides the DLA (leading) term as $z\to0$,
while the second and third terms correspond to higher powers of
$\sqrt{\alpha_s}$, that is MLLA and NMLLA corrections respectively.
The  $z$-dependence of $\alpha_s$ in (\ref{eq:nmlla2}) has only been taken
into account in the singular (DLA) part dominated by small $z$.
On the contrary, for the non-singular parts corresponding 
to branching processes in which $z \sim 1-z = \cO{1}$, $\alpha_s$
has been taken out of the $z$ integral~\footnote{Furthermore, 
if one formally expands $\alpha_s =
2\pi/[4N_c\beta_0(\ln z + \ldots)]$ in powers of $\ln
z$, the resulting first non-leading term
$\propto 1/(\ln z + \ldots)^2$
yields an extra $\alpha_s^2$ which only starts contributing at
next-to-next-to-MLLA, out of the scope of the present work.}
as done in \cite{RPR2}.
The dependence on the other variables, $\kt$, $\Theta$, is of course
unchanged.

Likewise, the term proportional to $Q(1-z)-Q$ in (\ref{eq:qpr}) 
can be expanded according to (\ref{eq:logic1}), leading to
\begin{eqnarray}
\int_0^1 \dd{z}\ \frac{\alpha_s}{\pi}\ \Phi_q^g(z)\ 
\Big(Q(1-z)-Q\Big)&\approx&
\int_0^1 \dd{z}\ \frac{\alpha_s}{\pi}
Q_\ell\ \Phi_q^g(z)\ \ln(1-z)\nonumber  \\
&\approx&\left(\frac{C_F}{N_c}\right)^2
\gamma_0^2\left(\frac58-\frac{\pi^2}6\right)G_\ell.\label{eq:nmlla3}
\end{eqnarray}
In the second line of (\ref{eq:nmlla3}), we have used the
approximated formula $Q_\ell\approx C_F/N_c\ G_\ell+{\cal O}(\gamma_0^2)$
that holds at DLA because subleading terms would give
${\cal O}(\gamma_0^4)$ corrections which
are beyond NMLLA (see also appendix \ref{section:Q1}).
Finally, plugging (\ref{eq:nmlla2}) and (\ref{eq:nmlla3})
into (\ref{eq:qpr}), we obtain
\begin{eqnarray}
Q_y=
\frac{C_F}{N_c}
\left\{\left(\int_0^1\frac{\dd{z}}z\gamma_0^2G(z)\right) - \frac34\gamma_0^2G
+\left[\frac78+\frac{C_F}{N_c}\left(\frac58-\frac{\pi^2}6
\right)\right]\gamma_0^2G_\ell
\right\},
\label{eq:nmllaquark}
\end{eqnarray}
where the term proportional to $\gamma_0^2G_\ell={\cal O}(\gamma_0^3)$
constitutes the new NMLLA correction. It is quite sizable and should be
taken into account in the coming calculations.

\subsubsection{Gluon jet}
%%%%%%%%%%%%%%%%%%%%%%%%%

Along similar steps, we now evaluate NMLLA corrections to Eq.~(\ref{eq:gpr}). The first term in the integral can be cast in the form
\begin{eqnarray}
\label{eq:nmllagg}
&& \int_0^1 \dd{z}\> \frac{\alpha_s}{\pi}\ 
\Phi_g^g(z)(1-z) \Big(G(z) + G(1-z) - G\Big)\cr
&&\approx\left(\int_0^1\frac{\dd{z}}z\gamma_0^2G(z)\right)
-\frac{11}{12}\gamma_0^2G+\left(\frac{67}{36}-\frac{\pi^2}{6}\right)
\gamma_0^2G_\ell,
\end{eqnarray}
and the second  into
\begin{eqnarray}
n_f\int_0^1 \dd{z}\> \frac{\alpha_s}{\pi}
\; \Phi_g^q(z)\, \Big(2Q(z)-G\Big) 
\approx\frac23\frac{n_fT_R}{2N_c}\gamma_0^2\left(2Q-G\right)
-\frac{13}{18}\frac{n_fT_R}{N_c}\gamma_0^2Q_\ell.
\label{eq:nmllagq}
\end{eqnarray}
Summing (\ref{eq:nmllagg}) and (\ref{eq:nmllagq}),
replacing like before $Q$ by its DLA formula $Q\approx C_F/N_c\ G$ 
(see appendix \ref{section:Q1} for further details),
the evolution equation for particle spectra inside a gluon jet reads
\begin{eqnarray}\label{eq:nmllagluon}
G_y&=& \left(\int_0^1\frac{\dd{z}}z\gamma_0^2G(z)\right)
-\left[\frac{11}{12}+\frac{n_fT_R}{3N_c}\left(1-2\ \frac{C_F}{N_c}\right)\right]
\gamma_0^2\ G\nonumber\\
&&\qquad\qquad\qquad\qquad+\left(\frac{67}{36}-\frac{\pi^2}6
-\frac{13}{18}\ \frac{n_fT_R}{N_c}\ \frac{C_F}{N_c}\right)
\gamma_0^2\ G_\ell.
\end{eqnarray}
The first term in parenthesis in (\ref{eq:nmllaquark}) and
(\ref{eq:nmllagluon}) is, as stressed before, the main
(double logarithmic) contribution.
According to the Low-Barnett-Kroll theorem \cite{LBK}, the  $\dd{z}/z$
term, which is of classical origin, is universal, that is, independent
of the process and of the partonic quantum numbers. The other two
(single logarithmic) contributions, which arise from hard parton splitting,
are quantum corrections. It should also be noticed that, despite the large
size of NMLLA corrections coming from
$g\to gg$ and $g\to q\bar q$ splittings, a large cancellation occurs
in their sum (\ref{eq:nmllagluon}).
The coefficients of the terms proportional to $G_\ell$ in
(\ref{eq:nmllaquark}) and in
(\ref{eq:nmllagluon}) are in agreement with \cite{DreminNechitailo}.

\subsubsection{NMLLA system of evolution equations}
\label{subsub:eveq}
%%%%%%%%%%%%%%%%%%%%%%%%%%%%%%%%%%%%%%%%%%%%%%%%%%%%%%%

Once written in terms of $\ell'=\ln(z/x)$ and $y'=\ln{(xE\Theta/Q_0)}$, the system of two-coupled evolution equations
(\ref{eq:nmllaquark}) and (\ref{eq:nmllagluon}) finally reads, 
\begin{eqnarray}
Q(\ell,y)=\delta(\ell)+\frac{C_F}{N_c}
\int_0^{\ell} \dd\ell'\int_0^{y} \dd y' \gamma_0^2(\ell'+y')\Big[
1-\tilde a_1 \delta(\ell'-\ell) + \tilde a_2\delta(\ell'-\ell)
\psi_\ell(\ell',y')  \Big]G(\ell',y'),\nonumber\\ &&
\label{eq:solq}
\end{eqnarray}
\begin{eqnarray}
G(\ell,y) = \delta(\ell)
+\int_0^{\ell} \dd\ell'\int_0^{y} \dd y' \gamma_0^2(\ell'+y')\Big[
 1 - a_1\delta(\ell'-\ell) + a_2\delta(\ell'-\ell)\psi_\ell(\ell',y')\Big] G(\ell',y'),\nonumber\\ &&
\label{eq:solg}
\end{eqnarray}
with $\psi_\ell \equiv G_\ell/G$ and the MLLA and
NMLLA coefficients~\footnote{In
 practice, these coefficients can be safely estimated by using
the leading order formula $Q/G = C_f/N_c$ instead of (\ref{eq:ratioqg})
below. We checked that this approximation only marginally affects their
values.} given by:
\begin{subequations}
\begin{eqnarray}
\tilde a_1&=&\frac34,\label{eq:tildea1}\\
a_1 &=&\frac{11}{12}+\frac{n_fT_R}{3N_c}\left(1-2\frac{C_F}{N_c}\right)
\stackrel{n_f=3}{\approx}0.935,\label{eq:a1}\\
\tilde a_2&=&\frac78+\frac{C_F}{N_c}\left(\frac58-\frac{\pi^2}6
\right)\approx0.42,\label{eq:tildea2}\\
a_2&=&\frac{67}{36}-\frac{\pi^2}6-\frac{13}{18}\frac{n_fT_R}{N_c}\frac{C_F}{N_c}
\stackrel{n_f=3}{\approx}0.06.\label{eq:a2}
\end{eqnarray}
\end{subequations}
As can be seen, the NMLLA coefficient $a_2$ is very small
This may explain {\it a posteriori} why the MLLA ``hump-backed plateau''
agrees very well with experimental data~\cite{HBP,DFK}.
Therefore, the NMLLA solution of (\ref{eq:solg}) can be approximated
by the MLLA solution of $G$ ({\it i.e.} taking $a_2=0$), which will be used
in the following to compute the inclusive $\kt$-distribution as well
as two-particle correlations inside a jet~\footnote{Finding the 
analytical solution of these equations
including NMLLA corrections is beyond the scope of this paper.}.
The MLLA gluon inclusive spectrum is given by \cite{Basics}:
\begin{equation}
G(\ell,y) = 2\ \frac{\Gamma(B)}{\beta_0}\
 \int_0^\frac{\pi}{2}\
  \frac{\dd\tau}{\pi}\,e^{-B\alpha}
\  {\cal F}_B(\tau,y,\ell),
\label{eq:ifD}
\end{equation}
where the integration
is performed with respect to $\tau$ defined by
$\displaystyle \alpha = \frac{1}{2}\ln\frac{y}{\ell}  + i\tau$ and with
\begin{eqnarray*}
{\cal F}_B(\tau,y,\ell) &=& \left[ \frac{\cosh\alpha
-\displaystyle{\frac{y-\ell}{y+\ell}}
\sinh\alpha} 
 {\displaystyle \frac{\ell +
y}{\beta_0}\,\frac{\alpha}{\sinh\alpha}} \right]^{B/2}
  I_B(2\sqrt{Z(\tau,y,\ell)}), \cr
&& \cr
&& \cr
 Z(\tau,y,\ell) &=&
\frac{\ell + y}{\beta_0}\,
\frac{\alpha}{\sinh\alpha}\,
 \left(\cosh\alpha
%+ (1-2\zeta)
-\frac{y-\ell}{y+\ell}
\sinh\alpha\right),
\label{eq:calFdef}
\end{eqnarray*}
$B=a_1/\beta_0$ and $I_B$ is the modified Bessel function of the first kind.
To get a quantitative idea on the difference between MLLA and NMLLA
gluon inclusive spectrum, the reader is reported
to Appendix~\ref{section:roleofnmlla} where a simplified
NMLLA equation (\ref{eq:solg}) with a frozen coupling constant is solved.
The magnitude of $\tilde a_2$, however, indicates that the NMLLA
corrections to the inclusive quark jet spectrum may not be negligible
and should be taken into account. 
After solving (\ref{eq:solg}), the solution of (\ref{eq:solq}) reads
\begin{equation}\label{eq:ratioqg}
Q(\ell,y)=\frac{C_F}{N_c}\left[G(\ell,y)
+\Big(a_1-\tilde a_1\Big)G_\ell(\ell,y)+\left
(a_1\Big(a_1-\tilde a_1\Big)+\tilde a_2-a_2\right)G_{\ell\ell}(\ell,y)
\right]+{\cal O}(\gamma_0^2).
\end{equation}
It differs from the MLLA expression given in \cite{PerezMachet} by the term
proportional to $G_{\ell\ell}$, which can be deduced from the subtraction of 
$(C_F/N_c)\times$(\ref{eq:solg}) to Eq.~(\ref{eq:solq}).

%%%%%%%%%%%%%%%%%%%%%%%%%%%%%%%%%%%%%%%%%%%%%%%%%%%%%%%%%%%%%%
\section{Single-inclusive $\kt$-distribution in the limiting spectrum}
\label{section:kperp}
%%%%%%%%%%%%%%%%%%%%%%%%%%%%%%%%%%%%%%%%%%%%%%%%%%%%%%%%%%%%%%

While MLLA calculations show that, asymptotically, the shape of the 
inclusive spectrum becomes independent of $\lambda$
 \cite{Basics,finitelambda},
setting the infrared cutoff $Q_0$ of cascading processes as low as
the intrinsic QCD scale $\lqcd$ is a daring
hypothesis, since it is tantamount to assuming that a perturbative
treatment can be trusted in regions of large running $\alpha_s$.
However, it turns out that,
experimentally, this shape is  very well described by
$\lambda=0$.  We shall show below that this remarkable property
is also true for  the single-inclusive $k_\perp$-distribution.
This will be further confirmed in section \ref{section:numer}
in which non-vanishing values of $\lambda$ are considered.

\subsection{Double-differential distribution}
\label{subsec:dd1p}
%%%%%%%%%%%%%%%%%%%%%%%%%%%%%%%%%%%%%%%%%%%%%

The double differential distribution $\dd^2N/(\dd x\, \dd\ln\theta)$
for the production of a single hadron $h$ at angle $\Theta$ in a high
energy jet of total energy $E$ and opening angle $\Theta_0\geq\Theta$,
carrying the energy fraction $x$, is obtained by integrating
the inclusive double differential 2-particle cross 
section (see~\cite{PerezMachet})~\footnote{The 
basic process under consideration is accordingly
 the emission of two
hadrons, $h_1$ and $h_2$, inside a jet produced in a high-energy collision.
Since the notations and kinematics are identical to the ones used in
\cite{PerezMachet}, the reader is referred to this work for a more detailed
description.}. Then,
the single-inclusive $k_\perp$-distribution of hadrons inside a jet is
obtained by integrating 
$\dd^2N/(\dd{x}\,\dd\!\ln\theta)$ over all energy-fractions $x$:
\begin{equation}
\left(\frac{\dd{N}}{\dd\ln k_\perp}\right)_{q\, {\rm or}\, g} =
\int dx \left(\frac{\dd^2N}{\dd{x}\, \dd\ln k_\perp}\right)_{q\, {\rm or}\, g}
\equiv \int_{\lmin}^{Y_{\Theta_0}-y} d\ell \left(\frac{\dd^2N}{\dd\ell\,
\dd\ln k_\perp}\right)_{q\, {\rm or}\, g}.
\label{eq:includef}
\end{equation}
As in \cite{PerezMachet},  a lower bound of integration, $\lmin$,
is introduced since the present
calculation is only valid in the small-$x$ region, and therefore cannot be
trusted  when $\ell\equiv\ln(1/x)$ becomes ``too'' small.
We shall discuss this in more detail in
Section~\ref{subsec:ktdistributionslimspec} and Appendix
\ref{section:positivity}.

\begin{figure}[ht]
\begin{center}
 \includegraphics[height=3.5cm,width=10cm]{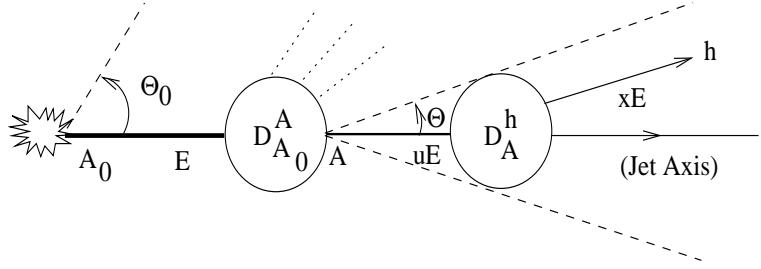}
\caption{\label{fig:distri} Inclusive production of hadron $h$
at an angle $\Theta$
inside a high energy jet of total opening angle $\Theta_0$.}
\end{center}
\end{figure}

According to~\cite{PerezMachet}, $\frac{\dd^2N}{\dd x\,\dd\ln{\Theta}}$ can
be expressed as
\begin{equation}
\frac{\dd^2N}{\dd x\,\dd\ln{\Theta}}=
\frac{\dd}{\dd\ln\Theta}F_{{\tt A_0}}^{h}\left(x,\Theta,E,\Theta_0\right),
\label{eq:DD}
\end{equation}
where $F_{{\tt A_0}}^{h}$, which represents the inclusive production
of $h$ in the sub-jet of opening angle
$\Theta$ inside the jet $A_0$ of opening angle $\Theta_0$,
is given by a convolution product of two fragmentation
functions ~\cite{PerezMachet}:
\begin{equation}
 F_{{\tt A_0}}^{h}\left(x,\Theta,E,\Theta_0\right)
\equiv \sum_{\tt A}\int_x^1 \dd{u}\
 D_{{\tt A_0}}^{\tt A}\left(u,E\Theta_0,uE\Theta\right)D_{\tt A}^{h}\left(\frac{x}
 {u},uE\Theta,Q_0\right).
\label{eq:F}
\end{equation}
The convolution expresses the
correlation between the energy flux of the jet and one particle within it.
Eq.~(\ref{eq:F}) is schematically depicted in Fig.~\ref{fig:distri}:
$u$ is the energy-fraction of the intermediate parton
${\tt A}$, $D_{{\tt A_0}}^{\tt A}$ describes the probability to emit
{\tt A} with energy $uE$ off the parton $\tt{{\tt A_0}}$
(which initiates the jet) taking into account the evolution of the jet
between $\Theta_0$ and $\Theta$, and $D_{\tt A}^{h}$ describes
the probability to produce the hadron $h$ off {\tt A} with energy fraction
$x/u$ and transverse momentum $k_\perp\approx uE\Theta\geq Q_0$; $k_\perp$ is defined with respect to the jet axis which is, in this context,
identified with the direction of the energy flux.

As discussed in \cite{PerezMachet}, the convolution (\ref{eq:F})
is dominated by $u=\cO{1}$. Therefore,
$D_{{\tt A_0}}^{\tt A}\left(u,E\Theta_0,uE\Theta\right)$
is given by DGLAP evolution equations.  On the contrary, since
$x\ll{u}=\cO{1}$ in the small-$x$ limit where MLLA evolution equations
are valid, $D_{\tt A}^{h}$ behaves as (see (\ref{eq:DBh}))
\begin{equation}
D_{\tt A}^{h}\left(\frac{x}{u},uE\Theta,Q_0\right)\ \stackrel{x\ll u}{\approx}\
\frac{u}{x}\rho_{\tt A}^{h}\left(\ln\frac{u}{x},\ln u + Y_\Theta\right).
\label{eq:Dlowx2}
\end{equation}
Since $Y_\Theta + \ln u=\ell + \ln u + y$, the hump-backed plateau
$\rho_{\tt A}^{h}$ depends on the two variables $\ell+\ln u$ and $y$,
and we conveniently define $\tilde D$ as:
\begin{equation}
\tilde D_{\tt A}^{h}(\ell + \ln u,y) \equiv \frac{x}{u} D_{\tt A}^{h}\left(\frac{x}{u},uE\Theta,Q_0\right). 
\end{equation}
%.
The Taylor expansion of $\rho_{\tt A}^{h}$ to the second order in $\ln u$
for  $u\sim 1 \Leftrightarrow |\ln u| \ll 1$,
that is, one step further than in \cite{PerezMachet}, leads to
\begin{eqnarray}
x F_{{\tt A_0}}^{h}(x,\Theta,E,\Theta_0)
&\approx&
x \widetilde{F}_{{\tt A_0}}^{h}(x,\Theta,E,\Theta_0) \nonumber\\
&&+ \frac12\sum_{\tt A} \left[\int \dd{u}\, u(\ln^2\!u)
D_{{\tt A_0}}^{\tt A}(u,E\Theta_0,uE\Theta)\right]
\frac{\dd^2 \tilde D_{\tt A}^{h}(\ell,y)}{\dd\ell^2},
\label{eq:Fdev}
\end{eqnarray}
where
\begin{equation}
x \widetilde{F}_{{\tt A_0}}^{h}(x,\Theta,E,\Theta_0)
\approx\sum_{\tt A}\left[\int \dd{u}\, u\Big(1+(\ln u)
\psi_{{\tt A},\ell}(\ell,y)
\Big) D_{{\tt A_0}}^{\tt A}(u,E\Theta_0,uE\Theta)\right]
\tilde D_{\tt A}^{h}(\ell,y)
\label{eq:first}
\end{equation}
is the MLLA distribution calculated in \cite{PerezMachet}.
In (\ref{eq:first}) we have introduced first  logarithmic derivatives of
$\tilde D_{\tt A}^{h}$
\begin{equation}
\psi_{{\tt A},\ell}(\ell,y) = \frac{1}{\tilde D_{\tt A}^{h}(\ell, y)}
\frac{\dd \tilde D_{\tt A}(\ell, y)}{\dd\ell}={\cal O}(\sqrt{\alpha_s}).
\label{eq:psidef1}
\end{equation}
Thus, as in \cite{PerezMachet}, in the soft limit
the correlation disappears and the convolution (\ref{eq:F})
is reduced to the factorized expression in (\ref{eq:Fdev}).

The second
term in the r.h.s. of (\ref{eq:Fdev}) is the new NMLLA correction
calculated in this paper.
Since $x/u$ is small, the inclusive spectrum $\tilde D_{\tt A}^{h}(\ell,y)$
occurring in (\ref{eq:Fdev}) should be taken as the next-to-MLLA
solution of the  evolution equations
(\ref{eq:solq}) and (\ref{eq:solg}). However, as already mentioned
and shown in Appendix \ref{section:roleofnmlla}, the
MLLA inclusive spectrum for a gluon jet can be used as a good approximation
for (\ref{eq:solg}) (with $a_1\ne0, a_2=0$)
such that, in  (\ref{eq:first}), it is enough to use
this level of approximation.
So, we shall therefore use Eqs.~(\ref{eq:ifD}) and (\ref{eq:calFdef})
in the following.

The NMLLA correction in (\ref{eq:Fdev}) globally decreases $|x
F_{{\tt A_0}}^{h}|$ in the perturbative region ($y\geq1.5$).
Indeed, while the MLLA part 
proportional to $\ln u$ in~(\ref{eq:first}) is negative 
\cite{PerezMachet}, it is instead, there, positive because of the 
positivity of $u$ and $\ln^2 u$ and 
$\frac{\dd^2 \tilde D_{\tt A}^{h}}{\dd\ell^2}\simeq\frac{\dd^2G}{\dd\ell^2}$ 
(see Fig.~\ref{fig:Gll} in Appendix~\ref{section:2ndder}).
The NMLLA contribution therefore tempers somehow the size of the
MLLA corrections when $y$ is large enough.

\subsection{Color currents}
\label{subsec:CC}
%%%%%%%%%%%%%%%%%%%%%%%%%%%

The function $F_{{\tt A_0}}^{h}$ is related to the inclusive
gluon distribution {\it via} the color currents defined
as~\cite{Basics,PerezMachet}
\begin{equation}
\label{eq:coldef}
x F_{{\tt A_0}}^{h} = \frac{\langle C\rangle_{{\tt A_0}}}{N_c}\ G(\ell,y).
\end{equation}
The  color current can be seen as the average color charge carried
by the parton {\tt A} due to the DGLAP evolution from ${\tt A_0}$ to {\tt A}.
Introducing the first and second logarithmic derivatives of
$\tilde D_{\tt A}^{h}$,
\begin{equation}
(\psi_{{\tt A},\ell}^2+\psi_{{\tt A},\ell\ell})(\ell,y)=\frac{1}{\tilde D_{\tt A}^{h}(\ell, y)}
\frac{\dd^2 \tilde D_{\tt A}(\ell, y)}{\dd\ell^2}={\cal O}(\alpha_s),
\label{eq:psidef2}
\end{equation}
which are MLLA and NMLLA corrections, respectively, Eq.~(\ref{eq:Fdev})
can now be written
\begin{equation*}
  x F_{{\tt A_0}}^{h} \approx 
\sum_{\tt A}\ \Big[ \langle u \rangle_{{\tt A_0}}^{\tt A}
+ \langle u \ln u\rangle_{{\tt A_0}}^{\tt A} \psi_{{\tt A},\ell}(\ell,y)
+ \frac12 \langle u \ln^2 u\rangle_{{\tt A_0}}^{\tt A}
(\psi_{{\tt A},\ell}^2+\psi_{A,\ell\ell})(\ell,y) \Big]\ \tilde D_{\tt
A}^{h}(\ell,y),
\end{equation*}
where
\begin{equation}
\langle u \ln^i u\rangle_{{\tt A_0}}^{\tt A}
\equiv \int_{0}^{1} \dd u\ (u\ \ln^i u)\ 
D_{{\tt A_0}}^{\tt A}\left(u,E\Theta_0,uE\Theta\right).
\label{eq:befscal}
\end{equation}
Unlike in \cite{PerezMachet} at MLLA, using the approximation $u=\cO{1}$
to replace in (\ref{eq:befscal}) $uE\Theta$ by $E\Theta$
requires here some care,
since the resulting scaling violation of the DGLAP fragmentation functions
also provides ${\cal O}(\alpha_s)$ corrections
to $\langle u\rangle$.
Explicit calculations (see Appendix \ref{section:scalviol}) show that
they never exceed $5\%$ of the leading term. Accordingly, we neglect them
in the following and replace (\ref{eq:befscal}) by
\begin{equation}
\langle u \ln^i u\rangle_{{\tt A_0}}^{\tt A}
\simeq \int_{0}^{1} \dd u\ (u\ \ln^i u)\ D_{{\tt A_0}}^{\tt A}
\left(u,E\Theta_0,E\Theta\right).
\label{eqLaftsval}
\end{equation}
The total average color current $\langle C \rangle_{{\tt A_0}}$ of
partons caught by the calorimeter decomposes accordingly into three terms
which can be written:
\begin{eqnarray}
  \label{eq:ccmlla}
  \langle C \rangle_{{\tt A_0}} &=&  \langle C \rangle^{\rm LO}_{{\tt A_0}} + \delta \langle C \rangle^{\rm MLLA-LO}_{{\tt A_0}} + \delta \langle C \rangle^{\rm NMLLA-MLLA}_{{\tt A_0}}.
\end{eqnarray}
The leading order (LO) $\cO{1}$ and MLLA $\cO{\sqrt{\alpha_s}}$
contributions to the color currents have been determined
in~\cite{PerezMachet}.
The new NMLLA $\cO{\alpha_s}$ correction evaluated in this paper reads
\begin{equation}
\label{eq:ccnmlla}
\delta\langle C \rangle_{{\tt A_0}}^{\rm NMLLA-MLLA}
= N_c\ \langle u \ln^2 u\rangle_{{\tt A_0}}^g\ 
(\psi^2_{g, \ell}+\psi_{g, \ell\ell})
 +\ C_{F}\ \langle u \ln^2 u\rangle_{{\tt A_0}}^q\ (\psi^2_{q, \ell}
+ \psi_{q, \ell\ell}),
\end{equation}
assuming $Q=C_F/N_c\ G$. We checked that using instead the  NMLLA exact
formula (\ref{eq:ratioqg}) for the quark inclusive spectrum $Q$ actually
leads to negligible corrections to the color currents
(see Appendix~\ref{section:exactcc}).
Eq.~(\ref{eq:ccnmlla}) can be obtained from the  Mellin-transformed
DGLAP fragmentation functions
$$
{\cal D}_{{\tt A_0}}^{\tt A}(j,\xi)=
\int_0^1\dd{u}\,u^{j-1} D_{{\tt A_0}}^{\tt A}(u,\xi),
$$
through the formula
\begin{equation}\label{eq:delta2}
\langle u \ln^2 u \rangle_{{\tt A_0}}^{\tt A}
=\frac{\dd^2}{\dd{j^2}}{\cal D}_{{\tt A_0}}^{\tt A}
(j,\xi(E\Theta_0)-\xi(E\Theta))\bigg|_{j=2}\equiv
\int_0^1\dd{u}\,u\ln^2u D_{{\tt A_0}}^{\tt A}(u,\xi).
\end{equation}
Given the rather lengthy expressions, the complete analytic results
for $\langle C\rangle^{\rm NMLLA-MLLA}_{{\tt A_0}}$ for quark and gluon
jets are given in Appendix \ref{section:delta2C}.

\begin{figure}[ht]
\begin{center}
  \includegraphics[height=6.5cm,width=8cm]{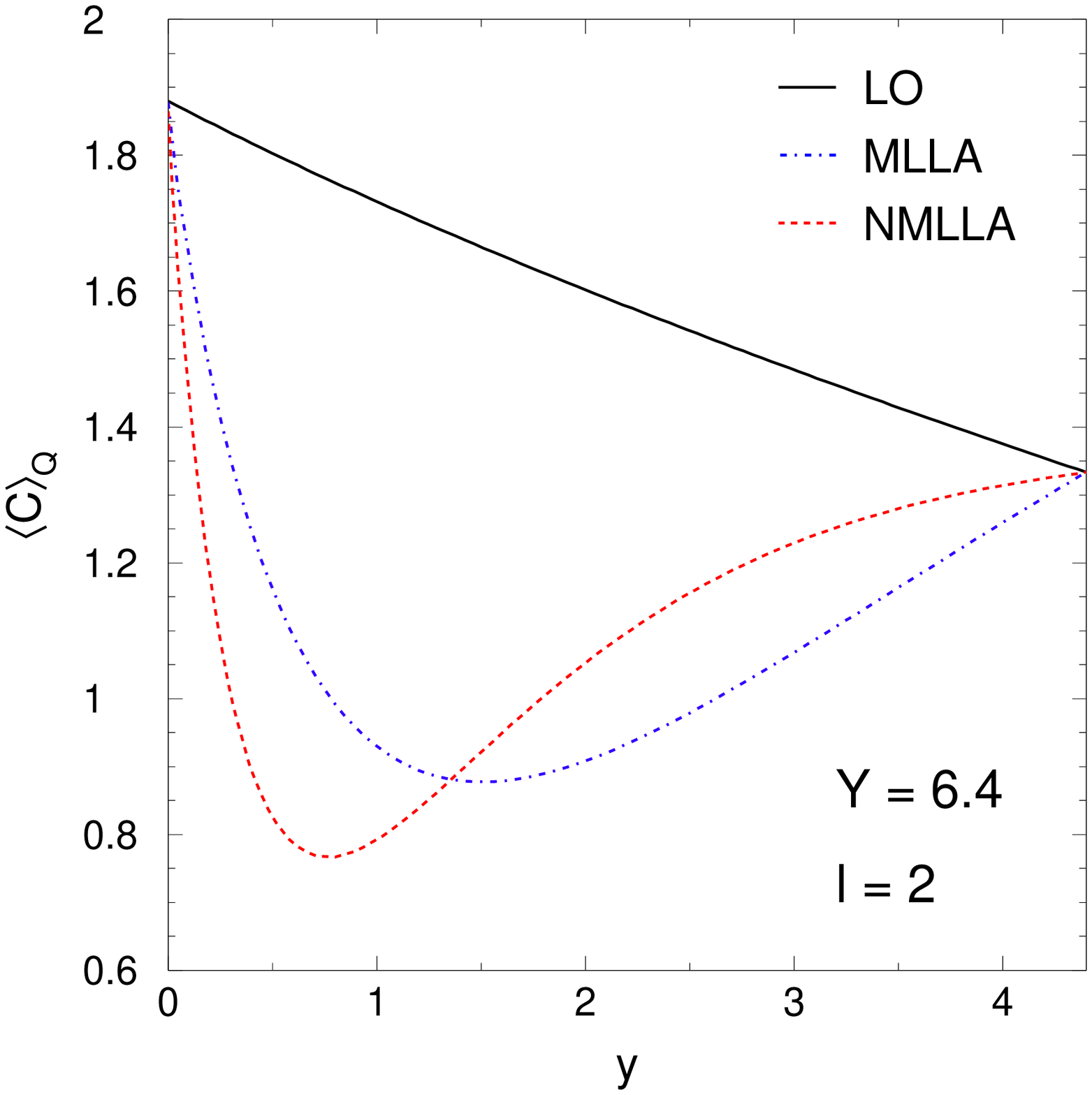}
\hskip 1cm
  \includegraphics[height=6.5cm,width=8cm]{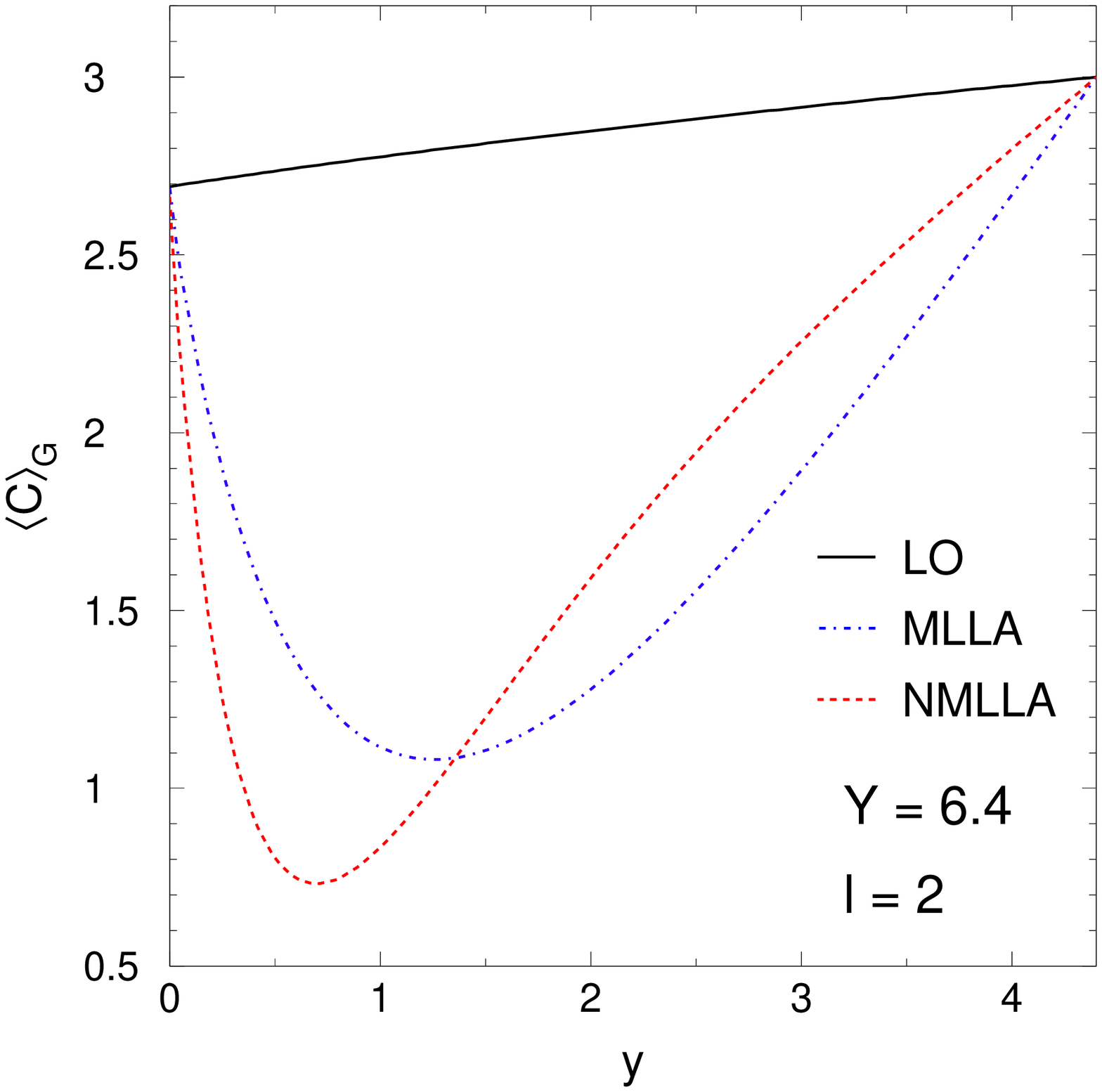}
\caption{\label{fig:CCGQ} Color currents at LO (solid lines), MLLA (dash-dotted), and NMLLA (dashed) 
for a quark (left) and gluon jet (right) with $Y_{\Theta_0}=6.4$ and $\ell=2$.}
\end{center}
\end{figure}

For illustrative purposes, the color currents are plotted in
Fig.~\ref{fig:CCGQ} in the limiting spectrum approximation ($\lambda=0$).
The LO (solid line), MLLA (dash-dotted) and NMLLA (dashed) currents
are computed for a quark (left) and for a gluon jet (right) with energy
$Y_{\Theta_0}=6.4$ --~corresponding to Tevatron energies~-- and at
fixed $\ell=2$. As can be seen in Fig.~\ref{fig:CCGQ}, NMLLA $\cO{\alpha_s}$
corrections to the MLLA color currents are clearly not negligible,
yet of course somewhat smaller than the MLLA $\cO{\sqrt{\alpha_s}}$
corrections to the LO result.
In the perturbative region ($y>1.5$), these corrections are positive
and consequently decrease the difference with the LO estimate.
On the contrary, at small $y\le 1.5$, the corrections are rather large
and negative coming from the negative sign of $G_{\ell\ell}(\ell, y)$
(see Fig.~\ref{fig:Gll} in Appendix~\ref{section:2ndder}).
However, it should be kept in mind that as $y$ goes to $0$, $\kt$ gets
closer to $\lqcd$ (remind that $Q_0=\lqcd$ in the limiting spectrum
approximation) and, thus, the present perturbative predictions may not be
reliable in this domain. 

Note also that both the MLLA and NMLLA
corrections vanish at $y=0$ (since $G_\ell = G_{\ell\ell} = 0$)
and when $\Theta = \Theta_0$. Another interesting property to mention
is the decrease of MLLA and NMLLA corrections as $\ell$ increases, that
is, when partons get softer and recoil effects more negligible.

\null
From the color currents, the NMLLA
double-differential 1-particle distribution at small $x$
(see Eq.~(\ref{eq:DD})),
\begin{equation}
\left(\frac{\dd^2N}{\dd\ell\, \dd y}
\right)_{\tt A_0}\ =\ \frac{1}{N_c}\ \langle C \rangle_{\tt A_0}\ 
\frac{\dd}{\dd{y}}G(\ell,y)\
+\ \frac{1}{N_c}\ G(\ell,y)\ \frac{\dd}{\dd{y}}\langle C \rangle_{\tt A_0},
\label{eq:DDD}
\end{equation}
can be determined for any value of $\lambda$.
The NMLLA behavior of $\dd^2N/\dd\ell \dd{y}$
is therefore easily deduced from $\langle C\rangle_{\tt A_0}$ and
its $y$-dependence, $\dd\langle C\rangle_{\tt A_0} / \dd{y}$.

\subsection{$\kt$-distributions}
\label{subsec:ktdistributionslimspec}
%%%%%%%%%%%%%%%%%%%%%%%%%%%%%%%%%%%%%%

\begin{figure}[ht]
\begin{center}
 \includegraphics[height=7cm,width=8cm]{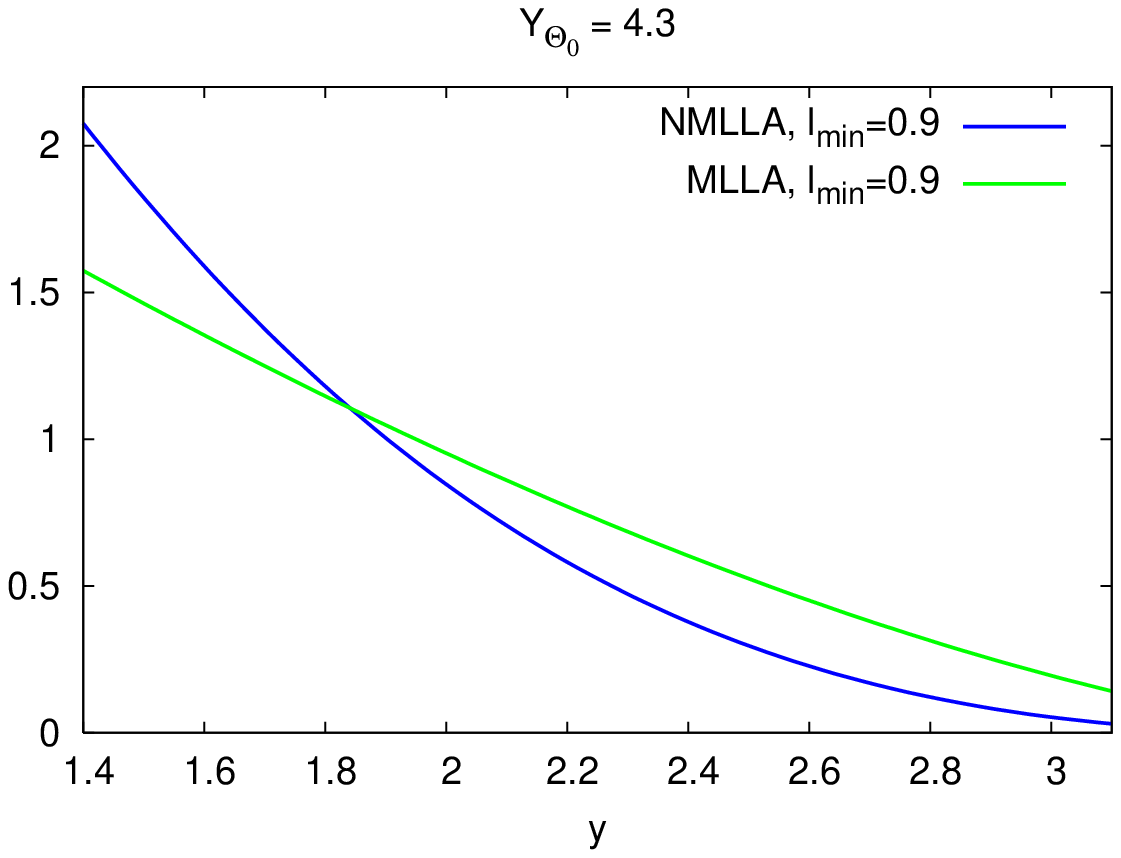}
\hskip 5mm
 \includegraphics[height=7cm,width=8cm]{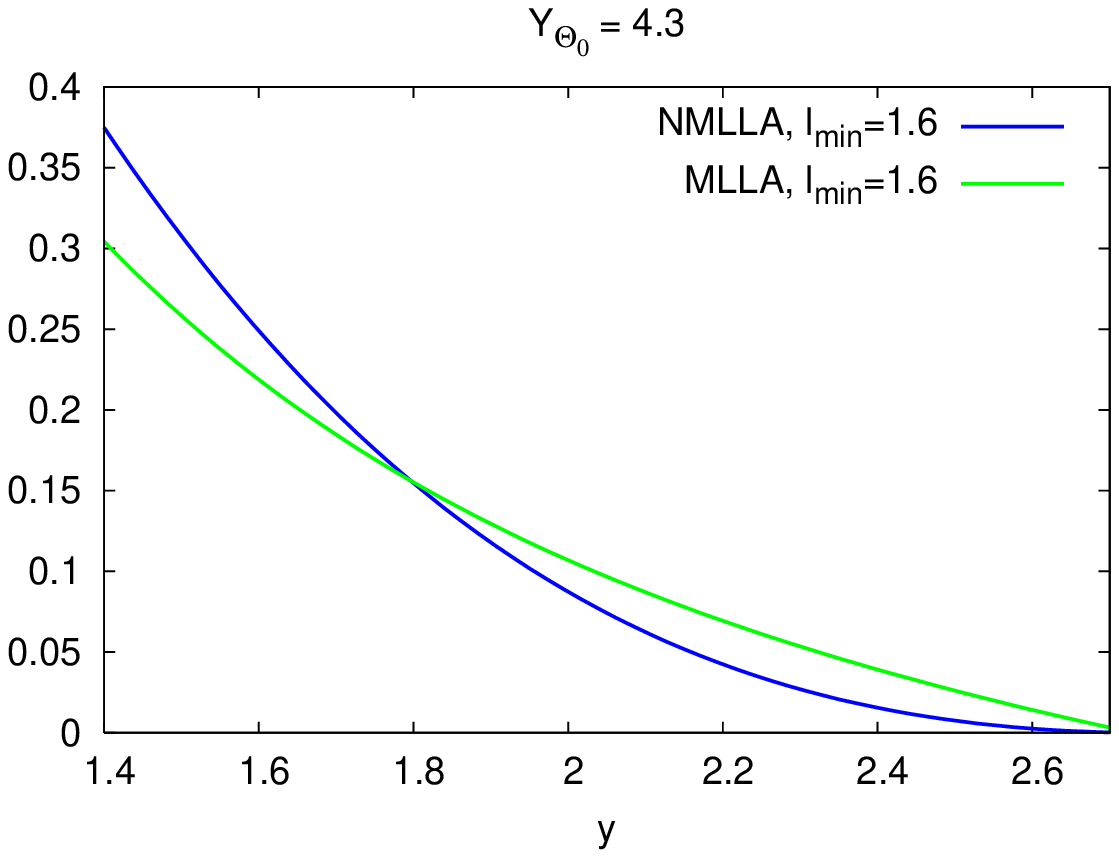}
\vskip .5cm
 \includegraphics[height=7cm,width=8cm]{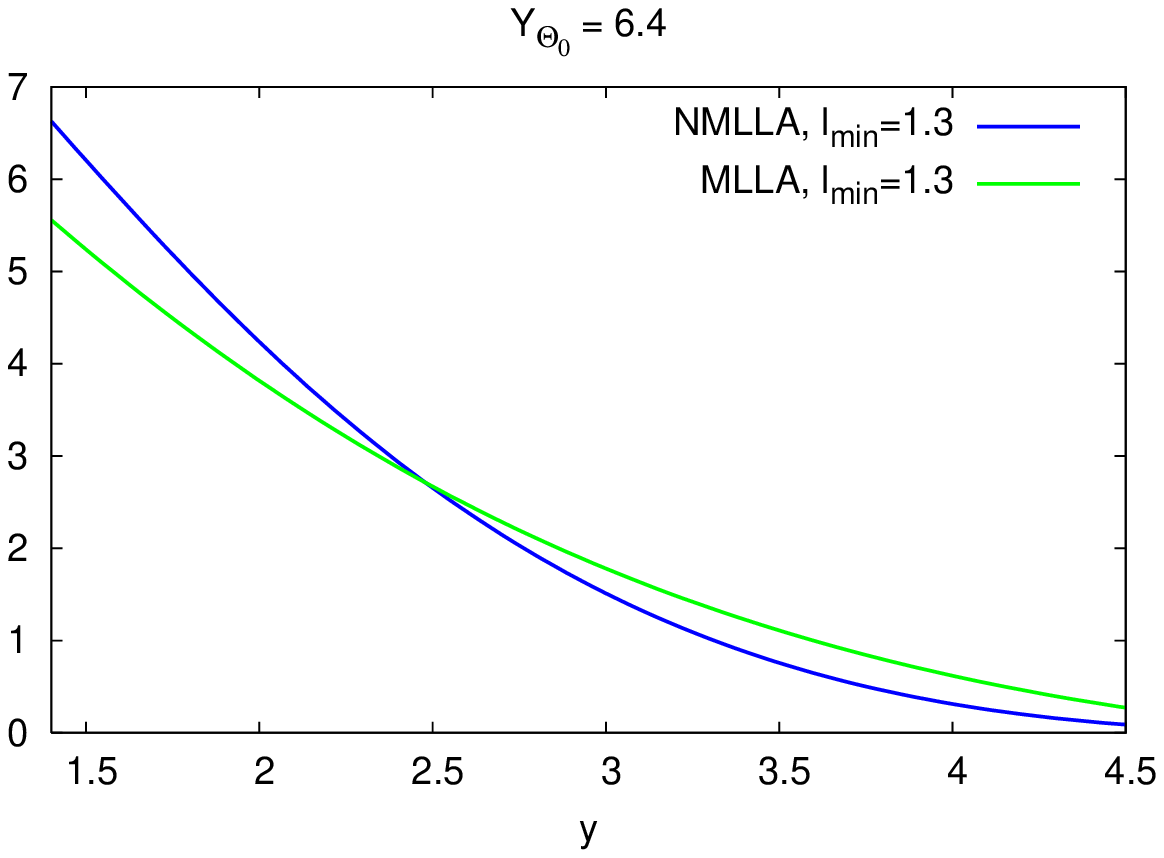}
\hskip 5mm
 \includegraphics[height=7cm,width=8cm]{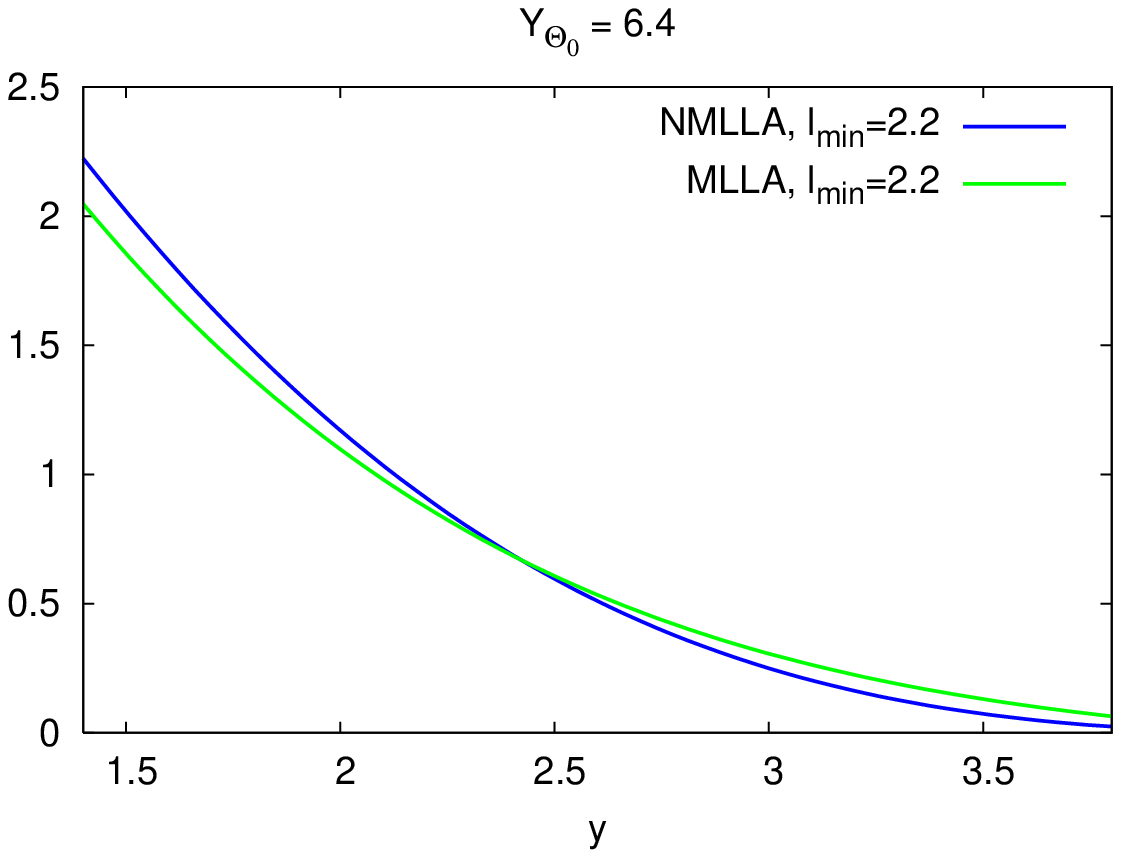}
\caption{\label{fig:TEV} MLLA (green) and NMLLA (blue)
inclusive $y$-distributions for a quark (left) and a gluon jet
(right) with $Y_{\Theta_0}=4.3$ ($Q=19$~GeV) and
$Y_{\Theta_0}=6.4$ ($Q=155$~GeV).}
\end{center}
\end{figure}

The $\kt$-distributions of hadrons are computed from the numerical
integration of the double-differential cross section, Eq.~(\ref{eq:DDD}).
On Fig.~\ref{fig:TEV} are shown the MLLA (dashed lines) and NMLLA
(solid lines) $\dd{N}/\dd{y}$ distributions for a quark (left) and a gluon jet
(right) with $Y_{\Theta_0}=4.3$ and $Y_{\Theta_0}=6.4$.
The size of NMLLA corrections proves quite substantial over
the whole $y$-range. We find in particular that at large $y$ (or $\kt$),
the distributions at NMLLA are lower than at MLLA (and
larger at small $y$).
This softening of the spectra can be understood physically 
by the role of energy conservation in jets.
With respect to DLA, MLLA and NMLLA take better into account
the recoil of the emitting parton
at each step of the cascading process. The fraction of
energy carried away by the emitted soft partons  gets reduced,
which finally damps the final emission of hadrons at large 
$k_\perp$~\footnote{Taking into account the running of $\alpha_s$, 
which increases with the evolution of the jet, goes {\it a priori}
in the opposite way of increasing the hadronic yield with respect to the
case where one freezes the coupling constant at the collision energy.}.
As already stressed in Section~\ref{subsec:dd1p}, the value of the lower
limit of integration $\lmin$ below which
the present small-$x$ calculation may not be trusted cannot be
directly predicted. In \cite{PerezMachet}, the appearance of
positivity problems in the double-differential distribution at small
$\ell$ led us to consider a minimal value $\lmin$ such
that $\dd^2N / \dd\ell\, \dd \ln\kt$ is kept positive for all
$\ell\ge \lmin$, leading to~\footnote{Its role 
was found very small in the calculation of the
$k_\perp$-distribution, such that, in  practice, it was taken to be
vanishing.}
$\lmin\simeq 2.5$. For consistency, the same criterion is used in
the present paper. We find that smaller values of $\ell$ actually
fulfill the positivity requirement, roughly $\lmin\simeq 2$ and
$\lmin\simeq 1$ for quark and gluon jets at Tevatron energies.

It is interesting to note that the range over which NMLLA calculations
appear sensible extends to smaller $\ell$, therefore to larger $x$,
than at MLLA; this also corresponds to larger $y$ at fixed $Y$.
One could therefore expect the present NMLLA predictions to agree with
experimental results in a larger domain of $k_\perp$.
This is discussed in the coming Section.

\subsection{Comparison with CDF preliminary data}
\label{subsec:compexp}
%%%%%%%%%%%%%%%%%%%%%%%%%%%%%%%%%%%%%%%%%%%%%%%%%

\begin{figure}[ht]
\begin{center}
  \includegraphics[height=14.6cm,width=15cm]{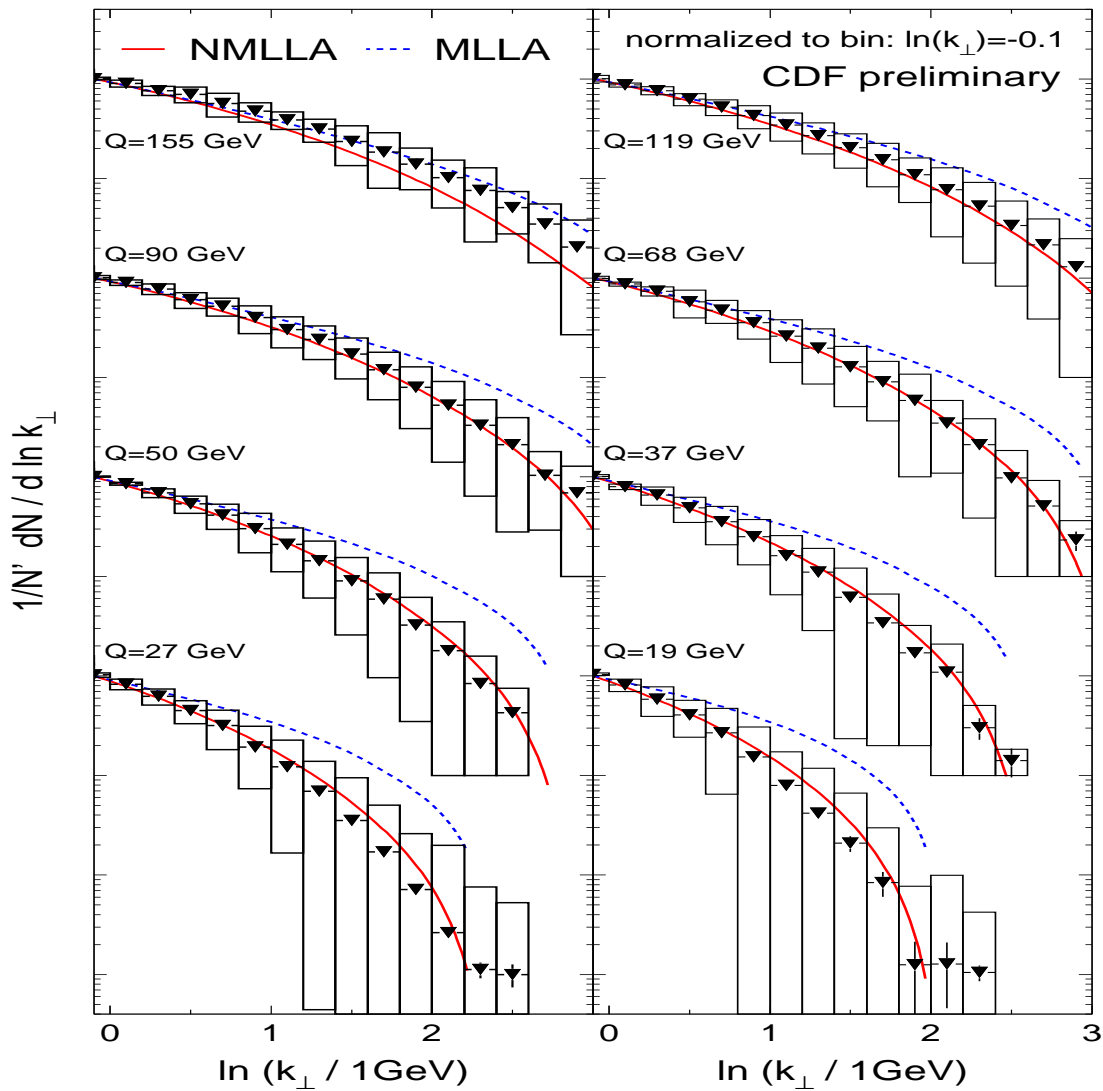}
\caption{\label{fig:CDF-NMLLA} CDF preliminary results on hadronic
single-inclusive $\kt$-distributions, compared with MLLA (dashed lines)
and NMLLA (solid lines) calculations at the limiting spectrum; the boxes
are the systematic errors (their lower limits are cut at large $k_\perp$
for the sake of clarity)}
\end{center}
\end{figure}

The CDF collaboration at Tevatron recently reported on preliminary data
of hadronic single-inclusive $\kt$-distributions inside jets produced
in $p\bar{p}$ collisions at $\sqrt{s}=1.96$~TeV~\cite{CDF}.
The measurements cover a wide domain of jet energies,
with hardness $Q=E\Theta_0$ ranging from $Q=19$~GeV to $Q=155$~GeV. 
The CDF results, including systematic errors,
 are plotted in Fig.~\ref{fig:CDF-NMLLA}  together with
the MLLA predictions of~\cite{PerezMachet} (dashed lines) and the
present NMLLA calculations (solid lines). Data and theory are normalized to
the same bin, $\ln \kt=-0.1$, because of presently too large
normalization errors in the CDF preliminary data.
The experimental measurements reflect a mixing of quark and gluon jets:
\begin{equation}
\left(\frac{\dd N}{\dd\ln k_\perp}\right)_{\rm mix}=
\omega\left(\frac{\dd N}{\dd\ln k_\perp}\right)_g+(1-\omega)
\left(\frac{\dd N}{\dd\ln k_\perp}\right)_q
\end{equation}
characterized by one $Q$-dependent mixing-parameter $\omega$,
estimated from PYTHIA~\footnote{See Ref.~\cite{CDF} and S.~Jindariani,
private communication.}, used in the theoretical calculation.
The agreement between the CDF results and the NMLLA distributions over
the whole $\kt$-range is excellent.
The NMLLA calculation is in particular able to capture the shape
of CDF spectra at all $Q$. Conversely, 
predictions at MLLA prove only reliable at not too large $\kt$.

The domain of validity of the predictions has been enlarged
to larger $k_\perp$ (and thus to larger $x$ since $Y$ is fixed)
computing from MLLA to NMLLA accuracy~\footnote{We recall that 
we only used the NMLLA solution
of the evolution equations for the inclusive spectrum.
All calculated NMLLA corrections to the $k_\perp$ distribution 
occur by the sole expansion of $\frac{x}{u}
D_{\tt A}^{h}\left(\frac{x}{u}, \ldots\right)$ at small
$x/u$ around $\ln u \approx \ln 1$ in the convolution (\ref{eq:F}).}
.
It is however to be mentioned that, due to the normalization at the first
bin, this extension of the domain of prediction only concerns, strictly
speaking, the shape of the distribution.
Equally importantly, the  agreement between NMLLA calculations and
experimental results brings further support to the
Local Parton Hadron Duality (LPHD) picture~\cite{LPHD}.
We indeed find it remarkable to observe that the 
entire $\kt$-domain probed experimentally can be very well
described by strict perturbation theory, 
leaving out only limited non-perturbative dynamics in the production
of hadrons inside a jet, at least for inclusive enough observable
like single-particle $\kt$-distributions.

\subsection{Theoretical uncertainties}
\label{subsec:uncertain}
%%%%%%%%%%%%%%%%%%%%%%%%%%%%%%%%%%%%%%%

The spectacular agreement between our NMLLA calculations and preliminary
data should not hide the theoretical uncertainties that affect the former.

First, we did not take into account all NMLLA corrections. 
While scaling violations have already been dealt with in subsection
\ref{subsec:CC} and Appendix \ref{section:scalviol}, other NMLLA
corrections arise from varying $\lqcd$ in the expression of $\alpha_s$.
In Figs.~\ref{fig:LQCD}  is plotted the inclusive $k_\perp$-distribution 
($Q=119$~GeV) at values of
 $\lqcd$ ranging from $150$ to $500$~MeV (left), as well as
the ratio to its value at the default $\lqcd=250$~MeV (right).
All curves have been normalized to the bin
$\ln(k_\perp/1\text{GeV})=-0.1$. In the largest bin
$\ln(k_\perp/1\text{GeV})= 3$, varying  $\lqcd$ varies from $150$ to $400$
MeV does not yield a relative variation larger than $20\%$.
The corresponding curves still fall within the
experimental systematic errors.
Note that the fact that variations seem only important at large $k_\perp$ only comes from the normalization procedure in the bin $\ln(k_\perp/1\text{GeV})=-0.1$. A more delicate matter concerns the dominance of the type of NMLLA
corrections that we have taken into account. Some remarks will concerning
this point are postponed to the general discussion in section
{\ref{sec:CONCL}.
\begin{figure}[ht]
\begin{center}
  \includegraphics[width=7.2cm]{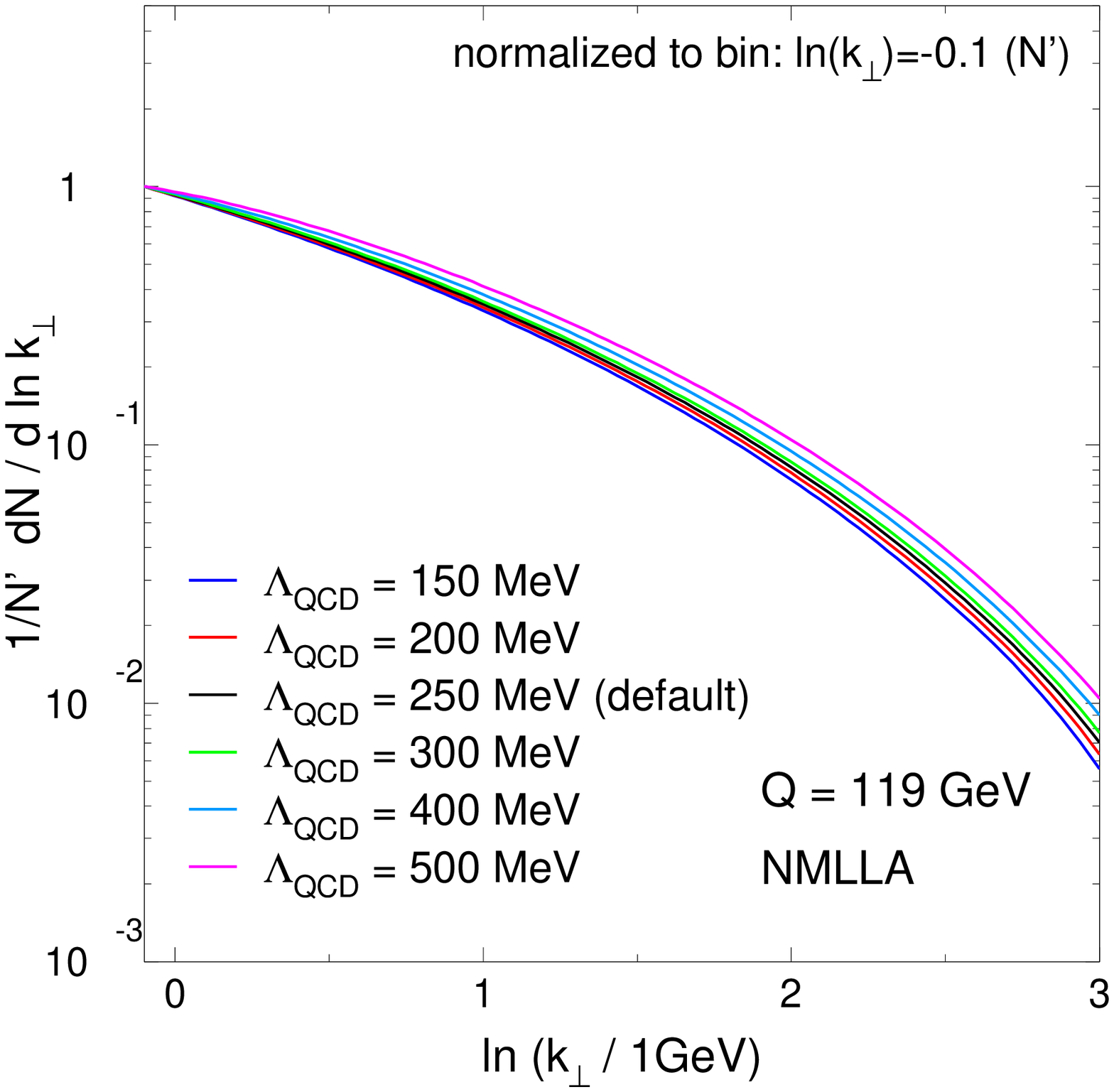}
\hskip 1.5cm
 \includegraphics[width=7.2cm]{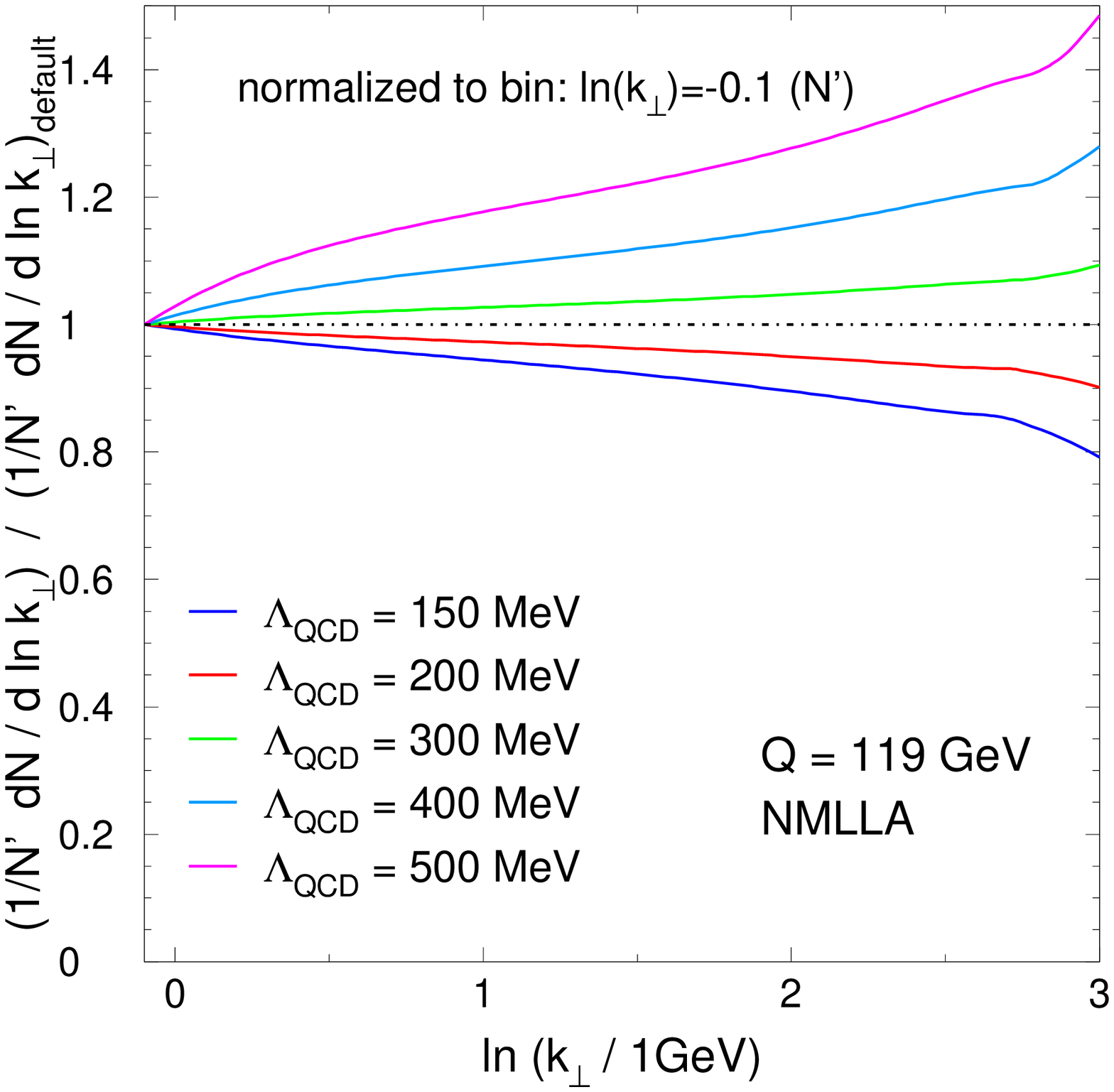}
\caption{\label{fig:LQCD} The dependence on $\lqcd$, absolute (left) and
relative (right).}
\end{center}
\end{figure}
The second point concerns the jet axis, which is defined
here as the direction of the energy flow. It is implicitly determined by
a summation over all secondary hadrons in energy-energy correlations. At
the opposite, the jet axis is experimentally determined exclusively 
from {\it all} particles inside the jet. Whether these two
definitions match within NMLLA accuracy, $\cO{\alpha_s}$, is a matter 
which deserves further investigation. This goes however 
beyond the scope of the present work.

Last, cutting the integral (\ref{eq:includef}) at small $\ell$ may look
somewhat arbitrary. However, at the end of  Appendix \ref{section:positivity},
we provide in Figs.~\ref{fig:lmin} 
 curves which show the variation of the inclusive
$k_\perp$-distribution at MLLA and NMLLA when $\ell_{\rm min}^g$ is changed.
Varying it from $1$ to
 to $1.75$ does not modify the NMLLA spectrum at large $k_\perp$ by more
than $20\%$. Variations are more dramatic at MLLA.

%%%%%%%%%%%%%%%%%%%%%%%%%%%%%%%%%%%%%%%%%%%%%%%%%%%%%%%%%%%%%%%%%%%%%%%%%%%
\section{Single-inclusive $\kt$-distributions beyond the limiting spectrum}
\label{section:numer}
%%%%%%%%%%%%%%%%%%%%%%%%%%%%%%%%%%%%%%%%%%%%%%%%%%%%%%%%%%%%%%%%%%%%%%%%%%%

\subsection{Inclusive spectrum}
%%%%%%%%%%%%%%%%%%%%%%%%%%%%%%%

So far, the calculations have been performed in the limiting spectrum
approximation, $Q_0=\lqcd$ or $\lambda=0$. This assumption, which cuts off
hadronic yield below $Q_0$ should
be valid as long as the mass of the produced hadrons  is not too large
as compared to $\lqcd$. This is the case when dealing mostly with pions.
We perform in this Section the exact calculation of
single-inclusive spectra as well as $\kt$-distributions beyond
this approximation, $\lambda\ne0$, that is for hadrons with mass
$m_h\simeq Q_0\ne\lqcd$~\cite{finitelambda}.

The inclusive gluon spectrum was given in~\cite{RPR2} a compact Mellin
representation:
\begin{equation*}
G\left(\ell,y\right)\ =\ \left(\ell+y+\lambda\right)\ \int\
\frac{\dd\omega\, \dd\nu}{\left(2\pi i\right)^2}\ e^{\omega\ell+\nu y}\ \int_{0}^{\infty}\frac{\dd s}{\nu+s}\ 
\left(\frac{\omega
\left(\nu+s\right)}{\left(\omega+s\right)\nu}\right)^{1/\beta_0
\left(\omega-\nu\right)}\left(\frac{\nu}{\nu+s}\right)^
{a_1/\beta_0}\,e^{-\lambda s},
\end{equation*}
from which an analytic approximated expression was found using the steepest
descent method~\cite{RPR3}. However, $G(\ell, y)$ is
here determined exactly from an equivalent representation in terms of a
single Mellin transform (which reduces to~(\ref{eq:ifD}) as $\lambda\to 0$)~\cite{finitelambda}
\begin{eqnarray}\label{eq:lambdadiff0}
G(\ell,y)&=&\frac{\ell+y+\lambda}{\beta_0\ B\ (B+1)}
\int_{\epsilon-i\infty}^{\epsilon+i\infty}\frac{\dd\omega}{2\pi i}\
e^{\omega\ell}\nonumber\\
&&\times\  \Phi(-A+B+1, B+2, -\omega(\ell+y+\lambda))\ \ {\cal K}(\omega, \lambda)
\end{eqnarray}
which is better suited for numerical studies. The function ${\cal K}$
appearing in Eq.~(\ref{eq:lambdadiff0}) reads
\begin{equation}
{\cal K}(\omega,\lambda)\ =\ \frac{\Gamma(A)}{\Gamma(B)}\ 
(\omega\ \lambda)^B\ \Psi(A, B+1,\omega\ \lambda),
\end{equation}
where $A=1/(\beta_0\ \omega)$, $B=a_1/\beta_0$, and $\Phi$ and $\Psi$
are the confluent hypergeometric function of the first and second kind,
respectively. The single-inclusive spectrum at MLLA is plotted
in Fig.~\ref{fig:gspec} for various values of $\lambda$,
 $\lambda=0,0.2,0.5,1$, for
a gluon jet with $Y_\Theta=6.4$. Increasing $\lambda$ reduces the emission
in the infrared region and therefore  favors hard particle production
at $\ell\ll Y/2$ (asymptotic position of the peak of the hump-backed
plateau).  Still, it is worth remarking
that the global shape of $G$ at finite $\lambda$ remains similar to
that obtained in the limiting spectrum approximation.
Note also that there is a
discrete part at finite $\lambda$, proportional to $\delta(\ell)$,
corresponding to the finite probability for no parton emission
when $Q_0\ne\lqcd$, the parton multiplicity becoming infrared
finite at $\lambda\ne 0$ (see the second reference in \cite{finitelambda}).
\begin{figure}[ht]
\begin{center}
  \includegraphics[height=7.2cm,width=8cm]{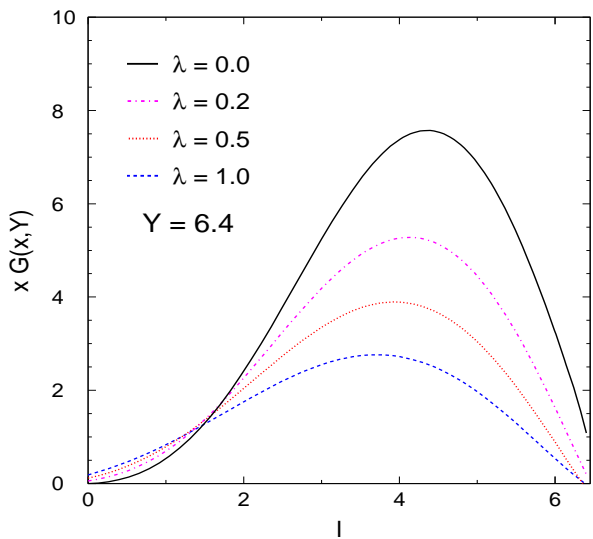}
\caption{\label{fig:gspec} Inclusive spectrum for a gluon jet ($Y_{\Theta_0}=6.4$) for different values of $\lambda$.}
\end{center}
\end{figure}

\subsection{Color currents}
%%%%%%%%%%%%%%%%%%%%%%%%%%%

The color currents, Eq.~(\ref{eq:ccmlla}), can now be determined
beyond the limiting spectrum from the inclusive spectrum calculated
in the previous section. In Fig.~\ref{fig:colcur} are displayed
the MLLA corrections to the LO color current,
$\delta \langle C \rangle^{\rm MLLA-LO}_{{\tt A_0}} / \langle C
\rangle^{\rm LO}_{{\tt A_0}}$ (left), and NMLLA corrections to
the MLLA color currents,
$\delta \langle C \rangle^{\rm NMLLA-MLLA}_{{\tt A_0}} / \langle C
\rangle^{\rm MLLA}_{{\tt A_0}}$ (right), for different values
$\lambda=0,0.5,1$. Fig.~\ref{fig:colcur} clearly indicates that
the larger the values of $\lambda$, the smaller the MLLA (and NMLLA)
corrections. In particular, MLLA (NMLLA) corrections can be as
large as $50\%$ ($30\%$) in the limiting spectrum but no more than
$20\%$ ($10\%$) for $\lambda=1$. This is not surprising since
$\lambda\ne 0$ ($Q_0\ne\lqcd$) reduces the parton emission in the
infrared sector and, consequently, higher-order corrections.
\begin{figure}[ht]
\begin{center}
  \includegraphics[height=7.2cm,width=8cm]{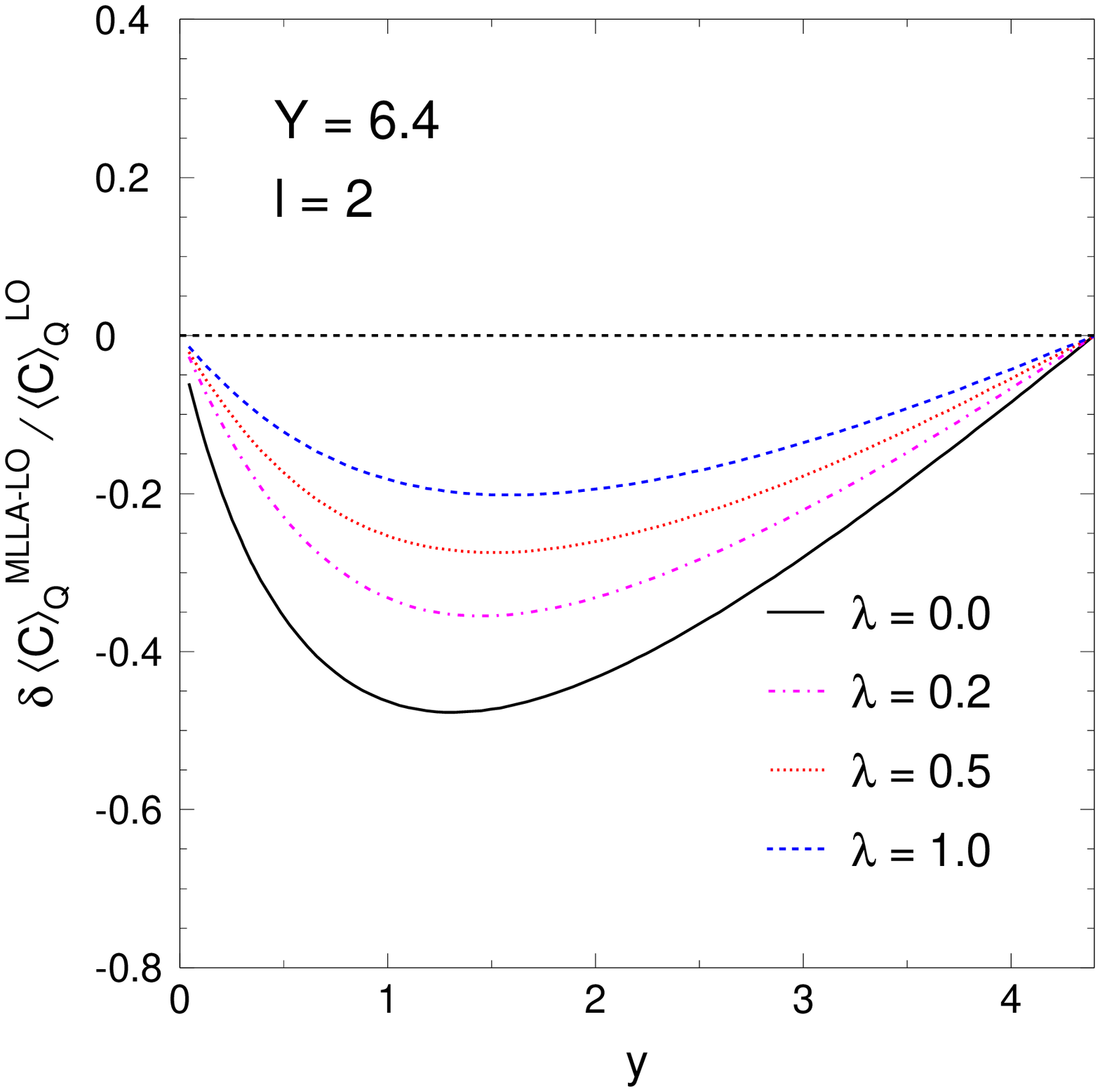}
\hskip 1cm
\includegraphics[height=7.2cm,width=8cm]{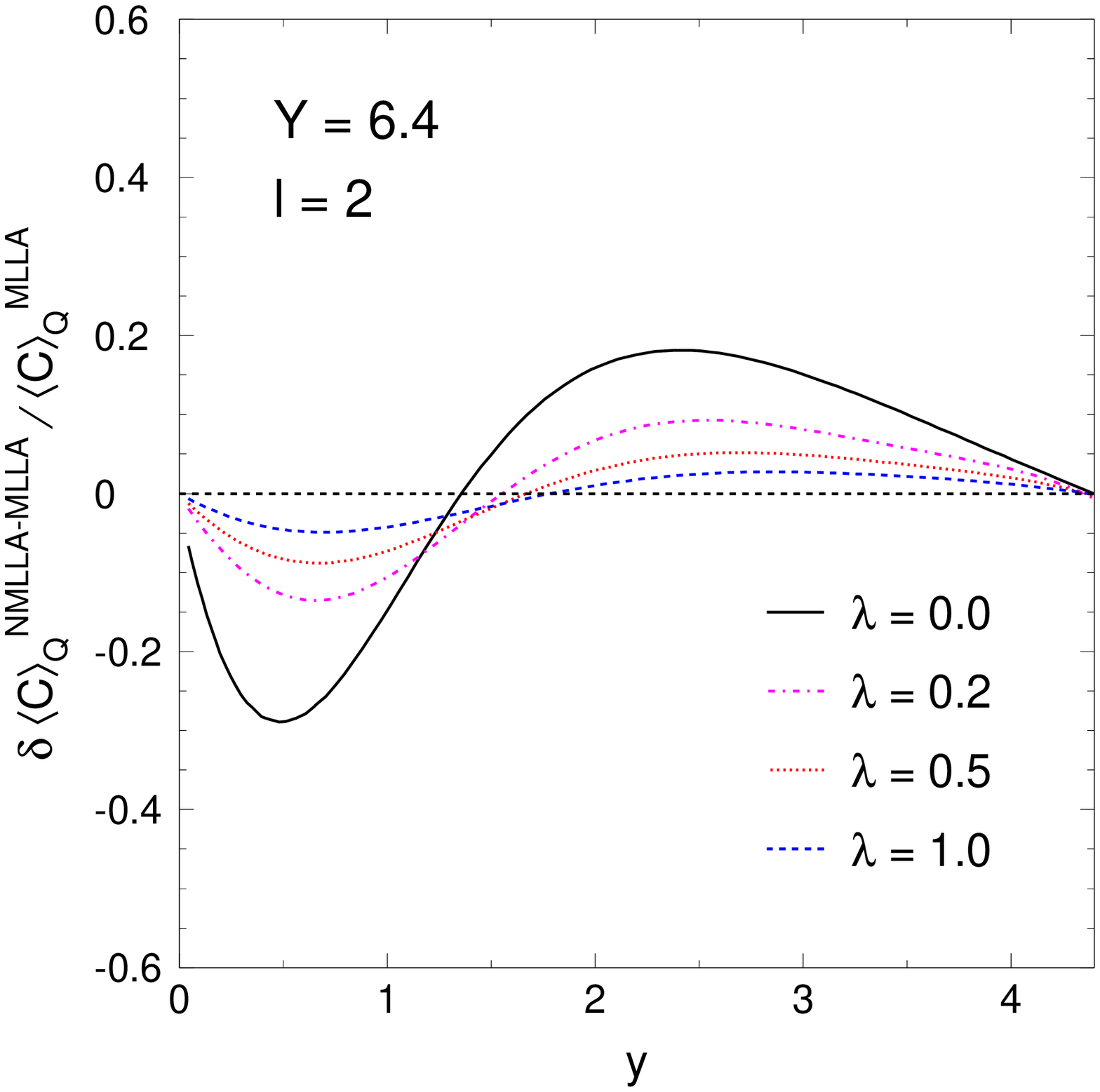}
\caption{\label{fig:colcur} MLLA (left) and NMLLA (right) normalized
corrections to the LO and MLLA color currents, respectively,
for different values of $\lambda$.}
\end{center}
\end{figure}
As discussed in Sect.~\ref{subsec:CC}, the large and negative
corrections to the color currents in the limiting spectrum lead to
negative double-differential spectra, $\dd^2N/\dd\ell \dd{y}$,
at small $y$.
Interestingly, at $\lambda\ne 0$, the infrared
sensitivity is somehow weakened.
As a consequence, $\dd^2N/\dd\ell \dd{y}$ is no longer negative at
finite $\lambda$, as illustrated in Fig.~\ref{fig:ddq}.
Another interesting consequence is the disappearance of the infrared
divergence at $y=0$ in the limiting spectrum, coming from the running
of $\alpha_s$: since $Q_0\ne\lqcd$, $\alpha_s$ and therefore
$\dd^2N/\dd\ell \dd{y}$ remain finite over the full momentum-space.
\begin{figure}[ht]
\begin{center}
  \includegraphics[height=7.2cm,width=8cm]{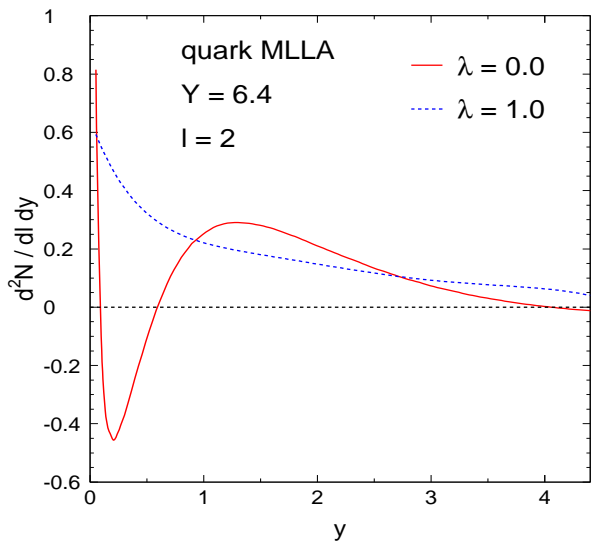}
 \caption{\label{fig:ddq} MLLA double differential distribution
for a quark jet $Y_{\Theta_0}=6.4$ computed at $\lambda=0$ (solid line)
and $\lambda=1$ (dashed line).}
\end{center}
\end{figure}

\subsection{$\kt$-distributions}
%%%%%%%%%%%%%%%%%%%%%%%%%%%%%%%%%%

The absolute $\kt$-distributions of ``massive'' hadrons is computed in
Fig.~\ref{fig:ktdis} (left) for various values of $\lambda$ for jets with
hardness $Q=119$~GeV.
As expected, as $\lambda$ gets larger, soft gluon emission
is strongly suppressed such that the distribution flattens
at small $k_\perp$, while more
hadrons are produced at large $\kt$, making in turn the distributions harder.
We also compare in Fig.~\ref{fig:ktdis} (right) these calculations
with CDF preliminary data, all normalized to the
$\log(\kt/1{\rm GeV})=-0.1$ bin
as before. The best description  is reached
in the limiting spectrum approximation, or at least for small values
of $\lambda\lesssim 0.5$. This is not too surprising since these
inclusive measurements mostly involve pions.

Predictions beyond the limiting spectrum were shown to describe very well the hump-backed 
shape of the inclusive spectra for various hadron species; in particular, the hadron-mass variation 
of the peak turned out to be in good agreement with QCD expectations~(see e.g. \cite{KhozeOchs}). The softening of 
the $\kt$-spectra with increasing hadron masses predicted in Fig.~\ref{fig:ktdis} is an observable worth to be measured, as this would provide an 
additional and independent check of the LPHD hypothesis beyond the limiting spectrum. 
This could only be achieved if the various species of hadrons inside a jet can be identified experimentally. Fortunately, it is likely to be the case at the LHC, where the ALICE~\cite{identificationalice}
and CMS~\cite{identificationcms}  experiments at the Large Hadron Collider have good 
identification capabilities at not too large transverse momenta.
\begin{figure}[ht]
\begin{center}
  \includegraphics[height=7.2cm,width=8cm]{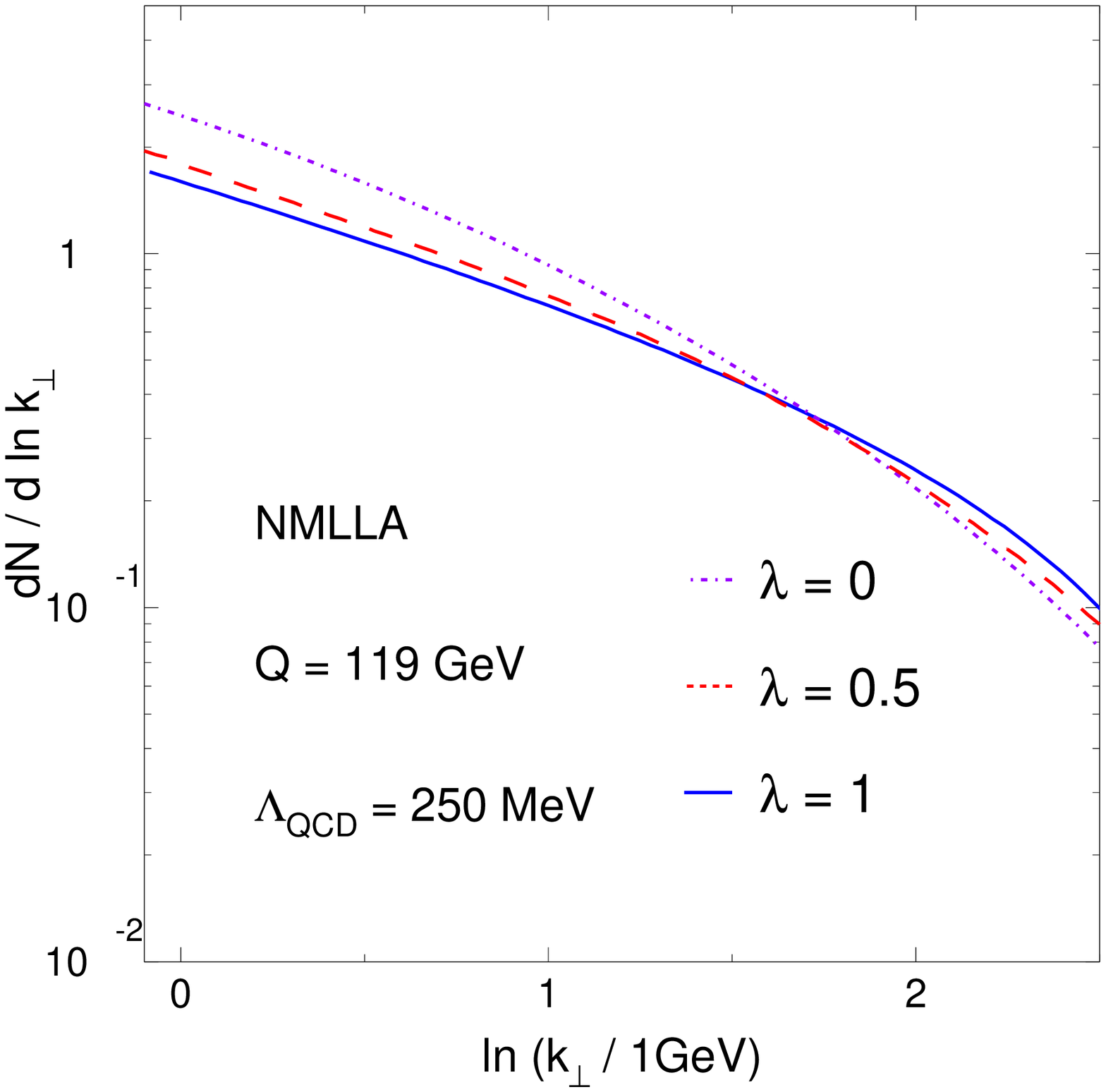}
\hskip 1cm
  \includegraphics[height=7.2cm,width=8cm]{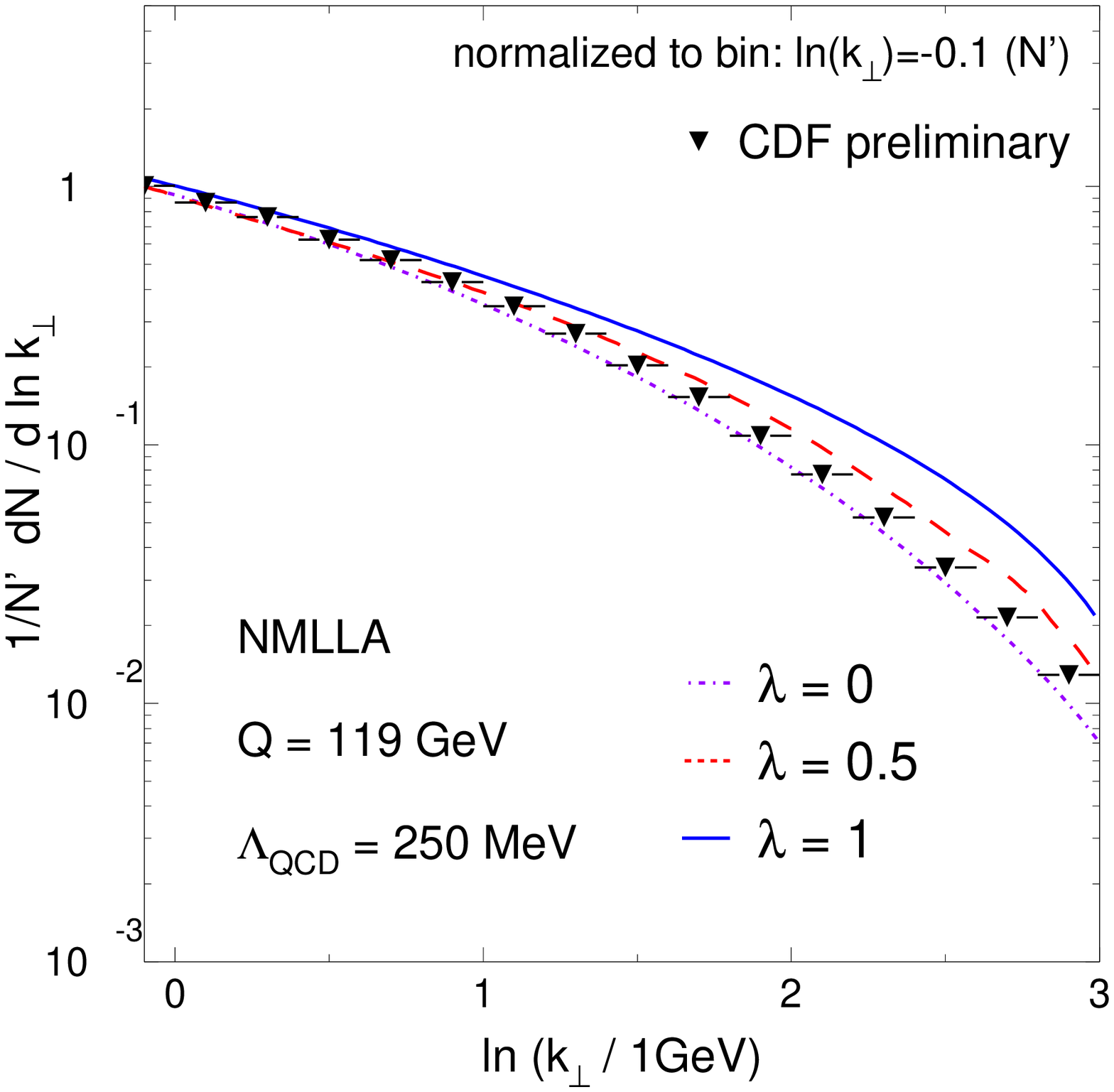}
 \caption{\label{fig:ktdis} Absolute (left) and normalized (right)
inclusive $\kt$-distribution beyond the limiting spectrum approximation
at NMLLA in a jet of hardness $Q=119$~GeV.}
\end{center}
\end{figure}
%

%%%%%%%%%%%%%%%%%%%%%%%%%%%%%%%%%%%%%%%%%%%%
\section{2-particle correlations}
\label{section:TPC}
%%%%%%%%%%%%%%%%%%%%%%%%%%%%%%%%%%%%%%%%%%%%

\subsection{Correlators and evolution equations}
\label{subsection:corr}
%%%%%%%%%%%%%%%%%%%%%%%%%%%%%%%%%%%%%%%%%%%%%%%%

We work, like in \cite{RPR2}, with the normalized correlators
\begin{equation}
{\cal C}_{g} = \frac{G^{(2)}}{G_1 G_2}, \quad
{\cal C}_{q} = \frac{Q^{(2)}}{Q_1 Q_2}
\end{equation}
where $G_i, Q_i, i=1,2$ are the inclusive spectra relative to the outgoing
hadrons $h_1$ and $h_2$, and $G^{(2)}, Q^{(2)}$ are the 2-particle
distributions in gluon and quark jets, respectively.
The former are obtained by a single
differentiation of the ``MLLA'' generating functional $Z$, and the latter 
by differentiating it twice \cite{RPR2} (see also the
discussion introduced in \ref{section:sis}). $Z$ satisfies the
evolution equation described in section (2) of \cite{RPR2}:
$\dd{Z_{\tt A}}/\dd\ln\Theta$ for the jet initiating parton {\tt A}
is expressed as an integral over $z$ involving the DGLAP splitting functions
$\Phi_{\tt A}^{\tt BC}(z)$ and $Z_{\tt B}$ and $Z_{\tt C}$
associated to the products {\tt B} and
{\tt C} of the splitting process; {\tt B} carries away the
fraction $z$ of the energy $E$ of {\tt A} and {\tt C}
the fraction $(1-z)$ (see Fig.~\ref{fig:correl}).
The topology of Fig.~\ref{fig:correl} respects the exact AO constraint
over the successive emission angles of
partons ($\Theta\geq\Theta_1\geq\Theta_2$).
\begin{figure}[ht]
\begin{center}
  \includegraphics[height=5cm,width=8cm]{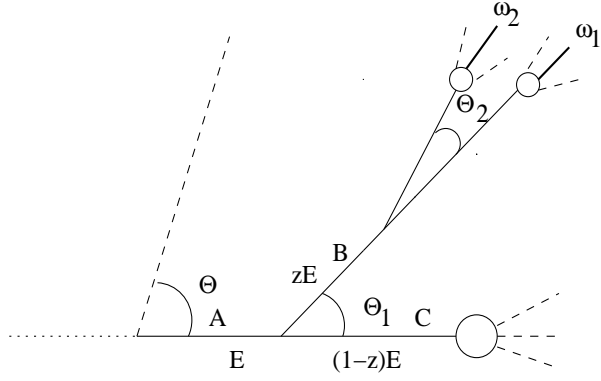}
\caption{\label{fig:correl} 2-particle correlations inside a jet}
\end{center}
\end{figure}
In practice, suitably differentiating the master evolution equation
for $Z_{\tt A}$, which arises as a consequence of exact AO in parton
cascades, yields, for the {\it correlation functions}~\cite{Basics}
\begin{equation}
G^{(2)} - G_1 G_2 \equiv ({\cal C}_g -1)\ G_1\ G_2,\quad
Q^{(2)} - Q_1\ Q_2 \equiv ({\cal C}_q -1)\ Q_1\ Q_2,
\end{equation}
the system of coupled evolution equations:

\vbox{
\begin{eqnarray}
\label{eq:Q2prsub}
(Q^{(2)}-Q_1Q_2)_{y}  &=&  \int_0^1 \dd{z}\>\frac{\alpha_s}{\pi}\
\> \Phi_q^g(z)\>\bigg[ G^{(2)}(z)+ \Big(Q^{(2)}(1-z)-Q^{(2)}\Big)\\
&& \hskip -1cm + \Big(G_1(z)-Q_1\Big)\Big(Q_2(1-z) - Q_2\Big) +  \Big(G_2(z)-
Q_2\Big)\Big(Q_1(1-z)- Q_1\Big)\bigg],\nonumber\\
\label{eq:G2prsub}
(G^{(2)}-G_1G_2)_{y} &=&
\int_0^1 \dd{z}\> \frac{\alpha_s}{\pi}
\Phi_g^g(z)\>\bigg[ \Big(G^{(2)}(z)-zG^{(2)}\Big)
  + \Big(G_1(z)-G_1\Big)\Big(G_2(1-z)-G_2\Big) \bigg] \cr
&&+  \int_0^1 \dd{z}\>\frac{\alpha_s}{\pi}\, n_f\Phi_g^q(z)\>
\bigg[ 2\Big(Q^{(2)}(z)-Q_1(z)Q_2(z)\Big) - \Big(G^{(2)}-G_1G_2\Big)\cr
&&+ \Big(2Q_1(z)-G_1\Big)\Big(2Q_2(1-z)-G_2\Big) \bigg].
\end{eqnarray}
}

The derivative is taken with respect to $y =Y -\ell$ rather than with
respect to $\ln \Theta$, since it is more convenient when a
collinear cutoff is imposed (see Section (2.1) of \cite{RPR2}).
Like for the inclusive
spectra, the notations have been lightened to a maximum, with $G^{(2)}$
standing for $G^{(2)}(z=1)$ and likewise for $Q^{(2)}$.
The notation $x_i, \ell_i, \ldots$ refers to the $\ell_i =\ln(1/x_i)$ of
the outgoing parton (hadron) $i$.

\subsection{Including NMLLA corrections}
\label{subsection:expandeveq}
%%%%%%%%%%%%%%%%%%%%%%%%%%%%%%%%%%%%%%%

We follow the same logic, exposed in Section~\ref{subsection:logic},
for the 2-particle distributions $Q^{(2)}, G^{(2)}$, as the one used for the
inclusive spectra $B$ in Section~\ref{subsec:dd1p}.
Therefore, the expansion at small $x_1,x_2$ is performed for $\frac{x_1}{z}
Q_1\left(\frac{x_1}{z}\right)\frac{x_2}{z} Q_2\left(\frac{x_2}{z}\right)$
and  $\frac{x_1}{z} G_1\left(\frac{x_1}{z}\right)\frac{x_2}{z} 
G_2\left(\frac{x_2}{z}\right)$ as well as for 
$\frac{x_1}{z}\frac{x_2}{z} Q^{(2)}\left(\frac{x_1}{z},\frac{x_2}{z}\right)$
and
$\frac{x_1}{z}\frac{x_2}{x} G^{(2)}\left(\frac{x_1}{z},\frac{x_2}{z}\right)$,
 similarly to Eq.~(\ref{eq:logic}).

\subsubsection{Quark jet}
%%%%%%%%%%%%%%%%%%%%%%%%%
 
Operating like for (\ref{eq:nmlla2}) and (\ref{eq:nmlla3}),
the first (MLLA) term in the r.h.s. of (\ref{eq:Q2prsub}) can be cast in
the form

\vbox{
\begin{eqnarray}
&&\hskip -0.5 cm\int_0^1 \dd{z}\>\frac{\alpha_s}{\pi}\>
 \Phi_q^g(z)\>\left[ G^{(2)}(z)+ \Big(Q^{(2)}(1-z)-Q^{(2)}\Big)\right]=
\frac{C_F}{N_c}
\left[\int_0^1\frac{\dd{z}}z\gamma_0^2G^{(2)}(z)\right] - \frac34\frac{C_F}{N_c}\gamma_0^2G^{(2)}\cr
&&\hskip -0.3cm
+\frac{C_F}{N_c}\left[\frac78+\frac{C_F}{N_c}\left(\frac58-\frac{\pi^2}6
\right)\right]\gamma_0^2 G^{(2)}_\ell
+\left(\frac{C_F}{N_c}\right)^2\left(\frac{C_F}{N_c}-1\right)
\left(\frac58-\frac{\pi^2}6
\right)\gamma_0^2 (G_1G_2)_\ell,
\label{eq:corrnmlla1}
\end{eqnarray}
}

where we have plugged the DLA formula~\cite{DFK} 
\begin{equation}\label{eq:dlaQ2}
Q^{(2)}_\ell=\frac{C_F}{N_c}G^{(2)}_\ell
+\frac{C_F}{N_c}\left(\frac{C_F}{N_c}-1\right)(G_1G_2)_\ell
+{\cal O}(\gamma_0^2)
\end{equation}
in the r.h.s. of (\ref{eq:corrnmlla1}); the terms
in (\ref{eq:dlaQ2}) of relative order ${\cal O}(\gamma_0)$ 
are neglected because their contribution provide corrections
to (\ref{eq:corrnmlla1})
beyond NMLLA (see also appendix \ref{section:Q2}).
The second and third terms in the r.h.s. of (\ref{eq:Q2prsub}) provide the
NMLLA correction:

\vbox{
\begin{eqnarray}
\int_0^1 \dd{z}\>\frac{\alpha_s}{\pi}\>
 \Phi_q^g(z)\Big(G_1(z)-Q_1\Big)\Big(Q_2(1-z) - Q_2\Big)
&=&\frac{\alpha_s}{\pi}\left(\int_0^1\dd{z}\>\Phi_q^g(z)
\ln(1-z)\right)(G_1-Q_1)Q_{2,\ell}\cr
&=&\left(\frac{C_F}{N_c}\right)^2
\left(1-\frac{C_F}{N_c}\right)
\left(\frac58-\frac{\pi^2}{6}\right)\gamma_0^2G_1G_{2,\ell},\cr
&&
\label{eq:corrnmlla2}
\end{eqnarray}
}
where the DLA expression $Q_\ell=\frac{C_F}{N_c}G_\ell+{\cal O}(\gamma_0^2)$
is used~\cite{DFK}; further corrections $({{\cal O}(\gamma_0^2)})$
to this formula
are here again dropped out because their inclusion goes beyond
the present resummation logic.  Likewise, we have
\begin{eqnarray}
\int_0^1 \dd{z}\>\frac{\alpha_s}{\pi}\>
 \Phi_q^g(z)\Big(G_2(z)-Q_2\Big)\Big(Q_1(1-z) - Q_1\Big)
&=&\left(\frac{C_F}{N_c}\right)^2
\left(1-\frac{C_F}{N_c}\right)
\left(\frac58-\frac{\pi^2}{6}\right)\gamma_0^2G_{1,\ell}G_2.\cr
&&
\label{eq:corrnmlla3}
\end{eqnarray}
Gathering (\ref{eq:corrnmlla1}), (\ref{eq:corrnmlla2})
and (\ref{eq:corrnmlla3})
yields
\begin{equation}\label{eq:nmllaqeq}
\left(Q^{(2)}-Q_1Q_2\right)_y=
\frac{C_F}{N_c}
\left[\int_0^1\frac{\dd{z}}z\gamma_0^2G^{(2)}(z)\right] - \frac34\frac{C_F}{N_c}\gamma_0^2G^{(2)}
+\frac{C_F}{N_c}\left[\frac78+\frac{C_F}{N_c}\left(\frac58-\frac{\pi^2}6
\right)\right]\gamma_0^2 G^{(2)}_\ell,
\end{equation}
which is written in a form similar to (\ref{eq:nmllaquark}).

\subsubsection{Gluon jet}
%%%%%%%%%%%%%%%%%%%%%%%%%

The structure of (\ref{eq:G2prsub}) can be worked out in the same way.
The first integral term in its r.h.s. is the same as that in
(\ref{eq:nmllagg}), such that we can simply set
\begin{eqnarray}
&&\int_0^1 \dd{z}\> \frac{\alpha_s}{\pi}
\Phi_g^g(z)\>\Big(G^{(2)}(z)-zG^{(2)}\Big) =
\left[\int_0^1\frac{\dd{z}}z\gamma_0^2G^{(2)}(z)\right]
-\frac{11}{12}\gamma_0^2G^{(2)}+\left(\frac{67}{36}-\frac{\pi^2}{6}\right)
\gamma_0^2G^{(2)}_\ell.\cr
&&
\label{eq:nmllacorrg1}
\end{eqnarray}
The second term provides a contribution
$$
\frac{\gamma_0^2}{2N_c}
G_{1\ell}G_{2\ell}\int_0^1 \dd{z}\>\Phi_g^g(z)\ln z\ln(1-z)=
\left[\frac{11\pi^2}{36}-\frac{395}{108}
+2\zeta(3)\right]\gamma_0^2G_{1\ell}G_{2\ell}={\cal O}(\gamma_0^4),
$$
that is beyond NMLLA and therefore dropped out here.
The second line of (\ref{eq:G2prsub}) simplifies to
\begin{eqnarray}
&&\int_0^1 \dd{z}\>\frac{\alpha_s}{\pi}\, n_f\Phi_g^q(z)\>
\bigg[ 2\Big(Q^{(2)}(z)-Q_1(z)Q_2(z)\Big) - \Big(G^{(2)}-G_1G_2\Big)\bigg]
=\frac{n_fT_R}{3N_c}\gamma_0^2\cr
&&\times\bigg[ 2\Big(Q^{(2)}-Q_1Q_2\Big) - \Big(G^{(2)}-G_1G_2\Big)\bigg]
-\frac{13}{18}\frac{n_fT_R}{N_c}\gamma_0^2\Big(Q^{(2)}-Q_1Q_2\Big)_\ell,
\label{eq:nmllacorrg2}
\end{eqnarray}
and the third one gives

\vbox{
\begin{eqnarray}
&&\hskip -1cm\int_0^1 \dd{z}\>\frac{\alpha_s}{\pi}\, n_f\Phi_g^q(z)\>
\Big(2Q_1(z)-G_1\Big)\Big(2Q_2(1-z)-G_2\Big)=\frac{n_fT_R}{3N_c}\gamma_0^2
\Big(2Q_1-G_1\Big)\Big(2Q_2-G_2\Big)\cr
&&\hskip 4cm-\frac{13}{18}\frac{n_fT_R}{N_c}\gamma_0^2\Big[(2Q_1-G_1)Q_{2\ell}+
(2Q_2-G_2)Q_{1\ell}\Big].
\label{eq:nmllacorrg3}
\end{eqnarray}
}

Gathering (\ref{eq:nmllacorrg1}), (\ref{eq:nmllacorrg2}),
(\ref{eq:nmllacorrg3}) and setting 
(see appendix \ref{section:Q2} for further explanations)
\begin{equation}\label{eq:approxim}
Q\approx\frac{C_F}{N_c}G+{\cal O}(\gamma_0),\qquad Q^{(2)}=\frac{C_F}{N_c}G^{(2)}+\frac{C_F}{N_c}
\left(\frac{C_F}{N_c}-1\right)G_1G_2+{\cal O}(\gamma_0)
\end{equation}
in the subleading pieces, we obtain the NMLLA equation
for the gluonic correlator

\vbox{
\begin{eqnarray}
\left(G^{(2)}-G_1G_2\right)_y&=&\left[\int_0^1\frac{\dd{z}}z\gamma_0^2G^{(2)}(z)\right]
-\left[\frac{11}{12}+\frac{n_fT_R}{3N_c}\left(1-2\frac{C_F}{N_c}\right)\right]
\gamma_0^2G^{(2)}\cr
&+&\frac{2n_fT_R}{3N_c}\left(1-\frac{C_F}{N_c}\right)
\left(1-2\frac{C_F}{N_c}\right)\gamma_0^2G_1G_2\label{eq:nmllageq}+
\left(\frac{67}{36}-\frac{\pi^2}6-\frac{13}{18}\frac{n_fT_R}{N_c}\frac{C_F}{N_c}\right)
\gamma_0^2G^{(2)}_{\ell}\cr
&+&\left[\frac{13}{9}\frac{n_fT_R}{N_c}\frac{C_F}{N_c}
\left(1-\frac{C_F}{N_c}\right)
\right]\gamma_0^2(G_1G_2)_\ell.\label{eq:nmllagcorreq}
\end{eqnarray}
}

The way to get the  equations for the correlators ${\cal C}_g$ and ${\cal
C}_q$, to be solved iteratively, proceeds like in Section~4 and
Appendices A and B of \cite{RPR2}.

\subsection{ NMLLA correlators}
\label{subsection:GC}
%%%%%%%%%%%%%%%%%%%%%%%%%%%%%%%

\subsubsection{Gluon correlator ${\cal C}_g$}
%%%%%%%%%%%%%%%%%%%%%%%%%%%%%%%%%%%%%%%%%%%%

The differential expression for (\ref{eq:nmllagluon}) reads
\begin{equation}\label{eq:gsdiffeq}
G_{\ell y}=\gamma_0^2G-a_1\gamma_0^2\left(\psi_\ell-\beta_0\gamma_0^2\right)G
+a_2\gamma_0^2\left(\psi_\ell^2+\psi_{\ell\ell}-\beta_0\gamma_0^2\psi_\ell\right)G.
\end{equation}
Differentiating (\ref{eq:nmllageq}) with respect to $\ell$ gives
the following NMLLA differential equation

\vbox{
\begin{eqnarray}
\left(G^{(2)}-G_1G_2\right)_{\ell y}&=&
\gamma_0^2G^{(2)}-a_1\gamma_0^2
\left(G_\ell^{(2)}-\beta_0\gamma_0^2G^{(2)}\right)+(a_1-b_1)\gamma_0^2
\left[(G_1G_2)_\ell-\beta_0\gamma_0^2G_1G_2\right]\cr\cr
&+&a_2\gamma_0^2\left(G^{(2)}_{\ell\ell}-\beta_0\gamma_0^2G^{(2)}_\ell
\right)+b_2\gamma_0^2\left[(G_1G_2)_{\ell\ell}-
\beta_0\gamma_0^2(G_1G_2)_{\ell}\right],\label{eq:nmllageq1}
\end{eqnarray}
}

where $a_1$, $a_2$ are given by (\ref{eq:a1}) and (\ref{eq:a2}), and with the following coefficients:
\begin{equation}
b_1=\frac{11}{12}-\frac{n_fT_R}{3N_c}\left(1-\frac{2C_F}{N_c}\right)^2
\stackrel{n_f=3}{=}0.915,\ \ 
b_2=\frac{13}{9}\frac{n_fT_R}{N_c}\frac{C_F}{N_c}\left(1-\frac{C_F}{N_c}\right)
\stackrel{n_f=3}{\approx}0.18.
\label{eq:eqb}
\end{equation}
Noting $\psi=\ln G$ and $\chi=\ln {\cal C}_g$, the second line
of (\ref{eq:nmllageq1}) can be rewritten
in terms of logarithmic derivatives of $G$ and of
${\cal C}_g$ (see Appendix \ref{sec:logderiv})
from which Eq.~(\ref{eq:nmllageq1}) is solved iteratively.
Setting $G^{(2)}={\cal C}_gG_1G_2$ in both members
and making use of (\ref{eq:gsdiffeq}) leads to the analytical solution
of (\ref{eq:nmllageq1}), valid for arbitrary $\lambda$
\begin{equation}
 {\cal C}_g-1
=\frac{1-\delta_1-b_1\left(\psi_{1,\ell}+\psi_{2,\ell}-
[\beta_0\gamma_0^2]\right)-\left[a_1\chi_{\ell}+\delta_2\right]
+\delta_3}
{1+\Delta
+\delta_1+\Big[a_1\left(\chi_{\ell}+
{[\beta_0\gamma_0^2]}\right)+\delta_2\Big]+\delta_4},
\label{eq:CGfull}
\end{equation}
where, like in \cite{RPR2}, we introduce $\eta = \ell_2 - \ell_1$.
 $\delta_3$ and $\delta_4$ are the new NMLLA corrections:
\begin{equation}
\label{eq:delta}
\begin{split}
\delta_3(\ell_1,\ell_2;\eta)\ =\ & a_2f_1(\ell_1,\ell_2;\eta)
            + b_2 f_2(\ell_1,\ell_2;\eta)\ =\ {\cal O}(\gamma_0^2),\\
\delta_4(\ell_1,\ell_2;\eta)\ =\ & -a_2 f_3(\ell_1,\ell_2;\eta)\
=\ {\cal O}(\gamma_0^2),
\end{split}
\end{equation}
and $f_1$, $f_2$ and $f_3$ are defined in (\ref{eq:fi}) of
appendix \ref{sec:logderiv}. Setting $\delta_3=\delta_4=0$ in
(\ref{eq:CGfull}), one recovers the exact analytical solution of the
corresponding MLLA gluon equation (with $a_2=b_2=0$ in (\ref{eq:nmllageq1})); 
to derive this formula we have used the same method
that was, for the first time, implemented in the appendix A of
\cite{RPR2}. The other quantities and their
order of magnitude are (see \cite{RPR2})
\begin{eqnarray}\label{eq:nota4bis}
\chi &=&  \ln {\cal C}_g,\ \ 
\chi_{\ell} = \frac{\dd\chi}{\dd\ell}={\cal O}(\gamma_0^2),\ \ 
\chi_{y}=\frac{\dd\chi}{\dd y}={\cal O}(\gamma_0^2),\\
\psi_{i} &=& \ln G_{i},\ \ 
\psi_{i,\ell}= \frac{1}{G_i}\frac{\dd G_i}{\dd\ell}={\cal O}(\gamma_0),\ \ 
\psi_{i,y}= \frac{1}{G_i}\frac{\dd G_i}{\dd y}={\cal O}(\gamma_0),\
(i=1,2),\\
\Delta &=& \gamma_0^{-2}
\Big(\psi_{1,\ell}\psi_{2,y}+\psi_{1,y}\psi_{2,\ell}\Big)={\cal O}(1),\\
\delta_1 &=& \gamma_0^{-2}\Big[\chi_{\ell}(\psi_{1,y}+\psi_{2,y}) +
   \chi_{y}(\psi_{1,\ell}+\psi_{2,\ell})\Big]={\cal O}(\gamma_0),\label{eq:delta1}\\
\delta_2 &=& \gamma_0^{-2}\Big(\chi_{\ell}\chi_{y} + \chi_{\ell\,y}\Big)
={\cal O}(\gamma_0^2).
\label{eq:nota4}
\end{eqnarray}
To evaluate (\ref{eq:nota4bis}) we consider the bare correlator:
$$
\chi=\ln\left[1+\frac{1-b_1\left(\psi_{1,\ell}+\psi_{2,\ell}\right)+
[b_1\beta_0\gamma_0^2]}
{1+\Delta+[a_1\beta_0\gamma_0^2]}\right].
$$
the derivatives of which are  calculated numerically to eventually 
determine (\ref{eq:delta1}) and (\ref{eq:nota4}). 

The analytical result (\ref{eq:CGfull}) for ${\cal C}_g$ will be
numerically displayed for the limiting spectrum $\lambda=0$
in section \ref{subsec:compar} by using (\ref{eq:ifD}).
For the case $\lambda \not=0$, we report the reader to \cite{RPR3} where
it has been treated in MLLA by the steepest descent method.

\subsubsection{Quark correlator ${\cal C}_q$}
%%%%%%%%%%%%%%%%%%%%%%%%%%%%%%%%%%%%%%%%%%%%
%
The differential expression of (\ref{eq:nmllaquark}) reads
\begin{equation}\label{eq:qsdiffeq}
Q_{\ell y}=\frac{C_F}{N_c}\left\{\gamma_0^2G-\frac34\gamma_0^2(\psi_\ell
-\beta_0\gamma_0^2)G+\tilde{a}_2\gamma_0^2(\psi_{\ell}^2+\psi_{\ell\ell}
-\beta_0\gamma_0^2\psi_{\ell})G\right\}.
\end{equation}
Differentiating (\ref{eq:nmllageq}) with respect to $\ell$ gives
the NMLLA differential equation
\begin{equation}
(Q^{(2)}-Q_1Q_2)_{\ell y}=\frac{C_F}{N_c}\left\{
\gamma_0^2G^{(2)}-\frac34\gamma_0^2\left(G^{(2)}_\ell-\beta_0
\gamma_0^2G^{(2)}\right)+
\tilde a_2\gamma_0^2\left(G^{(2)}_{\ell\ell}-\beta_0
\gamma_0^2G^{(2)}_{\ell}\right)
\right\},\label{eq:Q2eq}
\end{equation}
to be solved iteratively.
Setting $Q^{(2)}={\cal C}_qQ_1Q_2$ in both members and using
(\ref{eq:qsdiffeq}),
one gets the analytical solution  (\ref{eq:Q2eq}), valid
for arbitrary $\lambda$
{\small
\begin{equation}
{\cal C}_q-1	
=\displaystyle\frac{
     \frac {N_c}{C_F} {\cal C}_g
       \Big[ 1-\textstyle {\frac34}\Big(\psi_{1,\ell}
  +\psi_{2,\ell} +[\chi_{\ell}] - [\beta_0\gamma_0^2]\Big) + \tilde\delta_3\Big]
\frac{C_F}{N_c}\frac{G_1}{Q_1} \frac{C_F}{N_c}\frac{G_2}{Q_2}
-\tilde\delta_1 -[\tilde\delta_2]}
{\widetilde\Delta + \Big[1-\textstyle {\frac34}
  \big(\psi_{1,\ell}-[\beta_0\gamma_0^2]\big)+\tilde\delta_{4,1}
\Big]\frac{C_F}{N_c}\frac{G_1}{Q_1}
+\Big[1-\textstyle {\frac34} \big(\psi_{2,\ell}-[\beta_0\gamma_0^2]\big)
+\tilde\delta_{4,2}\Big]
\frac{C_F}{N_c}\frac{G_2}{Q_2}
+ \tilde\delta_{1}+[\tilde\delta_{2}]},
\label{eq:Qcorr}
\end{equation}}
where $\tilde\delta_3$ and $\tilde\delta_4$ are the new NMLLA
coefficients ($\tilde a_2$ is given by (\ref{eq:tildea2}))
\begin{equation}\label{eq:delta3}
\begin{split}
\tilde\delta_3(\ell_1,\ell_2;\eta)&=\tilde a_2f_1(\ell_1,\ell_2;\eta)={\cal O}(\gamma_0^2),\\
\tilde\delta_{4,i}(\ell_1,\ell_2;\eta)&=\tilde a_2 f_4(\ell_1,\ell_2;\eta).
={\cal O}(\gamma_0^2).
\end{split}
\end{equation}
Setting $\tilde\delta_3=\tilde\delta_{4,i}=0$ in
(\ref{eq:Qcorr}), one recovers the exact analytical solution of the
corresponding MLLA quark equation ($\tilde a_2=0$ in (\ref{eq:Q2eq}))
that was obtained in the appendix B of
\cite{RPR2}.
We have introduced (see ~\cite{RPR2})
\begin{eqnarray}
\widetilde\Delta &=&
\gamma_0^{-2}\Big(\varphi_{1,\ell}\varphi_{2,y}+\varphi_{1,y}\varphi_{2,\ell}\Big)
={\cal O}(1),\\
\tilde\delta_1&=&
\gamma_0^{-2}\Big[\sigma_{\ell}(\varphi_{1,y}+\varphi_{2,y}) +
\sigma_{y}(\varphi_{1,\ell}+\varphi_{2,\ell})\Big]={\cal O}(\gamma_0),\\
\tilde\delta_2 &=&
\gamma_0^{-2}\Big(\sigma_{\ell}\sigma_{y}+\sigma_{\ell\,y}\Big)=
{\cal O}(\gamma_0^2),
\label{eq:nota5}
\end{eqnarray}
with $\varphi_k = \ln Q_k$ and $\sigma= \ln {\cal C}_q$.
For the numerical computation of $\sigma$, we take
\begin{equation}
\sigma=\ln\left\{1+\displaystyle\frac{
     \frac {N_c}{C_F} {\cal C}_g
       \Big[ 1-\textstyle {\frac34}\Big(\psi_{1,\ell}
  +\psi_{2,\ell} +[\chi_{\ell} - \beta_0\gamma_0^2]\Big) \Big]
\frac{C_F}{N_c}\frac{G_1}{Q_1} \frac{C_F}{N_c}\frac{G_2}{Q_2}}
{\widetilde\Delta + \Big[1-\textstyle {\frac34}
  \big(\psi_{1,\ell}-[\beta_0\gamma_0^2]\big)\Big]\frac{C_F}{N_c}\frac{G_1}{Q_1}
+\Big[1-\textstyle {\frac34} \big(\psi_{2,\ell}-[\beta_0\gamma_0^2]\big)\Big]
\frac{C_F}{N_c}\frac{G_2}{Q_2}}\right\},
\end{equation}
in which one uses the NMLLA expression (\ref{eq:ratioqg}) for $G$ and $Q$
deduced from (\ref{eq:solq}) and (\ref{eq:solg}),
and the exact expression (\ref{eq:CGfull}) for ${\cal C}_g(\ell_1,y_2,\eta)$.

The numerical solution of (\ref{eq:Qcorr}) is given  in section 
\ref{subsec:compar} for $\lambda=0$. We make the approximation
$\varphi_{\ell}\approx\psi_\ell$, $\varphi_y\approx\psi_y$
that is justified in Appendix \ref{section:exactcc} through
(\ref{eq:varphi}). We can therefore also use (\ref{eq:ifD}).
The case  $\lambda\ne0$ was also dealt with at MLLA
for a quark jet  in \cite{RPR3}.

Finally, taking $x_1=x_2$ in (\ref{eq:CGfull},\ref{eq:Qcorr}) and going to
the asymptotic limit $Q\to\infty$ ($Y\to\infty$), one finds the
implicit overall normalization of these observables to be given by those
of the multiplicity correlators \cite{konishi}

$$
{\cal C}_g\stackrel{Y\to\infty}{\rightarrow}
\frac{\langle n_g(n_g-1)\rangle}{\langle n_g\rangle^2}=\frac{4}{3}, \qquad
{\cal C}_q\stackrel{Y\to\infty}{\rightarrow}
\frac{\langle n_q(n_q-1)\rangle}{\langle n_q\rangle^2}
=1+\frac{N_c}{3C_F},
$$
for the gluon and quark jets respectively. The statement above can 
be easily explained; the asymptotic expressions of 
(\ref{eq:CGfull},\ref{eq:Qcorr}) are respectively the DLA 
formul\ae (see \cite{Basics})
$$
{\cal C}_g(x_1,x_2)\stackrel{Y\to\infty}{\approx}1+
\frac{1}{1+\Delta(x_1,x_2)},\qquad 
{\cal C}_q(x_1,x_2)\stackrel{Y\to\infty}{\approx}1+\frac{N_c}{C_F}
\frac{1}{1+\Delta(x_1,x_2)},
$$  
and $\Delta(x_1,x_2)=2$ for $x_1=x_2$ in the same limit.

\subsection{NMLLA corrections versus MLLA}
\label{subsec:compar}
%%%%%%%%%%%%%%%%%%%%%%%%%%%%%%%%%%%%%%%%%%

Throughout this analysis, we have consistently incorporated
a set of NMLLA corrections. These were not calculated
in the previous work~\cite{RPR2} which was done at MLLA accuracy for 
$\lambda=0$.
The philosophy and the basic technique are, however, the same
(as well as in~\cite{Dremin4}).
We comment below on the role of these corrections for 2-particle correlations.
Both $\delta_3$ and $\tilde\delta_3$ are dominated
by their leading term, such that
$$
\delta_3\approx(a_2+b_2)(\psi_{1,\ell}+\psi_{2,\ell})^2={\cal O}(\gamma_0^2),
\qquad\tilde\delta_3\approx \tilde a_2(\psi_{1,\ell}+\psi_{2,\ell})^2
={\cal O}(\gamma_0^2).
$$
Since both $a_2+b_2$ and $\tilde a_2$ are positive and
$\psi_\ell$ increases as $\ell\to0$, NMLLA corrections are expected to
increase the  MLLA solution of \cite{RPR2} in the limit $\ell_1+\ell_2\to0$,
as can be seen in (\ref{eq:CGfull}) and (\ref{eq:Qcorr}).
Thus, as found for the single-inclusive $\kt$-distribution,
the  $(x_1,x_2)$ domain in which the two particles are ``correlated'',
{\it i.e.} ${\cal C}_{g,q}-1>0$,  becomes larger  than at MLLA.
In the limit $\ell_1+\ell_2\to 2Y$, the role of the new corrections is,
on the contrary, expected to vanish since $\psi_\ell\to0$ when $\ell\to Y$.

This is indeed what appears on Figs.~\ref{fig:fw}
and \ref{fig:fw2}, which compare the MLLA and NMLLA solutions
at the Tevatron energy scale ($Q=155$ GeV).
While Eqs.(\ref{eq:CGfull}) and (\ref{eq:Qcorr}) are general analytical
solutions of the evolution equations at $\lambda\not= 0$, the numerical
results displayed below are calculated at the limiting spectrum
$\lambda=0$, by plugging
the formula (\ref{eq:ifD}) for the inclusive spectrum into (\ref{eq:CGfull})
and (\ref{eq:Qcorr}).
\begin{figure}[ht]
\begin{center}
  \includegraphics[height=7cm,width=8cm]{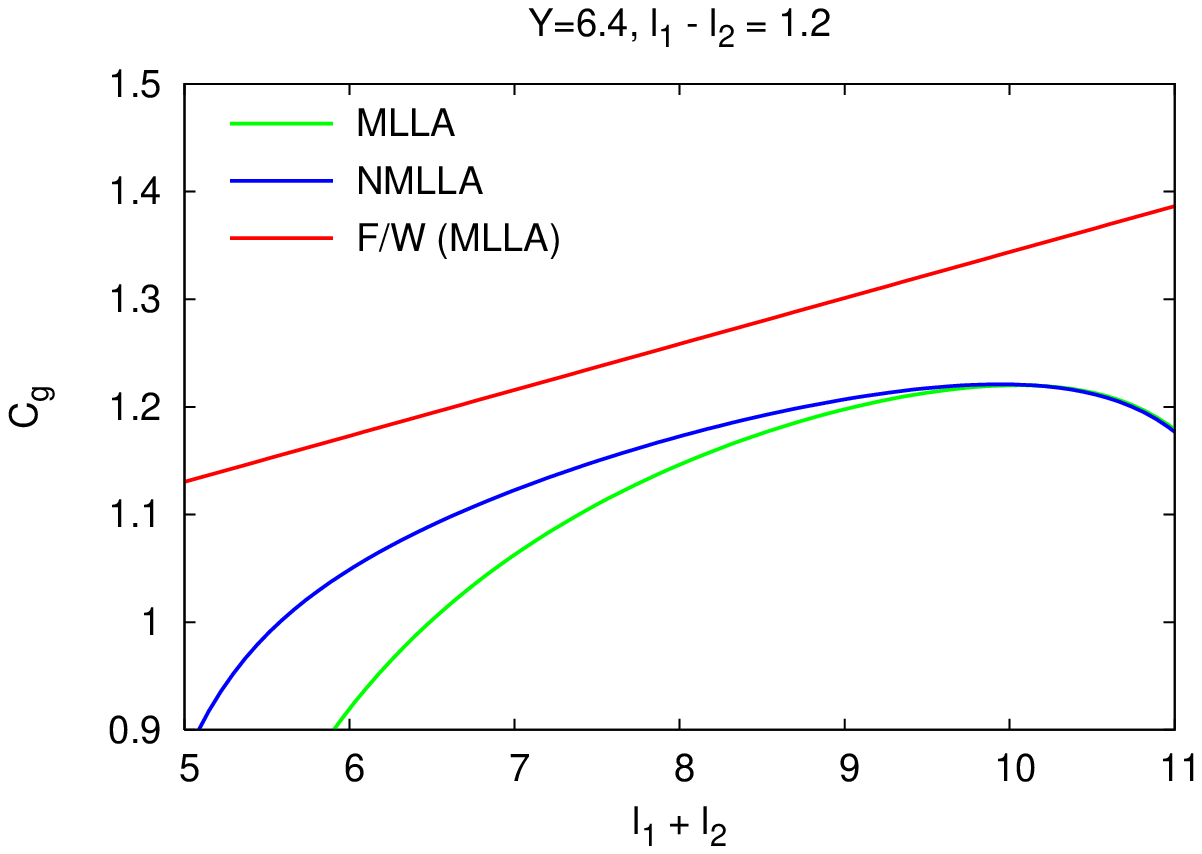}
\hskip 1cm
  \includegraphics[height=7cm,width=8cm]{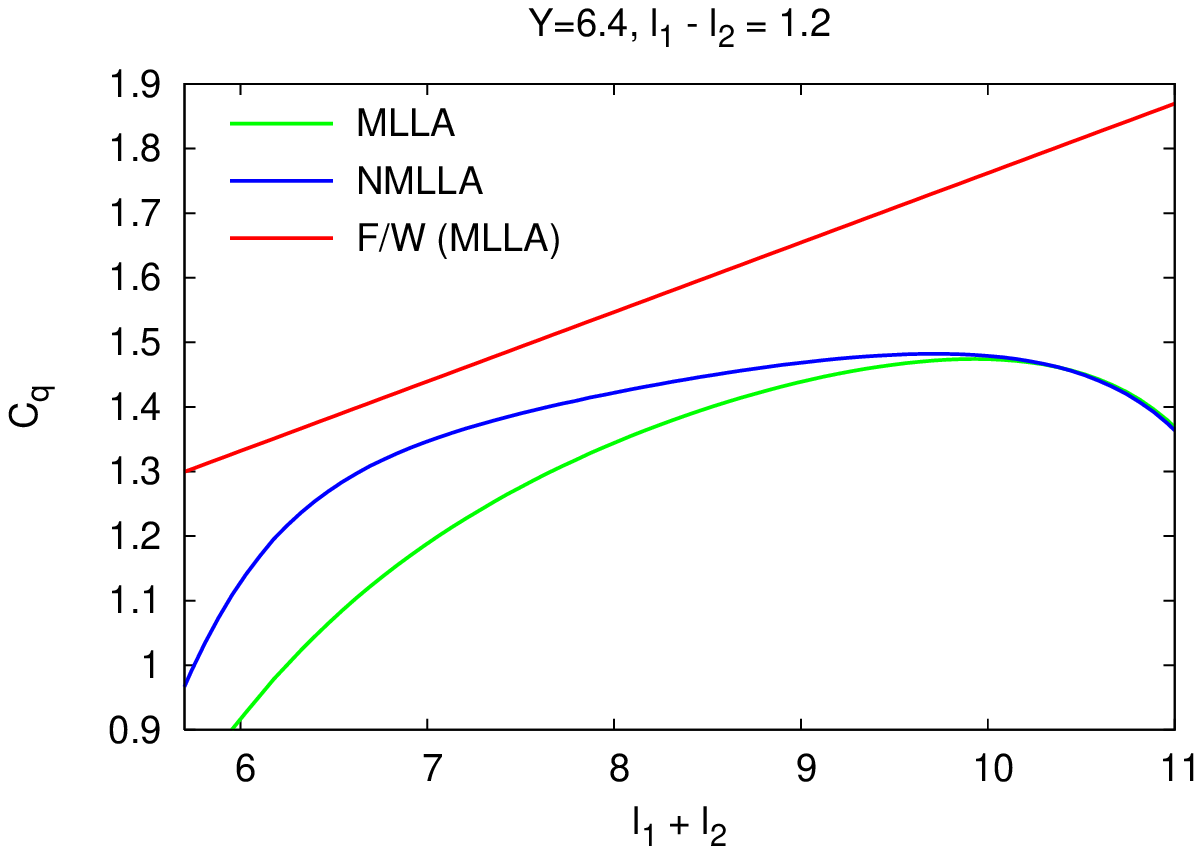}
\vskip 5mm
  \includegraphics[height=7cm,width=8cm]{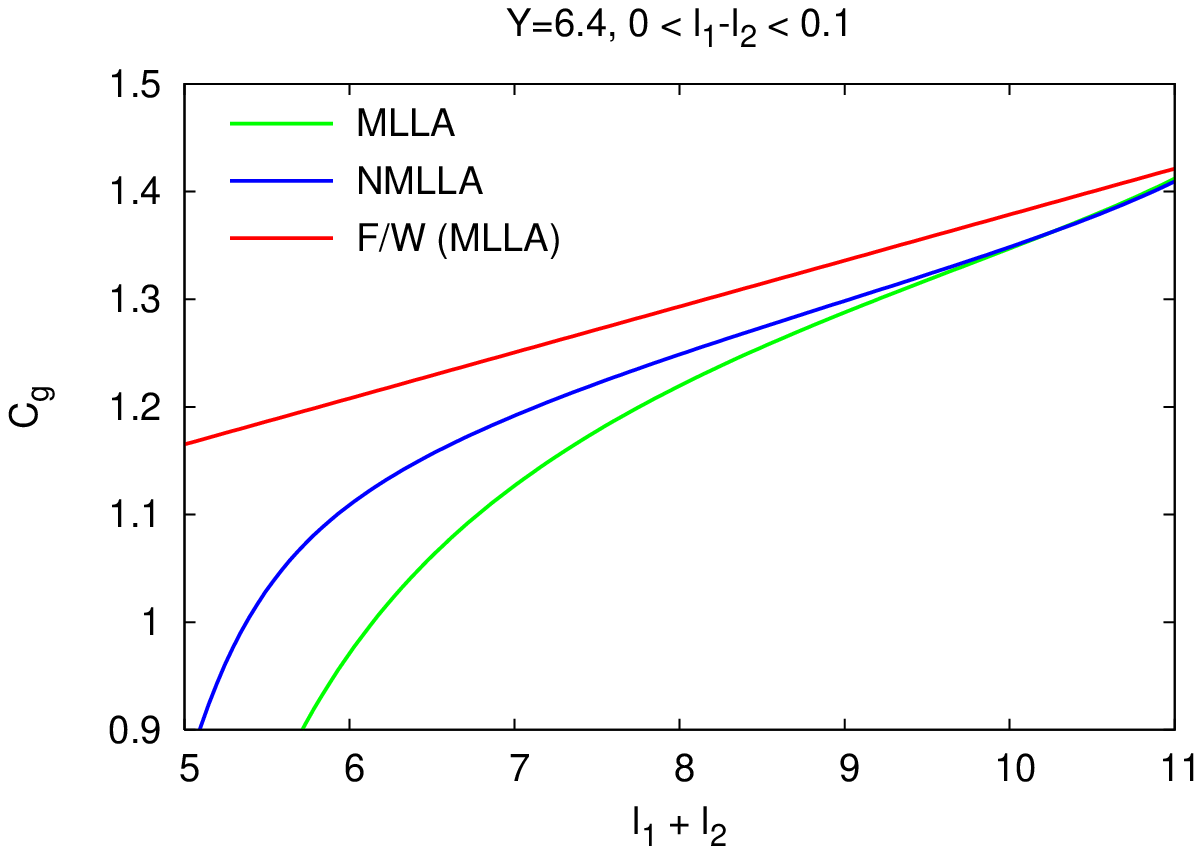}
\hskip 1cm
  \includegraphics[height=7cm,width=8cm]{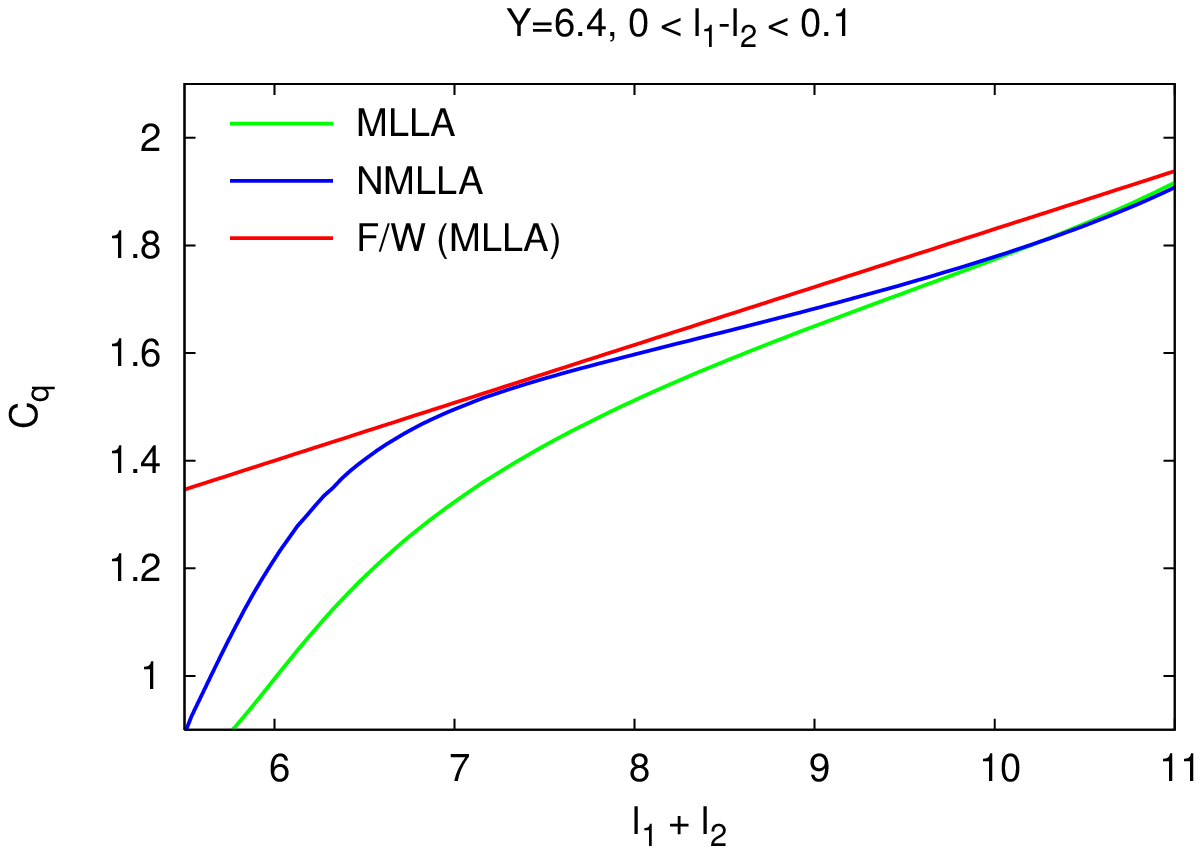}
\caption{\label{fig:fw} 2-particle correlations for a quark jet (left) and
a gluon jet (right) as a function of $\ell_1 +
\ell_2$ for $\ell_1 = \ell_2$; the MLLA, NMLLA and Fong and Webber~\cite{FW} predictions are shown as solid lines.}
\end{center}
\end{figure}
The four lines in Fig.~\ref{fig:bands} show the positions in
$(\ell_1, \ell_2)$ space corresponding to the curves of Figs \ref{fig:fw}
and \ref{fig:fw2}. The two upper curves
of Fig.~\ref{fig:fw} correspond to line 2, its two lower curves to line 1;
the two upper curves of Fig.~\ref{fig:fw2} correspond to line 3 and its two
lower curves to line 4.
\begin{figure}[ht]
\begin{center}
  \includegraphics[height=5cm,width=6cm]{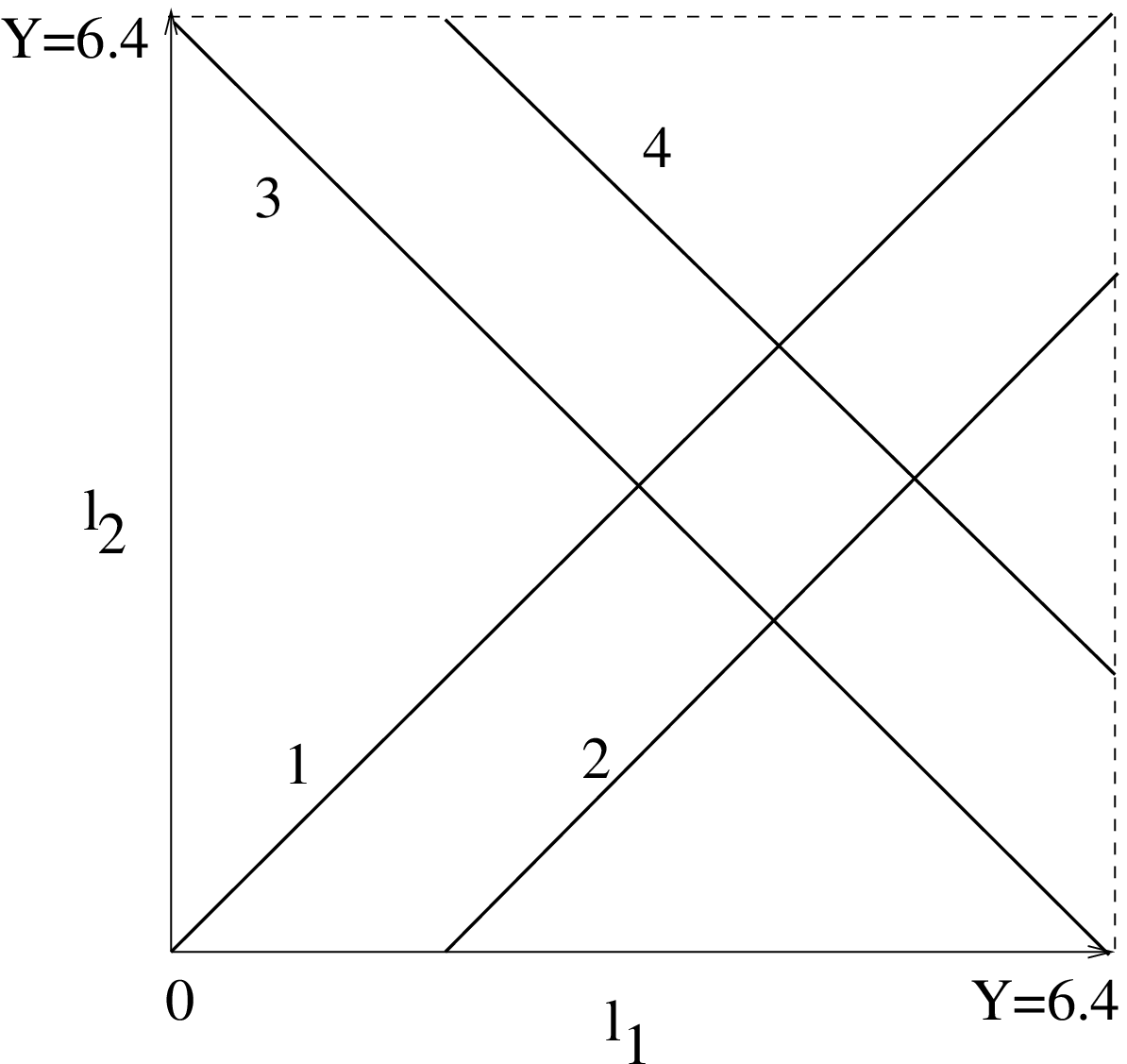}
\caption{\label{fig:bands} Positions in $(\ell_1,\ell_2)$ space of Figs.
\ref{fig:fw} and \ref{fig:fw2}.}
\end{center}
\end{figure}
\begin{figure}[ht]
\begin{center}
  \includegraphics[height=7cm,width=8cm]{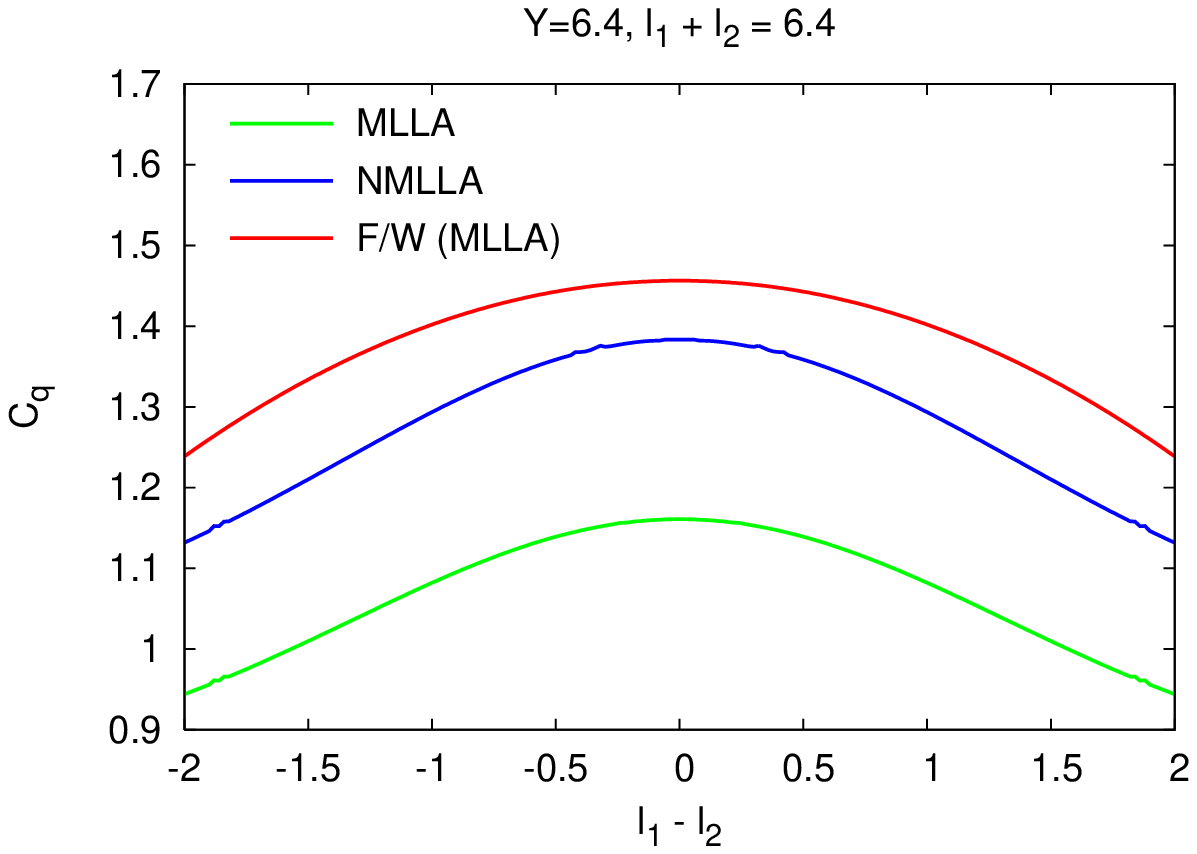}
\hskip 1cm
  \includegraphics[height=7cm,width=8cm]{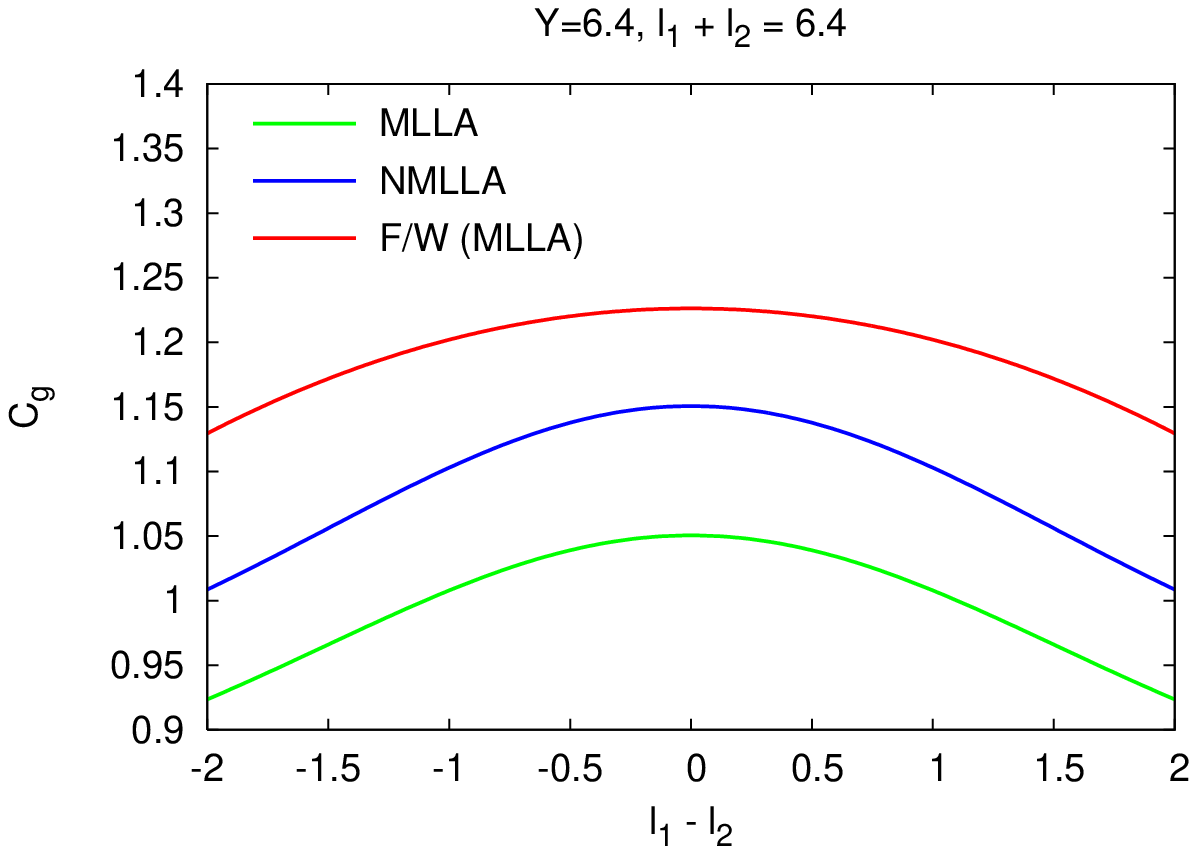}
\vskip 5mm
  \includegraphics[height=7cm,width=8cm]{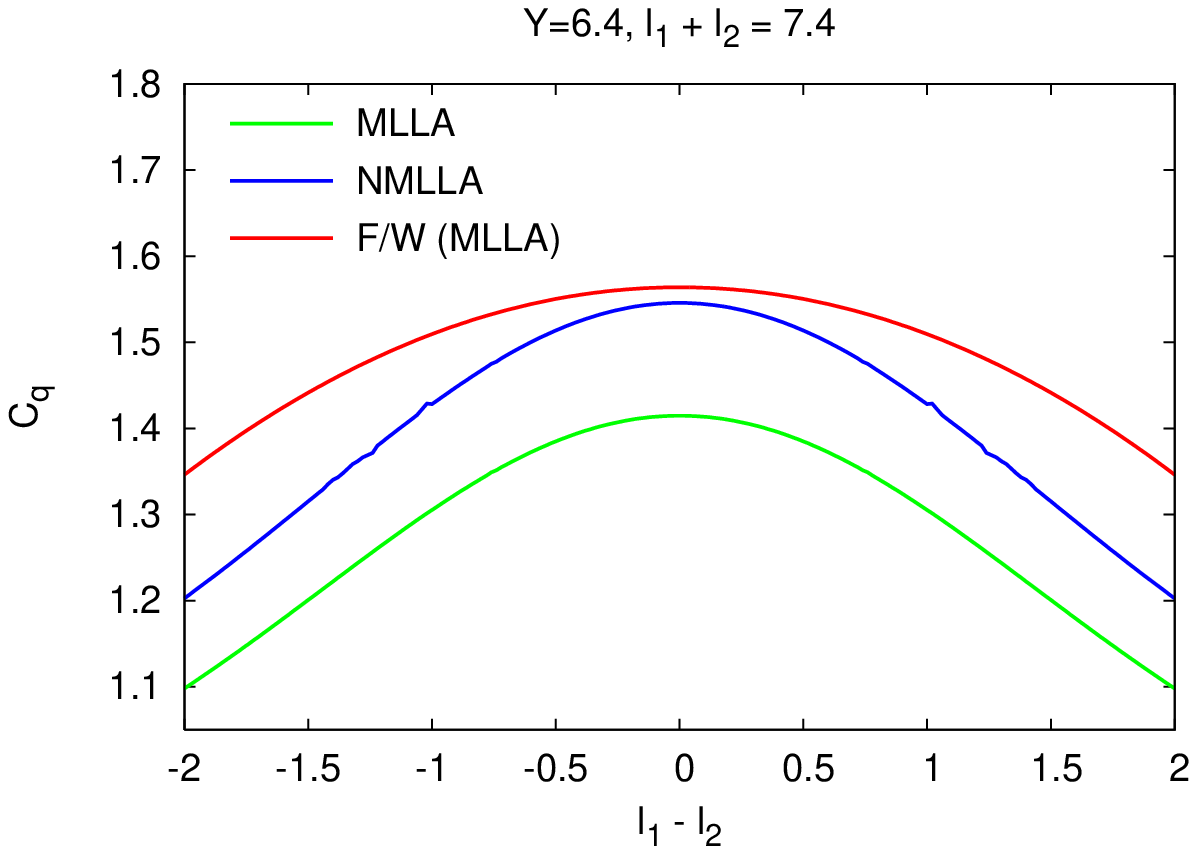}
\hskip 1cm
  \includegraphics[height=7cm,width=8cm]{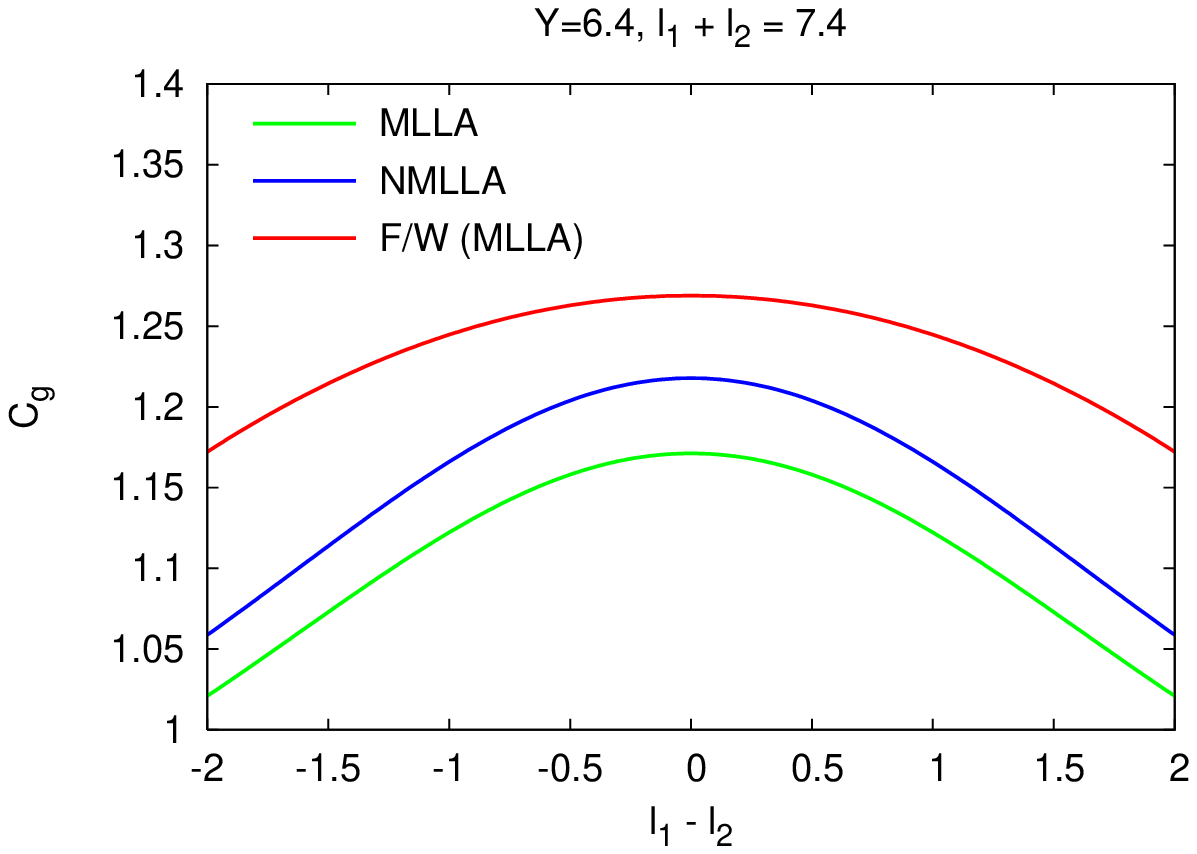}
\caption{\label{fig:fw2} 2-particle correlations for a quark jet (left) and
a gluon jet (right) as a function of $\ell_1 - \ell_2$;
MLLA, NMLLA and Fong-Webber prediction.}
\end{center}
\end{figure}
The correlations displayed in Fig.~\ref{fig:fw} and \ref{fig:fw2} appear more
important in NMLLA than in MLLA. Physically, because
the recoil of each emitting parton is better taken into account in the
former approximation, less energy becomes available and the multiplicity of
emitted particles is expected to decrease. Consequently, inside a bunch
of a fewer number of particles, two among them get more correlated.

\subsection{Dependence on $\boldsymbol{\Lambda_{QCD}}$}
\label{subsec:Lambdacorr}
%%%%%%%%%%%%%%%%%%%%%%%%%%%%%%%%%%%%%%%%%%%%%%%%%%%%%%%%

We have tested the dependence of the gluonic correlation function
${\cal C}_g$ on $\lqcd$, by varying it from $150$ MeV to $500$ MeV.
The results are displayed on Fig.~\ref{fig:Lambdacorr}, as a function
of $\ell_1 + \ell_2$ (left) and $\ell_1 -\ell_2$ (right).
Variations are seen to stay below $10\%$.
\begin{figure}[ht]
\begin{center}
  \includegraphics[height=7cm,width=8cm]{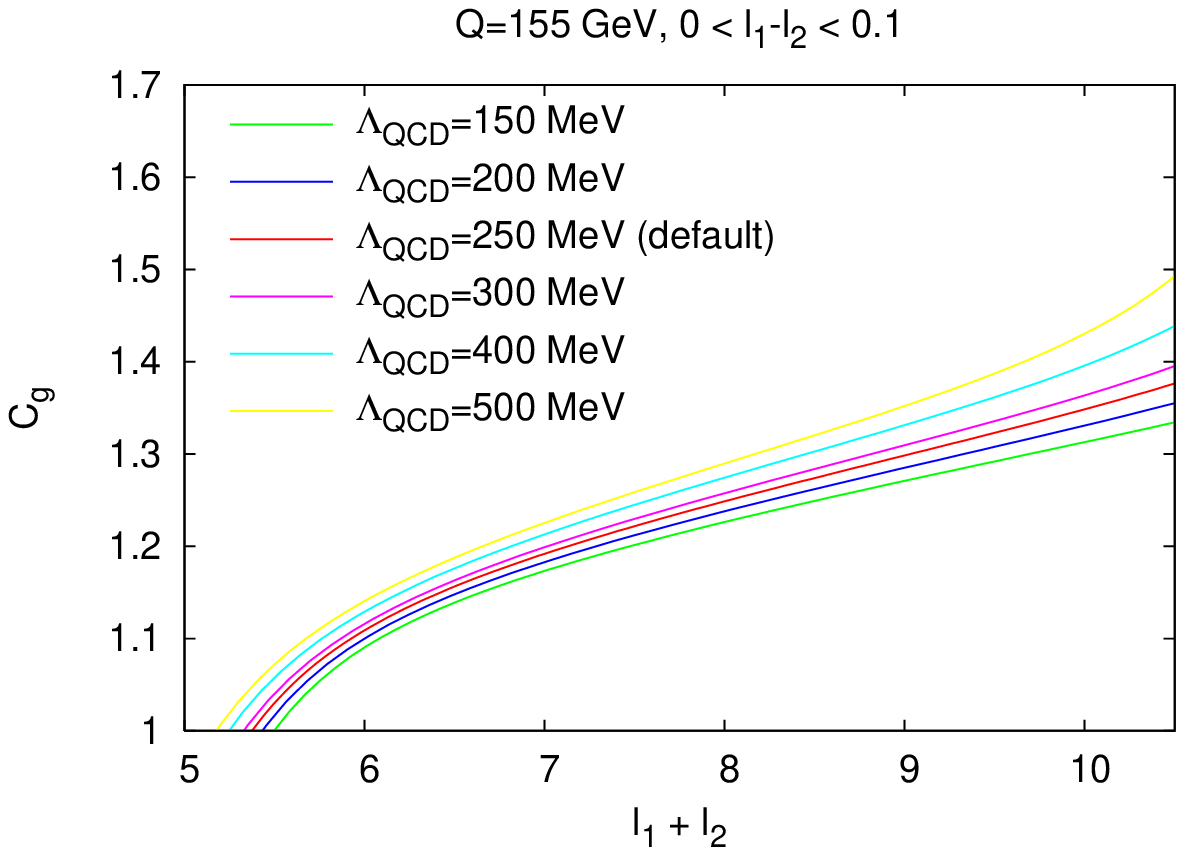}
\hskip 1cm
  \includegraphics[height=7cm,width=8cm]{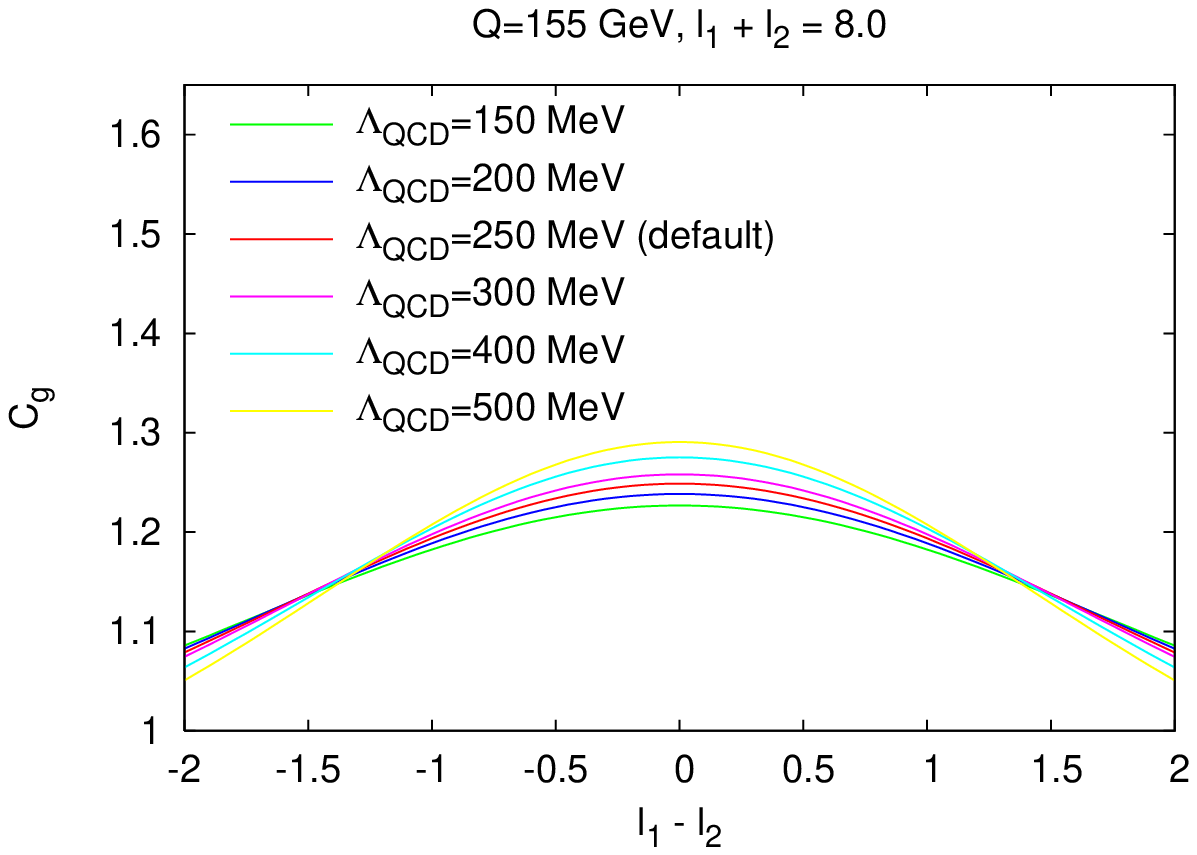}
\caption{\label{fig:Lambdacorr} dependence of the gluonic 2-particle
correlation function ${\cal C}_g$ on $\Lambda_{QCD}$.}
\end{center}
\end{figure}

\subsection{Comparison with Fong and Webber MLLA predictions}
%%%%%%%%%%%%%%%%%%%%%%%%%%%%%%%%%%%%%%%%%%%%%%%%%%%%%%%%%%%%%%

The comparison with the predictions by Fong and Webber \cite{FW}
is also instructive.
Let us recall that their calculation is done at MLLA, yet obtained
from the exact result of \cite{RPR2} when the two outgoing partons are
taken to be close to the peak of the inclusive spectrum, and when
the exact solution is expanded at first order in
$\sqrt{\alpha_s}$. From the present results and that of~\cite{RPR2},
we can conclude that:
\begin{itemize}
\item the convergence of the series obtained by expanding the exact MLLA
result in powers of $\sqrt{\alpha_s}$ is very bad; if one proceeds in this
way, NMLLA corrections may be as large as MLLA, making the series
meaningless;  note that similar
conclusions have been obtained in \cite{CuypersTesima} when dealing with
recoil effects and, more precisely, with the role of exact kinematics in
the bounds of integrations;
\item instead, in the procedure that has been adopted here,  {\it i.e.}
finding exact NMLLA solutions of the (approximate) MLLA evolution
equations, NMLLA corrections turn out to be under control and their 
inclusion  brings the predictions closer to Fong and Webber's.
\end{itemize}
The present study, together with \cite{RPR2}, consequently
stresses out the
importance of dealing with {\it exact} solutions of the evolution
equations in jet calculus.

%%%%%%%%%%%%%%%%%%%%%%%%%%%%%%%%
\section{Discussion and summary}
\label{sec:CONCL}
%%%%%%%%%%%%%%%%%%%%%%%%%%%%%%%%

\subsection{Discussion}
\label{se:discussion}
%%%%%%%%%%%%%%%%%%%%%%%

Energy conservation is a fundamental issue in jet calculus.
While it is well known that the complete neglect of the recoil of the
emitting parton leads to  DLA (taking only into
account the singular parts of the splitting functions),
the MLLA, in which ``single logarithms'' are added to DLA,
takes partial account of the recoil.
Corrections  appearing at higher orders in an expansion in powers of
$\sqrt{\alpha_s}$ come from (i) the shifts by $\ln z$ and $\ln(1-z)$ in the
arguments of the hadronic fragmentation functions; (ii) the non-singular
terms in the splitting functions; (iii) the running of $\alpha_s$.
Our line of approach in this paper was accordingly the following:
\begin{itemize}
\item we considered MLLA evolution equations as kinetic equations of QCD, and
expanded their (exact) analytical solutions in powers of $\sqrt{\alpha_s}$ 
up to the order $\cO{\alpha_s}$. Contributions that do not fit into such an
framework are discarded;
\item we stuck to the logic advocated in \cite{Dremin1,Dremin2,Dremin4}
that, at small $x$ and
for $|\ln z| \sim |\ln(1-z)| \ll \ln (1/x)$, the successive corrections,
MLLA, NMLLA \ldots , which better and better account for energy
conservation, are taken care of by a systematic expansion in powers of $\ln z$
and $\ln(1-z)$.
\end{itemize}

The size of the NMLLA terms depends on the precise definition of
$\lqcd$: a rescaling of $\lqcd$ would change the terms at this
order. Systematically solving this problem would require a 2-loop
calculation which has not been obtained so far for any
multiplicity-related observable. We therefore have to consider here
$\lqcd$ as a phenomenological parameter. The sensitivity of our results
to variations of $\lqcd$ have been studied and found moderate ($20\%$ for
inclusive $k_\perp$-distribution and less than $10\%$ for correlations)
when $\lqcd \to 2\lqcd$.

We  left aside the  question of the
matching of the two definitions of the jet axis, ``inclusive'' direction of
the energy flow in this work, and ``exclusive'' fixing from all outgoing
hadrons in experiments.

Last, hints that  NMLLA corrections that has been considered here
are the dominant one can already be found in the work \cite{DokKNO} 
where this type of NMLLA recoil effects was  shown to drastically affect
particle multiplicities  and particle correlations through a factor
proportional to the number of partons involved in the process.
This however only concerns {\em a priori} 2-particle correlations.
Spanning a gate between KNO phenomenon and the techniques that we have used
here stays a challenging task which we hope to achieve in the future.

Since calculated NMLLA corrections proved to be quite substantial, a
natural question arises concerning the magnitude of higher order
corrections. There, in correlation with the remarks at the end of the
introduction of section \ref{section:sis}, it seems legitimate to consider
that, since this observable is mainly sensitive to soft particles, the
corrections are expected to be moderate. This can be different for
integrated quantities like multiplicities.

\subsection{Summary}
\label{sec:summary}
%%%%%%%%%%%%%%%%%%%%%%%

In this work, we have computed next-to-MLLA (NMLLA) corrections to the
single-inclusive $\kt$-distributions as well as 2-particle momentum
correlations inside a jet at high energy colliders. It comes as a
natural extension of \cite{PerezMachet} and \cite{RPR2} in which MLLA
results are provided.  In particular, it exploits the same logic of
using, at small energy fraction $x$ of the emitted hadron, exact
solutions to (approximate) evolution equations for the inclusive
spectrum.  The technique used is based on a systematic expansion in powers of
$\sqrt{\alpha_s}$ which neglects non-perturbative effects.
Nevertheless, it proves to be remarkably efficient to describe the
preliminary measurements of (the shape of)
 the $\kt$-differential inclusive cross
section performed by CDF~\cite{CDF}. This is an indication that
non-perturbative contributions play a small role
in this observable, and concentrates in the overall normalization
(LPHD hypothesis is tantamount to stating that in this universal factor
lies the trace of the (soft and local) hadronization process).
 The transition from MLLA to next-to-MLLA enlarges
considerably the domain where the computations agree with the
experimental data, both in the transverse momentum of hadrons and
in their energy fraction $x$.

 In our analysis, single-inclusive
$x$-distributions as well as
$\kt$-spectra have been determined exactly beyond the limiting
spectrum approximation, {\it i.e.} for arbitrary $Q_0\ne\lqcd$. This
should in particular be relevant when dealing with distributions of
rather massive hadrons~\cite{finitelambda}. In this respect,
experimental identification of outgoing hadrons could provide precious
additional tests of LPHD and of the physical interpretation
 of the infrared cutoff $Q_0$ as the ``hadronization scale''.
 As far as 2-particle
correlations inside a jet are concerned, future results from 
LHC, in addition to the ones of OPAL \cite{opal} and
recent ones from CDF \cite{CDFcorr},
are waited for to be compared with the  NMLLA predictions
presented in this study.

The limitations of the method are in particular:
\begin{itemize}
\item neglecting non-perturbative contributions may prove less justified
for not so inclusive observables. In that respect, forthcoming data on
2-particle correlations from  LHC promise to be very
instructive. While incorporating some  non-perturbative contributions
is not excluded {\it a priori}, a systematic way to handle them is of
course still out of reach;
\item the absolute normalization of the distributions, which involve
non-perturbative effects (hadronization)  is not predicted;
\item the calculation is performed in the small-$x$ limit and extrapolation
to larger $x$ may become problematic. The transition to larger $x$,
or from MLLA to DGLAP evolution equations, is
undoubtedly also a very important issue. It may be tempting to
proceed in this direction by going to higher orders in the expansion
initiated in \cite{PerezMachet,RPR2,RPR3} and extended
in the present work. However, the universality of MLLA evolution
equations as kinetic equations of QCD should be cast on firmer grounds.
\end{itemize}

\subsection{Perspective: going to larger $\boldsymbol x$}
\label{subsection:meaning}
%%%%%%%%%%%%%%%%%%%%%%%%%%%%%%%%%%%%%%%%%%%%%%%%%%%%%%%%%

A Taylor expansion, when used inside evolution equations, was already
advocated long ago to better account for energy conservation
\cite{Dremin4,Dremin1}.
It appears fairly easy to realize that pushing it at higher and higher
orders of $\ln u$ at small $x$ inside the convolution
integral (\ref{eq:F})  should play a role in
it extending the domain of reliability of the solution
to larger and larger values of $x$.
Indeed, in  (\ref{eq:F}), one integrates from $u=x$ to $u=1$ a certain
function $F(\ln u - \ln x)$. $F$ is expanded at large $|\ln x|$
around $|\ln u|=0$, which corresponds to $u=1$. If one increases $x$, the
domain of integration shrinks closer and closer to its upper bound
$u=1$. Suppose that we set $x = 1-\epsilon$. The integral
becomes $\int_{1-\epsilon}^1 du F(\ln u - \ln (1-\epsilon))$. Now, in the
argument of $F$, for all $u$ in the domain of integration $|\ln u| \sim
|\ln x|$, such that a reliable expansion of $F$, if it exists (it depends of
its radius of convergence), must involve a large number of terms. This is like
expanding a function $f(t-a)$ around $f(-a)$: for $|a| \sim |\ln x|
\gg t \sim \ln u \approx \ln 1$, a few powers of $t$ provide a good
approximation to $f(t-a)$, but when $a$ decreases,
expanding $f(t-a)$ around $f(-a)$ uses an expansion parameter
$t$ of the same order of magnitude as $a$ itself.  We  conclude that
{\it increasing} $x$ requires going to higher and higher orders in the
expansion of $F$ in powers of $|\ln u|$. Conversely, going to higher and
higher order in this expansion is expected to yield a solution valid in a
larger and larger domain of $x$.

When applied to the evolution equations themselves, and to the similar
expansion in powers of $(\ln z)$ that we did in section \ref{section:sis}, the
same kind of arguments apply, which are not unrelated with the link between
MLLA and DGLAP evolution equations. Since NMLLA corrections to 2-particle
correlations, unlike the ones for the inclusive $k_\perp$ distribution, are
directly connected with NMLLA corrections to the evolution equations
themselves, it is worth giving a few comments concerning this issue.

a/ That MLLA evolution equations (\ref{eq:qpr}) and (\ref{eq:gpr}) are,
at least  for inclusive enough observables,  valid in a much broader $x$
domain than expected has been known for a long time
\cite{Basics}.
It was furthermore noticed some years ago \cite{LO} that, for
parton multiplicities, the exact numerical solution of MLLA evolution
equations perfectly matched experimental results in a very large
domain, and that, accordingly, the MLLA evolution equations contain more
information than expected
and  the problems of finding their
analytical solutions are essentially of technical nature;

b/ at small $x$ MLLA evolution equations are identical to DGLAP evolution
equations but for a shift by $\ln z$ ($z$ is the integration variable) of
the variable $Y = y + \ell$ which controls the evolution of the jet hardness
\cite{RPR2,Basics};

c/ for soft outgoing hadrons ($x$ small $\Leftrightarrow
|\ell| \equiv |\ln x|$ large), this shift is negligible in the hard
parton region ($|\ln z| < |\ln x|$). However,
when going to harder hadrons, that is when $x$ grows, $|\ell|$ decreases and
$|\ln z|$ is no longer negligible. When it is so, the function to integrate
is no longer safely approximated by its $0\raise 4pt\hbox{\tiny th}$ order expansion
(corresponding to $\ln z =0$) and higher
powers of $\ln z$ are needed. This provides, in addition to the
argumentation at the beginning of this subsection, another link between
this expansion at higher orders and going to larger $x$;

d/ accordingly, the Taylor expansion that we used inside
MLLA evolution equations, which  extends their ``validity''
to  larger $x$,
may contribute to spanning a bridge between them and DGLAP evolution
equations (see for example \cite{AKKO,DMS}).

\vskip .5cm

\noindent \underline{{\it Acknowledgments:}} It is pleasure to thank
Yu.L.~Dokshitzer, I.M.~Dremin and W.~Ochs for illuminating discussions.
We also thank S.~Jindariani (CDF) for an invaluable exchange of
information concerning CDF data. 

%%%%%%%%%%%%%%%%%%%%%%%%%%%%%%%%%%%%%%%%%%%%%%%%%%%%%%%%%%
\newpage
\appendix

%%%%%%%%%%%%%%%%%%%%%%%%%%%%%%%%%%%%%%%%%%%%%%%%%%%%%%%
\section{NMLLA corrections neglected in the derivation
of the approximate equations for the inclusive spectrum}
\label{section:Q1}
%%%%%%%%%%%%%%%%%%%%%%%%%%%%%%%%%%%%%%%%%%%%%%%%%%%%%%%

To get a self-contained equation for the inclusive spectrum
inside a gluon jet, one needs consistently to plug in
\begin{eqnarray}
Q&=&\frac{C_F}{N_c}\left[1+(a_1-\tilde a_1)\psi_\ell\right]G+
{\cal O}(\gamma_0^2)\label{eq:approxCF},\cr
Q_\ell&=&\frac{C_F}{N_c}G_\ell+{\cal O}(\gamma_0^2)
\end{eqnarray}
respectively, in the first and second terms
of the r.h.s. of (\ref{eq:nmllagq}). Taking into account
the correction proportional to $\psi_\ell$ in (\ref{eq:approxCF}) 
would provide an extra term
$$
\ldots+\frac{2}{3}\frac{n_fT_R}{N_c}\frac{C_F}{N_c}(a_1-\tilde a_1)G_\ell
$$
which adds to the r.h.s. of (\ref{eq:nmllagluon})
and slightly changes the value of $a_2$ (\ref{eq:a2}) from $0.06$ to 
$0.08$; this number is also small, such that the
approximation that we justify in  Appendix 
\ref{section:roleofnmlla} keeps valid.

%%%%%%%%%%%%%%%%%%%%%%%%%%%%%%%%%%%%%%%%%%%%%%%%%%%%%%%%%%%%%%
\section{Steepest descent evaluation of (\ref{eq:solg})
for constant $\boldsymbol{\gamma_0^2}$}
\label{section:roleofnmlla}
%%%%%%%%%%%%%%%%%%%%%%%%%%%%%%%%%%%%%%%%%%%%%%%%%%%%%%%%%%%%%%

We solve the self-contained gluon equation (\ref{eq:solg})
with frozen $\alpha_s$ by performing the Mellin's transform
\begin{equation}\label{eq:mellin}
G(\ell,y)=\iint_C\frac{\dd\omega\ \dd\nu}{(2\pi i)^2}\ e^{\omega\ell}\ e^{\nu y}\ {\cal G}(\omega,\nu).
\end{equation}
The contour of integration ($C$) lies to the right of all poles, and
${\cal G}(\omega,\nu)$ is the ``propagator'' in Mellin's space.
Plugging (\ref{eq:mellin}) into (\ref{eq:solg}) yields
\begin{equation}
G(\ell,y)=\int_C\frac{\dd\omega}{2\pi i}\exp\left[\omega\ell+\gamma_0^2\left(\frac1{\omega}-a_1\right)y
+a_2\gamma_0^2\omega y\right].
\end{equation}
The simplest way to estimate the previous Mellin's representation is by
substituting the DLA saddle point $\omega_0=\gamma_0\sqrt{\frac{y}{\ell}}$
into the MLLA ($\propto a_1$) and NMLLA ($\propto a_2$) terms. Doing so,
the steepest descent evaluation of the inclusive spectrum at fixed
$\alpha_s$ in the limit $\ell\gg1$ ($x\ll1$) leads to
\begin{equation}\label{eq:SDspec}
G(\ell,y)\approx\frac12\sqrt{\frac{\gamma_0\>y^{1/2}}
{\pi\>\ell}}
\exp\left(2\gamma_0\sqrt{\ell y}-a_1\gamma_0^2y+a_2\gamma_0^3\>y\sqrt{\frac{y}{\ell}}
\right).
\end{equation}
The result, plotted on Fig.~\ref{fig:SIS7.5} together with the DLA and MLLA
results (still at fixed $\alpha_s$), shows  no significant
difference between the MLLA and NMLLA solutions.
\begin{figure}[ht]
\begin{center}
  \includegraphics[height=7cm,width=8cm]{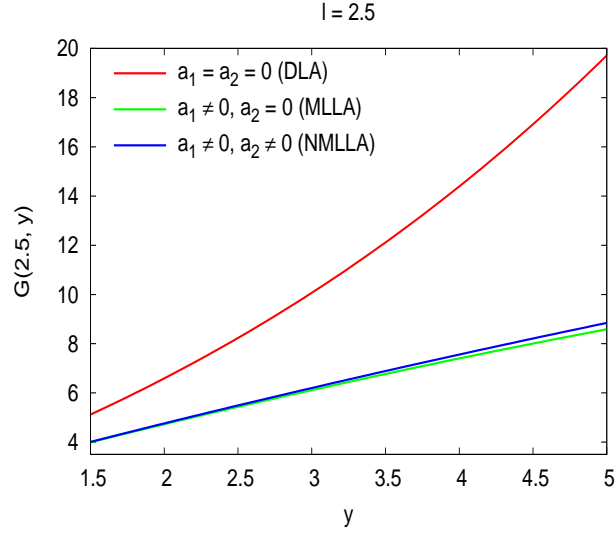}
\caption{\label{fig:SIS7.5} Single inclusive spectrum at fixed
$\alpha_s$ as a function of
$y$ at $Y_{\Theta_0}=7.5$ and $\ell=2.5$.}
\end{center}
\end{figure}
We can therefore safely use the exact MLLA solution (\ref{eq:ifD})
to compute the NMLLA inclusive $\kt$-distribution.
Likewise, the logarithmic derivatives $\psi_\ell=G_\ell/G$ and
$\psi_y=G_y/G$
\begin{equation}
\psi_\ell(\ell,y)=\gamma_0\sqrt{\frac{y}{\ell}}-\frac12a_2\gamma_0^3
\left(\frac{y}{\ell}\right)^{3/2},\qquad
\psi_y(\ell,y)=\gamma_0\sqrt{\frac{\ell}
{y}}-a_1\gamma_0^2+\frac32a_2\gamma_0^3
\sqrt{\frac{y}{\ell}},
\end{equation}
which are used to evaluate two-particle correlation, are displayed
on Fig.~\ref{fig:logder} as a function of $\ell= Y_\Theta -y$.
\begin{figure}[ht]
\begin{center}
  \includegraphics[height=7cm,width=8cm]{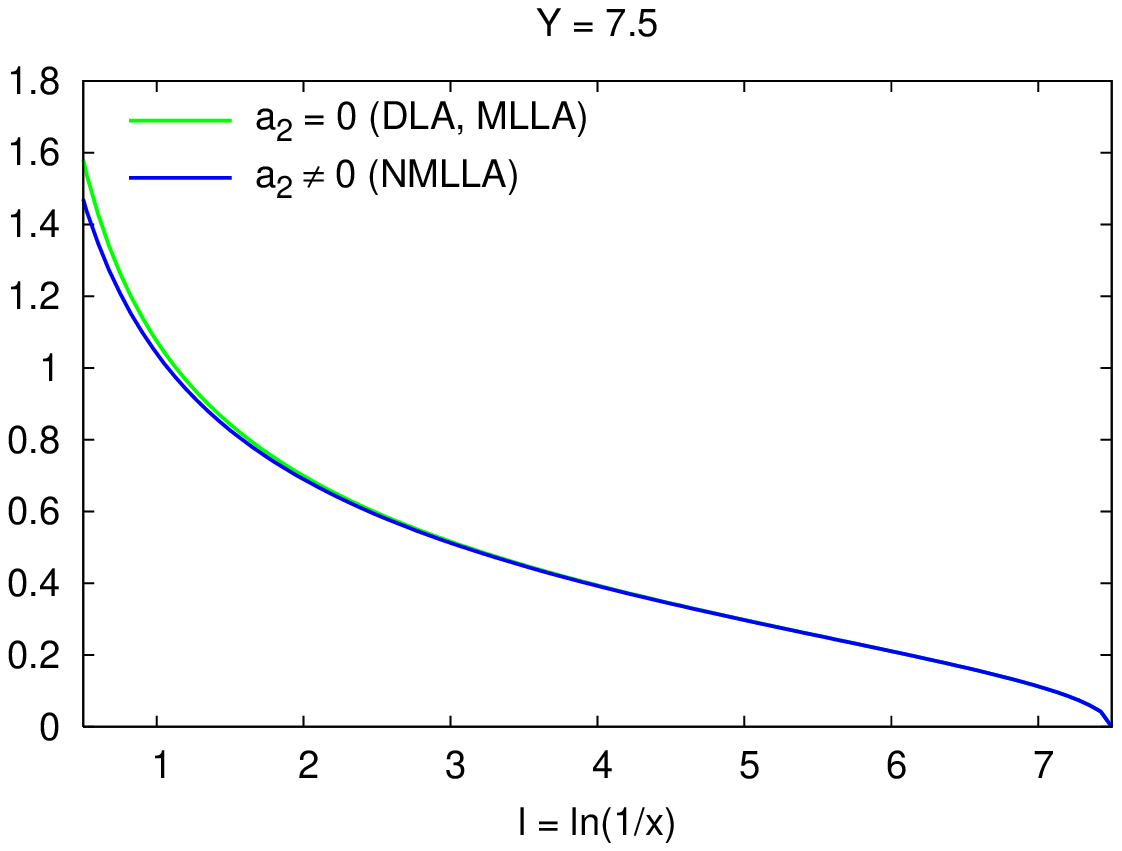}
\hskip  1cm
  \includegraphics[height=7cm,width=8cm]{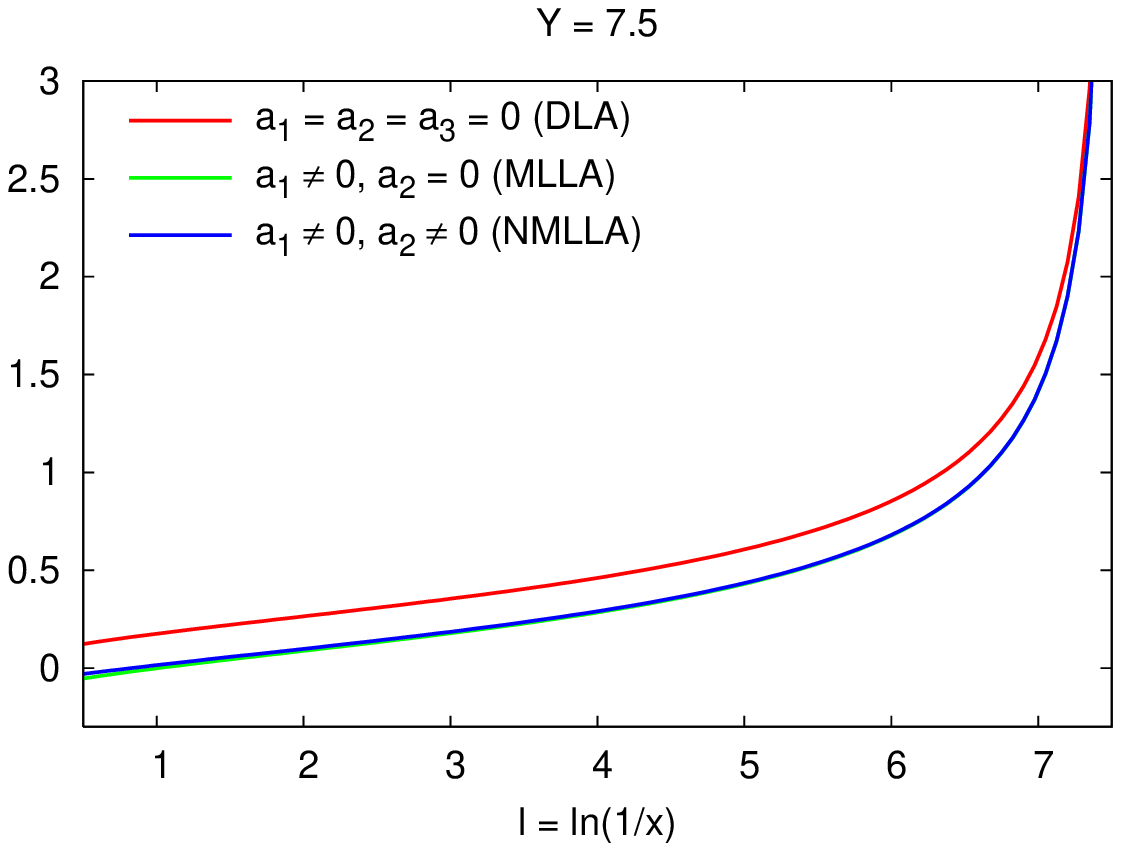}
\caption{\label{fig:logder} Logarithmic derivatives $\psi_\ell$ (left) and
$\psi_y$ (right) of the inclusive
spectrum $G(\ell,Y_{\Theta_0})$ at $Y_{\Theta_0}=7.5$.}
\end{center}
\end{figure}
There, again, the difference between MLLA and NMLLA
is negligible, such that
the exact MLLA expression of the single inclusive distribution can be taken
as a good approximation in the evaluation of NMLLA two-particle
correlations.

%%%%%%%%%%%%%%%%%%%%%%%%%%%%%%%%%%%%%%%%%%%%%%%%%%%%%%%%%%%%%%%%%%%
\section{Second derivative of the spectrum
$\boldsymbol{G_{\ell\ell}}$ at $\boldsymbol{\lambda=0}$}
\label{section:2ndder}
%%%%%%%%%%%%%%%%%%%%%%%%%%%%%%%%%%%%%%%%%%%%%%%%%%%%%%%%%%%%%%%%%%%

The expression of the second derivative of the inclusive spectrum for a
gluon jet reads
\begin{eqnarray}
G_{\ell\ell}(\ell, y)&\equiv&G(\ell,y)(\psi_{g,\ell}^2+\psi_{g,\ell\ell})(\ell,y)
=\frac2{\ell+y}\left(G_{\ell}(\ell,y)-
\frac1{\ell+y}G(\ell,y)\right)\\
&+&\frac{\Gamma(B)}{\beta_0}\int_{-\frac{\pi}{2}}^{\frac{\pi}{2}}
\frac{\dd\alpha}{\pi}e^{-(B-2)\alpha}
\left[\frac1{\beta_0^2}{\cal F}_{B+2}+\frac6{\beta_0(\ell+y)}
\sinh\alpha{\cal F}_{B+1}+\frac8{(\ell+y)^2}\sinh^2\alpha{\cal F}_B\right].\nonumber
\end{eqnarray}
$I_B$ is the modified Bessel function of the first kind.
$G_{\ell\ell}$ is displayed
in Fig.~\ref{fig:Gll} as a function of $y$ for three values
of $\ell$. We notice that it is negative at small values of
$y$ and gets positive at larger $y$.
\begin{figure}[ht]
\begin{center}
  \includegraphics[height=7cm,width=8cm]{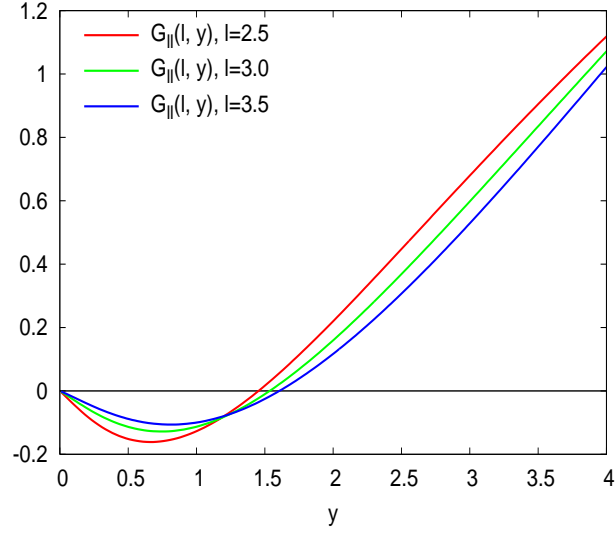}
\caption{\label{fig:Gll} Second derivative of the 
single inclusive spectrum as a function of
$y$ at $Y_{\Theta_0}=7.5$.}
\end{center}
\end{figure}

%%%%%%%%%%%%%%%%%%%%%%%%%%%%%%%%%%%%%%%%%%
\section{Scaling violations}
\label{section:scalviol}
%%%%%%%%%%%%%%%%%%%%%%%%%%%%%%%%%%%%%%%%%%

Varying $uE\Theta$ to $E\Theta$ in the argument of the DGLAP splitting
function $D_{A_0}^A$ in (\ref{eq:befscal}) yields corrections of relative
magnitude ${\cal O}(\alpha_s)$ which are accordingly NMLLA.
We need to estimate
\begin{equation}\label{eq:DA0}
\int_x^1du\,u\,D_{A_0}^A(u,E\Theta_0,uE\Theta)\equiv\int_x^1du\,u\,D_{A_0}^A\Big(u,\xi(u)\Big)
\end{equation}
where
$$
\xi(u)=\frac1b\ln\left(\frac{\ln\frac{E\Theta_0}{\Lambda}}
{\ln\frac{uE\Theta}{\Lambda}}\right) \quad ; \quad b=4N_c\beta_0.
$$
Writing
$$
D_{A_0}^A(u,E\Theta_0,uE\Theta)
=e^{\displaystyle{\ln u\frac{d}{d\ln(E\Theta)}}}
D_{A_0}^A(u,E\Theta_0,E\Theta),
$$
where
$$
\frac{d}{d\ln(E\Theta)}
=\frac{d}{d\xi}\frac{d\xi}{d\ln(E\Theta)}
=-\frac1b\frac1{\ln(E\Theta)}\frac{d}{d\xi},
$$
leads to
$$
e^{\displaystyle{\ln u\frac{d}{d\ln(E\Theta)}}}
=1-\frac1b\frac{\ln u}{\ln(E\Theta)}\frac{d}{d\xi}
+{\cal O}(\alpha_s^2).
$$
Finally, (\ref{eq:DA0}) can be approximated by

\vbox{
\begin{eqnarray*}
\int_x^1du\, u\, D_{A_0}^A(u,E\Theta_0,uE\Theta)
&\approx& \int_x^1du\, u\, D_{A_0}^A(u,E\Theta_0,E\Theta)\cr
&& -\frac1b\frac1{\ln\frac{E\Theta}{\Lambda}}
\int_x^1 du\, u\,\ln u\frac{\partial D_{A_0}^A}{\partial \xi}
\Big(u,\xi(u=1)\Big)+{\cal O}(\alpha_s^2).
\end{eqnarray*}
}

We can now estimate the order of magnitude of this correction,
taking, for example, the analytic form of $D_q^q(u)$
(non-singlet combination of quark distributions) in the $u\to1$
limit~\cite{Basics,ESW}
$$
D_q^q(u)\sim(1-u)^{-1+4C_F\xi}
$$
with $\xi=\xi(u=1)=\frac1b\ln\left(\frac{Y_{\Theta_0}}{Y_\Theta}\right)$.
We need to  compare 
$$
I=\int_x^1du\, u(1-u)^{-1+4C_F\xi}
$$
with
$$
\delta I=\frac{4C_F}{b(\ell+y+\lambda)}
\int_x^1 du\, u\,\ln u\,\ln(1-u)\,(1-u)^{-1+4C_F\xi(u=1)}.
$$
Taking, for instance, $\xi(Y_{\Theta_0}=6,Y_\Theta=3)=0.08$, which
 is a typical value at LEP or Tevatron energy scales,
one finds $\delta I/I\approx 0.04$. When $Y_\Theta\to Y_{\Theta_0}$,
this ratio tends very fast to $0$, such that the role of this correction
at larger $k_\perp$ is negligible. 

%%%%%%%%%%%%%%%%%%%%%%%%%%%%%%%%%%%%%%%%%%%%%%%%%%%
\section{Exact versus approximate NMLLA color currents}
\label{section:exactcc}
%%%%%%%%%%%%%%%%%%%%%%%%%%%%%%%%%%%%%%%%%%%%%%%%%%%

Using (\ref{eq:ratioqg}) yields the following exact (in the
sense that it takes into account all subleading corrections
coming from (\ref{eq:ratioqg})) expression for the color
currents

\vbox{
\begin{eqnarray}
\langle C\rangle_i^{\rm exact}(\ell,y)&=&\langle C\rangle_i^{\rm approx}(\ell,y)+\langle u\rangle_i^q(\ell,y)
\frac{C_F}{N_c}\left[\Big(a_1-\tilde a_1\Big)\psi_{g,\ell}(\ell,y)\right.\cr
&+&\left.\left
(a_1\Big(a_1-\tilde a_1\Big)+\tilde a_2-a_2\right)\left(\psi_{g,\ell}^2+
\psi_{g,\ell\ell}\right)(\ell,y)
\right]\cr
&+&\delta\langle u\rangle_i^q(\ell,y)
\frac{C_F}{N_c}\Big(a_1-\tilde a_1\Big)\psi_{g,\ell}^2(\ell,y)+{\cal O}(\gamma_0^2),
\end{eqnarray}
}

where $i=g,q$, and  $\langle u\rangle_i^q$ is given in \cite{RPR2}.
The approximate expression, used in the core of the paper,
only keeps $(C_F/N_c)\ G$ in (\ref{eq:ratioqg}).

On Fig.~\ref{fig:nmllacc}, the exact and approximate color currents are
shown to be in practice indistinguishable, which justifies the use of the
latter in the core of the paper.
\begin{figure}[ht]
\begin{center}
  \includegraphics[height=7cm,width=8cm]{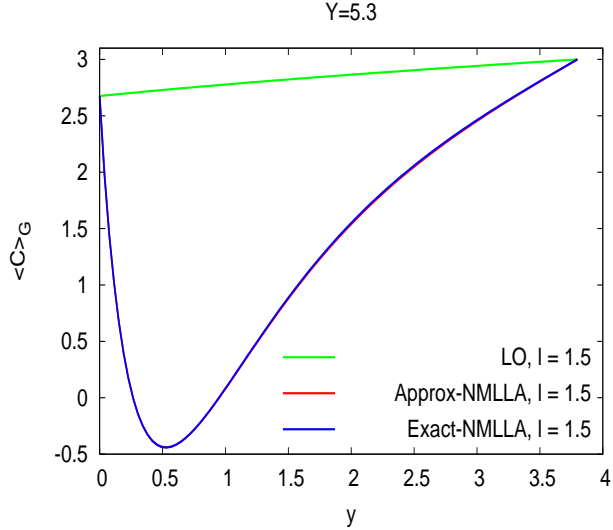}
\caption{\label{fig:nmllacc} Approximate (used in the core if the
paper) and exact color currents for a gluon jet.}
\end{center}
\end{figure}

We also performed the following approximation to evaluate the color current:
\begin{equation}\label{eq:varphi}
\varphi_{\ell}\equiv\psi_{q,\ell} = \psi_{g,\ell}\left[ 1 + \Big(a_1 - \tilde a_1\Big)
\left( \frac{G_{\ell\ell}}{G_\ell} - \frac{G_\ell}{G}\right)\right] + {\cal O}
(\gamma_0^2)\approx\psi_{g,\ell}\equiv\psi_{\ell}.
\end{equation}
In fact, $(a_1-\tilde a_1)\approx 0.18$ and $G_{\ell\ell}/G_\ell\sim G_\ell/G=\cO{\gamma_0}$.
These approximations were also made and numerically tested in \cite{PerezMachet,RPR2} (see,
for example, Fig.~20 in \cite{RPR2}). 

%%%%%%%%%%%%%%%%%%%%%%%%%%%%%%%%%%%%%%%%%%%%%%%%%%%%%%%%%%%%
\section{Expression of
$\boldsymbol{\delta \langle C\rangle_{\tt A_0}^{\rm NMLLA-MLLA}}$}
\label{section:delta2C}
%%%%%%%%%%%%%%%%%%%%%%%%%%%%%%%%%%%%%%%%%%%%%%%%%%%%%%%%%%%%

A straightforward calculation that follows from (\ref{eq:delta2})
gives respectively, for the gluon and quark jets, the following results:
{\footnotesize
\begin{eqnarray*}
\delta
\langle C\rangle_{g}^{\rm NMLLA-MLLA}&=&\frac12\left[12.7394\left(-1.49751-\frac19
\ln\frac{\ell+y+\lambda}{Y_{\Theta_0}+\lambda}\right)
\left(-0.260721 - \frac19
\ln\frac{\ell+y+\lambda}{Y_{\Theta_0}+\lambda}\right)
\left(\frac{\ell+y+\lambda}{Y_{\Theta_0}+\lambda}\right)
^{\frac{50}{81}}\right.\cr
 &+&\left.
  356.711\left(-0.0369486-\frac19
\ln\frac{\ell+y+\lambda}{Y_{\Theta_0}+\lambda}\right)\left(0.377382
-\frac19\ln\frac{\ell+y+\lambda}{Y_{\Theta_0}+\lambda}\right)\right]
(\psi_{g,\ell}^2+\psi_{g,\ell\ell}),\\\notag\\
\delta\langle C\rangle_{q}^{\rm NMLLA-MLLA}&=&
\frac12\left[-22.6479
\left(-0.936071-\frac19\ln\frac{\ell+y+\lambda}{Y_{\Theta_0}+\lambda}\right)
\left(0.164816-\frac19\ln\frac{\ell+y+\lambda}{Y_{\Theta_0}+\lambda}\right)
\left(\frac{\ell+y+\lambda}{Y_{\Theta_0}+\lambda}\right)
^{\frac{50}{81}}\right.\cr
&+&\left.
356.711\left(-0.0635496-
\frac19\ln\frac{\ell+y+\lambda}{Y_{\Theta_0}+\lambda}\right)\left(0.154028-
\frac19\ln\frac{\ell+y+\lambda}{Y_{\Theta_0}+\lambda}\right)\right]
(\psi_{g,\ell}^2+\psi_{g,\ell\ell}),
\end{eqnarray*}}
where the expression and behavior of the function
$(\psi_{g,\ell}^2+\psi_{g,\ell\ell})$
 are given in Appendix~\ref{section:2ndder}.

%%%%%%%%%%%%%%%%%%%%%%%%%%%%%%%%%%%%%%%%%%
\section{Fixing and varying $\boldsymbol{\ell_{\rm min}}$}
\label{section:positivity}
%%%%%%%%%%%%%%%%%%%%%%%%%%%%%%%%%%%%%%%%%%
%
Our small $x$ calculation cannot be trusted below a certain
$\lmin$, otherwise, as shown in Fig.~\ref{fig:doublediff}, 
$\dd^2N/\dd\ell \dd{y}$ gets negative in the perturbative domain.
\begin{figure}[ht]
\begin{center}
  \includegraphics[height=7cm,width=8cm]{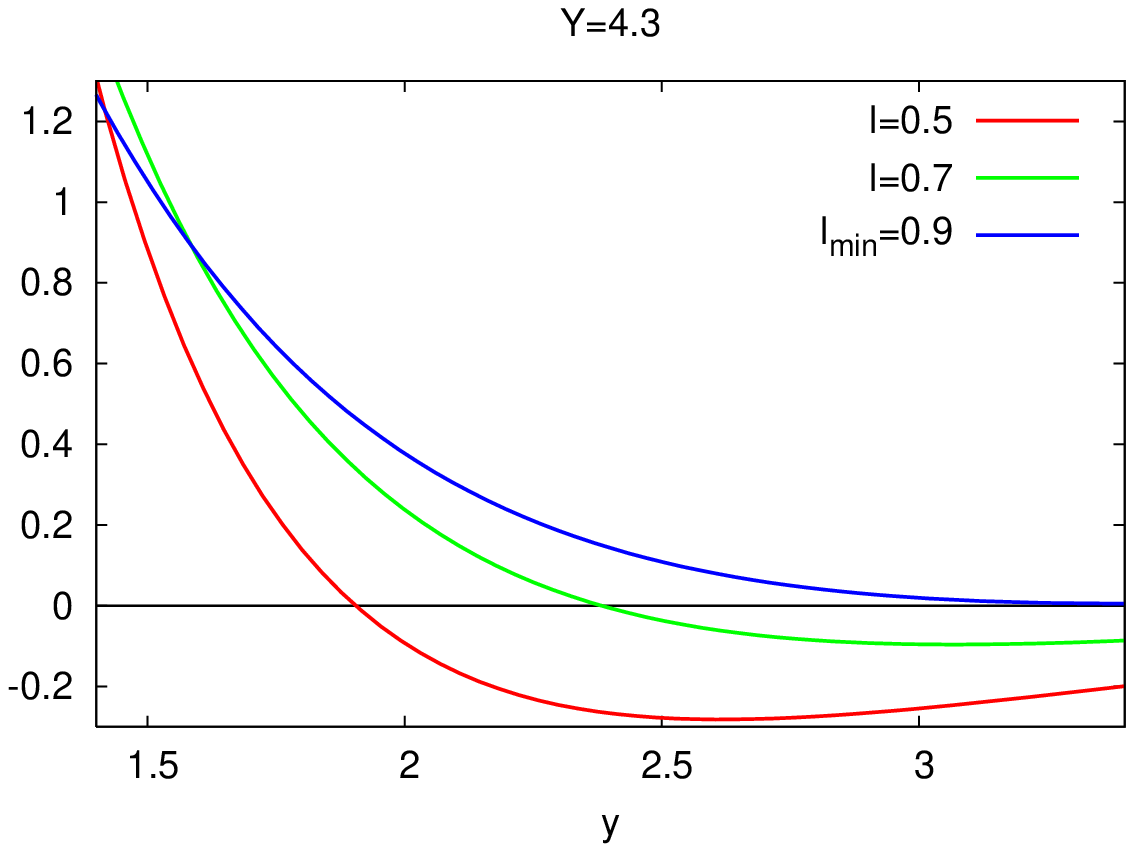}
\hskip 1cm
  \includegraphics[height=7cm,width=8cm]{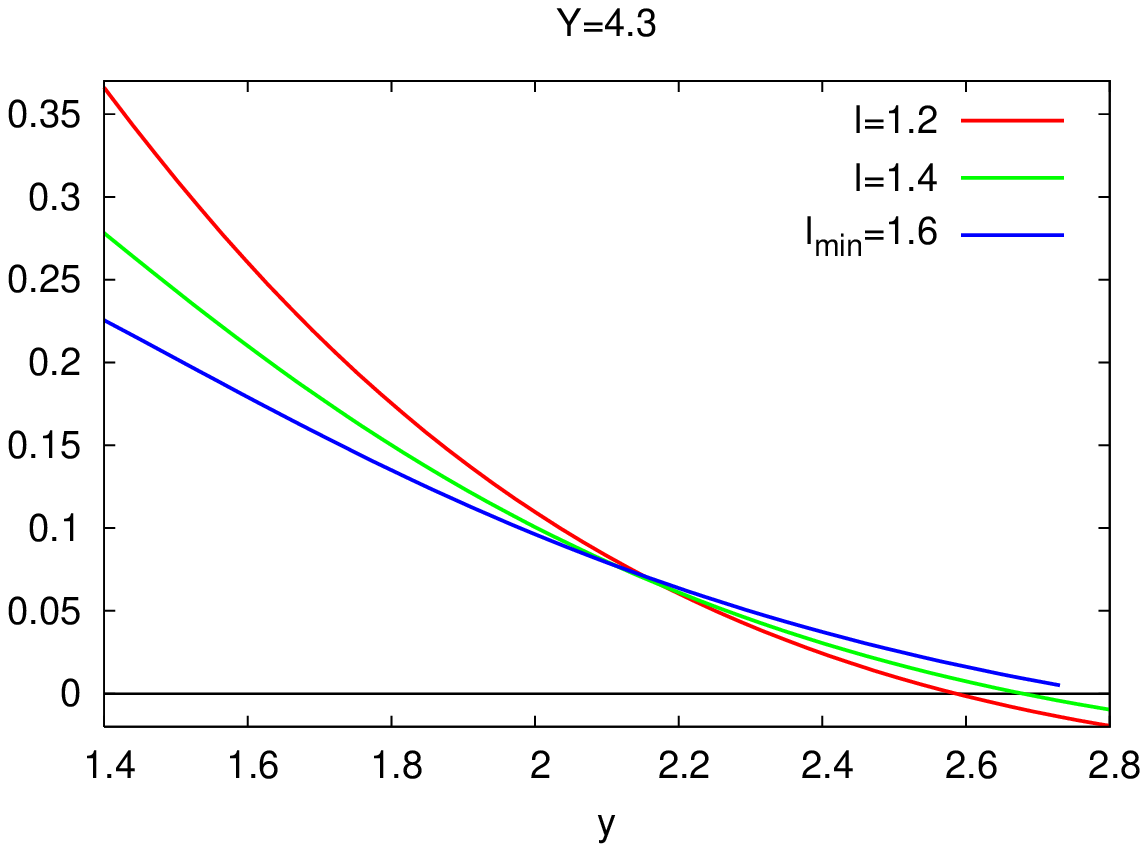}
\caption{\label{fig:doublediff} $\frac{d^2N}{d\ell dy}$ for a gluon jet (left)
and a quark jet (right) as a function  of $y$ for three values of $\ell$.}
\end{center}
\end{figure} 
We give in Table~1 values of $\lmin$ as they come out from the
requirement of positivity, for different values of the jet hardness (and
the corresponding maximal values of $y$).
\begin{table}[htb]
\begin{center}
\begin{tabular}{cccccc}
\hline
\hline
$Q$ (GeV) & $Y_{\Theta_0}$
& $\lmin^{\rm g}$ & $y_{\rm max}^{\rm g}$ 
& $\lmin^{\rm q}$ & $y_{\rm max}^{\rm q}$\\ \hline
&&&&&\\
19 (CDF)  & 4.3 & 0.9 & 3.4 & 1.6 & 2.7\\
27 (CDF)  & 4.7 & 1.0 & 3.8 & 1.7 & 3.1\\
37 (CDF)  & 5.0 & 1.0 & 4.1 & 1.8 & 3.4\\
50 (CDF)  & 5.3 & 1.1 & 4.4 & 1.9 & 3.7\\
68 (CDF)  & 5.6 & 1.1 & 4.7 & 2.0 & 4.0\\
90 (CDF)  & 5.9 & 1.2 & 5.0 & 2.0 & 4.3\\
119 (CDF) & 6.2 & 1.2 & 5.3 & 2.1 & 4.6\\
155 (CDF) & 6.4 & 1.3 & 5.4 & 2.2 & 4.7\\
450 (LHC) & 7.5 & 1.4 & 6.1 & 2.4 & 5.1\\ 
&&&&&\\
\hline\hline
\end{tabular}
\caption{Values of $\lmin$ and $y_{\rm max}$ for different values of the jet
hardness}
\label{table:CDF-NMLLA}
\end{center}
\end{table}
$\lmin$ is not an intrinsic
(physical) characteristic of the system under concern (gluon or quark jet),
it is only an ad-hoc parameter below which poor credibility can be attached
to the results.
One notices in Table~\ref{table:CDF-NMLLA} that, at a given $Q$, the $\lmin$ for a quark
jet is always larger than the one for a gluon jet; this only means that
our calculations can be pushed to larger $x$ for gluons than for quarks
without encountering problems of positivity.
The question then arises whether, in calculating the inclusive
$k_\perp$ distribution of a mixed jet, one should attach the same
$\lmin$ to each of its components, which can only be, of course, the
larger one, that is, the one of the quark component, or give to each
component its proper value of $\lmin$ as given in Table 1. The simple
answer to this question comes from the fact that the two choices give, in
practice, extremely close results. Deeper considerations on which
$\lmin$ should be chosen are thus irrelevant.

For the sake of completeness, we plot in Fig.~\ref{fig:lmin} the 
inclusive gluon $\kt$-distribution at $Y_{\Theta_0}=6$, for different values of
$\ell_{\rm min}^g$, both at MLLA (left) and NMLLA (right).
Changing $\ell_{\rm min}^g$ from $1$ to $1.5$ modifies the NMLLA spectrum by
no more than $20\%$ for $\log(\kt/1{\rm GeV})$=2.5. At MLLA, the dependence proves much more dramatic,
\begin{figure}[ht]
\begin{center}
\includegraphics[width=6.5cm,width=8cm]{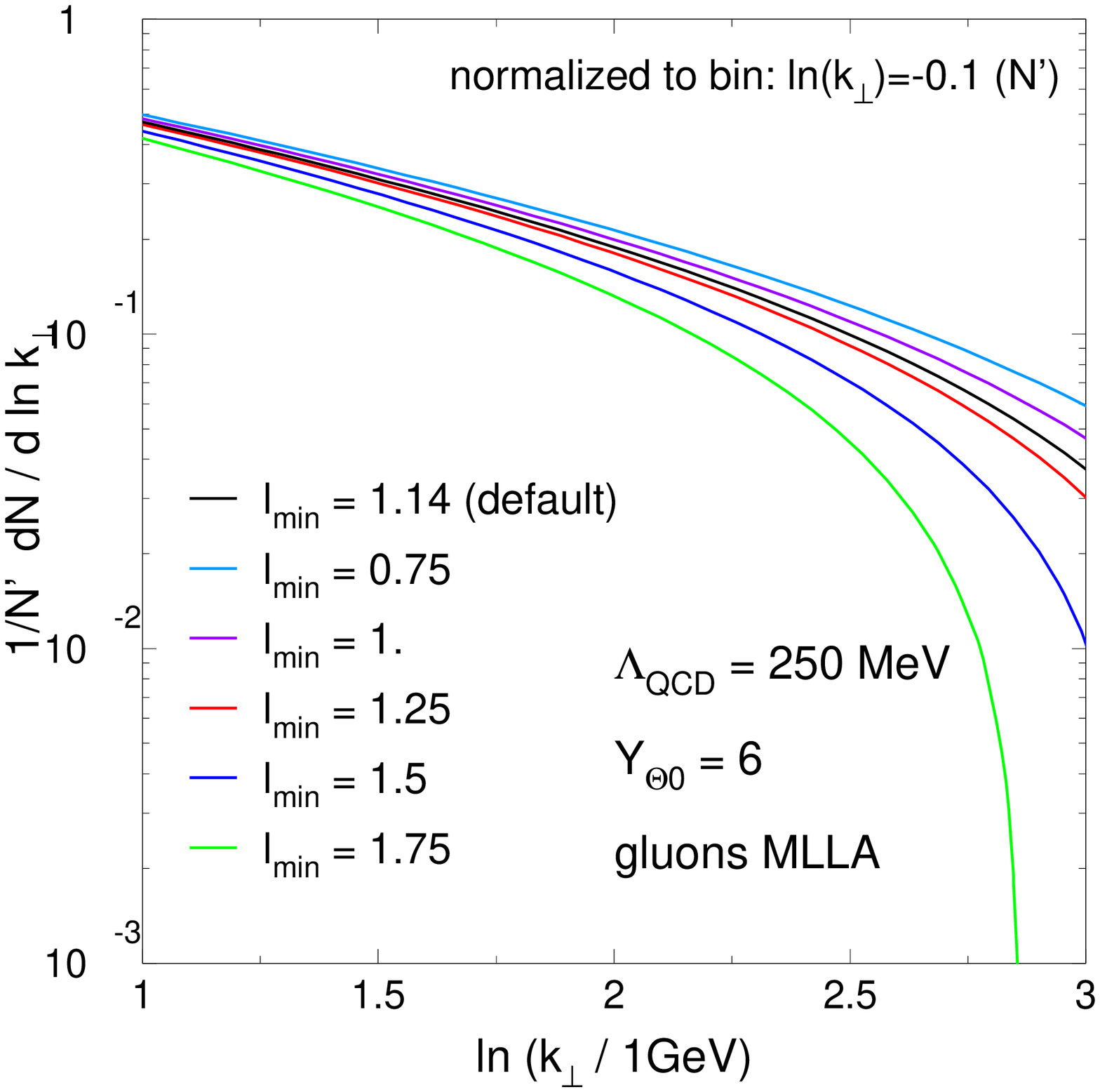}
\hskip 1cm
\includegraphics[width=6.5cm,width=8cm]{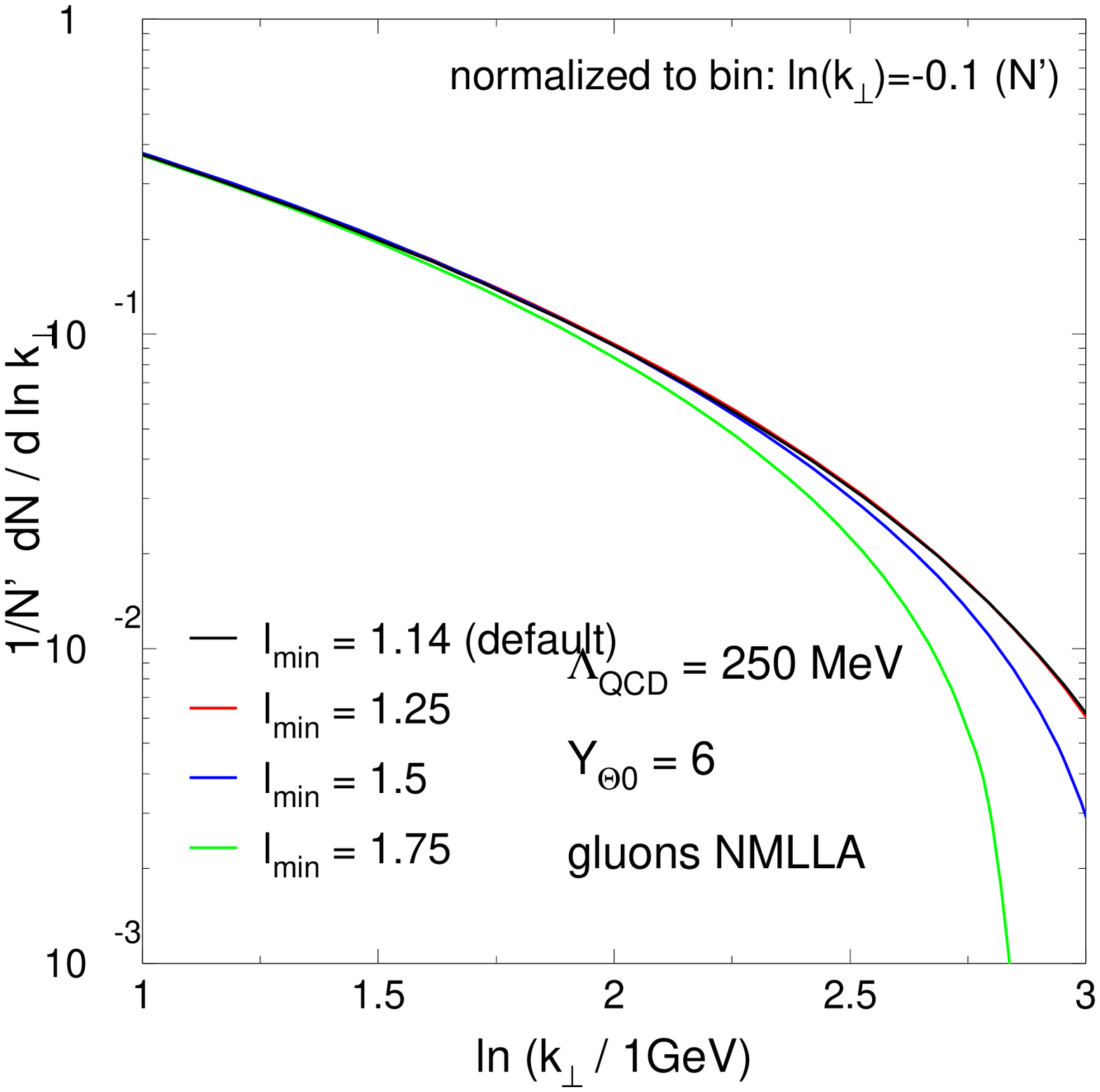}
\caption{The dependence of the inclusive gluon $\kt$ distribution at $Y_{\Theta_0}=6$ on $\ell_{\rm min}^g$.}
\label{fig:lmin}
\end{center}
\end{figure}
Like for the variation with $\lqcd$ in subsection
\ref{subsec:uncertain}, that the variations with $\lmin$ seem to increase
with $k_\perp$ is only an artifact due to the normalization at the first
bin.

%%%%%%%%%%%%%%%%%%%%%%%%%%%%%%%%%%%%%%%%%%%%%%%%%%%%%%%%%%%
\section{NMLLA terms neglected in \ref{subsection:expandeveq}}
\label{section:Q2}
%%%%%%%%%%%%%%%%%%%%%%%%%%%%%%%%%%%%%%%%%%%%%%%%%%%%%%%%%%%

The approximations we have made in (\ref{eq:approxim}) needs
further comments; one has indeed to replace $Q$ and
$Q^{(2)}$ by the full MLLA expressions
\begin{eqnarray}\label{eq:mllaqspec}
Q=\frac{C_F}{N_c}\left[1+(a_1-\tilde a_1)\psi_\ell\right]G+{\cal O}(\gamma_0^2)
\end{eqnarray},
\begin{eqnarray}\label{eq:mllaq2spec}
\displaystyle\frac{\displaystyle\frac{Q^{(2)}}{Q_1Q_2}-1}{\displaystyle\frac{G^{(2)}}{G_1G_2}-1}=
\frac{N_c}{C_F}\left[1+(b_1-a_1)(\psi_{\ell_1}+\psi_{\ell_2})
\frac{1+\Delta}{2+\Delta}\right]+{\cal O}(\gamma_0^2)
\end{eqnarray}
respectively. $b_1$ is defined in (\ref{eq:eqb}) and 
$$
\Delta=\gamma_0^{-2}
\Big(\psi_{1,\ell}\psi_{2,y}+\psi_{1,y}\psi_{2,\ell}\Big)={\cal O}(1),\qquad
\psi_\ell={\cal O}(\gamma_0).
$$

(\ref{eq:mllaq2spec}) was obtained in \cite{RPR2} 
and displayed later in \cite{RPR3}.
Working out the structure of (\ref{eq:mllaq2spec})
after we have inserted (\ref{eq:mllaqspec}), leads to
\begin{eqnarray}
Q^{(2)}&=&\frac{C_F}{N_c}G^{(2)}+\frac{C_F}{N_c}
\left(\frac{C_F}{N_c}-1\right)G_1G_2+\frac{C_F}{N_c}(b_1-a_1)
(\psi_{\ell_1}+\psi_{\ell_2})\frac{1+\Delta}{2+\Delta}
(G^{(2)}-G_1G_2)\notag\\
&+&\frac{C_F}{N_c}(a_1-\tilde a_1)(\psi_{\ell,1}+\psi_{\ell,2})
(G^{(2)}-G_1G_2)+\frac{C_F^2}{N_c^2}(a_1-\tilde a_1)(\psi_{\ell_1}+\psi_{\ell_2})
G_1G_2+{\cal O}(\gamma_0^2).\label{eq:fullmllaq2}
\end{eqnarray}
As  already mentioned in \cite{RPR3}, the coefficient ($b_1-a_1$),
which is color suppressed, is $\simeq10^{-2}$, 
$\psi_\ell\simeq10^{-1}$ and 
$\frac{1+\Delta}{2+\Delta}\simeq\frac{3}{4}$. Thus, the whole
correction is roughly $\simeq10^{-4}$. This is why it is not taken into
account here, which allows for analytic results. Introducing
the terms of (\ref{eq:fullmllaq2}) $\propto(a_1-\tilde a_1)$ in the r.h.s. of 
(\ref{eq:nmllacorrg2}) provides extra terms 
$$
\ldots+\frac{2n_fT_R}{3N_c}
\frac{C_F}{N_c}(a_1-\tilde a_1)(\psi_{\ell,1}+\psi_{\ell,2})
\gamma_0^2(G^{(2)}-G_1G_2)
+\frac{2n_fT_R}{3N_c}\frac{C_F^2}{N_c^2}(a_1-\tilde a_1)
\gamma_0^2(G_1G_2)_\ell
$$ 
which add to the r.h.s. of (\ref{eq:nmllagcorreq}). They are both,
in particular, color suppressed, the first one by a factor $\propto1/N_c^2$
and the second one, by $\propto1/N_c^3$.
Thus, for example, taking $\psi_\ell\simeq10^{-1}$,
taking into account that $2n_fT_R/3=1$ for $n_f=3$,
the coefficient $a_2$ defined in (\ref{eq:a2}) and which also
appears in the r.h.s. of (\ref{eq:nmllageq1})
would be modified to the close value 
$a_2\approx0.07$. Finally, since in the above
$$
\ldots+\frac{2n_fT_R}{3N_c}\frac{C_F}{N_c}\left(\frac{C_F}{N_c}
-1\right)(a_1-\tilde a_1)\gamma_0^2(G_1G_2)_\ell\approx
-0.01\times\gamma_0^2(G_1G_2)_\ell,
$$ 
$b_2$ defined in (\ref{eq:eqb}) would be changed to the
value $b_2\approx0.17$,
which only represents a $1\%$ variation.

The derivatives of (\ref{eq:mllaqspec}) and (\ref{eq:mllaq2spec})
with respect to $\ell$ are therefore respectively approximated by
\begin{eqnarray}
Q_\ell&=&\frac{C_F}{N_c}G_\ell+{\cal O}(\gamma_0^2),\cr
Q^{(2)}_\ell&=&\frac{C_F}{N_c}G^{(2)}_\ell+\frac{C_F}{N_c}
\left(\frac{C_F}{N_c}-1\right)(G_1G_2)_\ell+{\cal O}(\gamma_0^2),
\end{eqnarray}
because the inclusion of higher order contributions (coming from
the derivatives of the above ${\cal O}(\gamma_0)$ terms) in the
non-singular parts of the equations (such as (\ref{eq:nmlla3}), 
(\ref{eq:nmllagq}), (\ref{eq:corrnmlla2}) and (\ref{eq:nmllacorrg3}))
would yield corrections  beyond the precision of our
approach.

%%%%%%%%%%%%%%%%%%%%%%%%%%%%%%%%%%%%%%%%%%%%%%%%%%%%%%%%%%%
\section{Logarithmic derivatives of the inclusive spectrum}
\label{sec:logderiv}
%%%%%%%%%%%%%%%%%%%%%%%%%%%%%%%%%%%%%%%%%%%%%%%%%%%%%%%%%%%

The logarithmic derivatives of $G$, that are used in
Section~\ref{subsection:GC}, read
\begin{eqnarray*}
G_{\ell}^{(2)}&=&{\cal C}_gG_1G_2\left(\chi_\ell+\psi_{1,\ell}+\psi_{2,\ell}\right),
\quad
(G_1G_2)_\ell=G_1G_2(\psi_{1,\ell}+\psi_{2,\ell}),\\
G_{\ell\ell}^{(2)}&=&{\cal C}_gG_1G_2\left[
\left(\chi_{\ell}+\psi_{1,\ell}+
\psi_{2,\ell}\right)^2+\chi_{\ell\ell}+\psi_{1,\ell\ell}+\psi_{2,\ell\ell}\right],\\
(G_1G_2)_{\ell\ell}&=&G_1G_2\left[(\psi_{1,\ell}+\psi_{2,\ell})^2+\psi_{1,\ell\ell}
+\psi_{2,\ell\ell}\right].
\end{eqnarray*}
The functions introduced in (\ref{eq:delta}) and (\ref{eq:delta3}) are the following:
\begin{equation}
  \label{eq:fi}
  \begin{split}
f_1(\ell_1,\ell_2;\eta)&=(\psi_{1,\ell}+\psi_{2,\ell}+\chi_{\ell})^2+
\psi_{1,\ell\ell}+\psi_{2,\ell\ell}-\beta_0\gamma_0^2(\psi_{1,\ell}+\psi_{2,\ell}+\chi_\ell)
+\chi_{\ell\ell}={\cal O}(\gamma_0^2),\\
f_2(\ell_1,\ell_2;\eta)&=(\psi_{1,\ell}+\psi_{2,\ell})^2+\psi_{1,\ell\ell}+\psi_{2,\ell\ell}
-\beta_0\gamma_0^2(\psi_{1,\ell}+\psi_{2,\ell})={\cal O}(\gamma_0^2),\\
f_3(\ell_1,\ell_2;\eta)&=
2\psi_{1,\ell}\psi_{2,\ell}+2\chi_{\ell}(\psi_{1,\ell}+\psi_{2,\ell})
+\chi_{\ell\ell}+\chi_{\ell}^2
-\beta_0\gamma_0^2\chi_\ell={\cal O}(\gamma_0^2),\\
f_4(\ell_1,\ell_2;\eta)&=\psi_{i,\ell}^2+\psi_{i,\ell\ell}
-\beta_0\gamma_0^2\psi_{i,\ell}={\cal O}(\gamma_0^2).
  \end{split}
\end{equation}
%

%%%%%%%%%%%%%%%%%%%%%%%%%%%%%%%%%%%%%%%%%%%%%%%%%%%%%%%%%%%%%%%%%%%%%%%%%%%

%%%%%%%%%%%%%%%%%%%%%%%%%%%%%%%%%%%%%%%%%%%%%%%%%%%%%%%%%%%%%%%%%%%%%%%%%%%
\end{document}